\documentclass[twocolumn]{aastex62}

\newcommand{\arcs}{$^{\prime\prime}$} 
\newcommand{\beq}{\begin{equation}\begin{aligned}}
\newcommand{\eeq}{\end{aligned}\end{equation}}
\newcommand\ddfrac[2]{\frac{\displaystyle #1}{\displaystyle #2}}
\newcommand{\msun}{M$_\odot$}

\usepackage{scalerel,amssymb}
\usepackage{amsmath}

\usepackage[]{url}
\graphicspath{{./}{figures/}}

\accepted{\today}
\submitjournal{ApJ}

\shorttitle{Dwarf satellite systems in the Local Volume.}
\shortauthors{Carlsten et al.}

\begin{document}

\title{Luminosity Functions and Host-to-Host Scatter of Dwarf Satellite Systems in the Local Volume}

\correspondingauthor{Scott G. Carlsten}
\email{scottgc@princeton.edu}

\author[0000-0002-5382-2898]{Scott G. Carlsten}
\affil{Department of Astrophysical Sciences, 4 Ivy Lane, Princeton University, Princeton, NJ 08544}

\author{Jenny E. Greene}
\affil{Department of Astrophysical Sciences, 4 Ivy Lane, Princeton University, Princeton, NJ 08544}

\author[0000-0002-8040-6785]{Annika H. G. Peter}
\affiliation{Department of Physics, The Ohio State University, 191 W. Woodruff Ave., Columbus OH 43210, USA}
\affiliation{Department of Astronomy, The Ohio State University, 140 W. 18th Ave., Columbus OH 43210, USA}
\affiliation{Center for Cosmology and AstroParticle Physics (CCAPP), The Ohio State University, Columbus, OH 43210, USA}

\author[0000-0002-1691-8217]{Rachael L. Beaton}
\altaffiliation{Hubble Fellow}
\affiliation{Department of Astrophysical Sciences, 4 Ivy Lane, Princeton University, Princeton, NJ 08544}
\affiliation{The Observatories of the Carnegie Institution for Science, 813 Santa Barbara St., Pasadena, CA~91101\\}

\author[0000-0003-4970-2874]{Johnny P. Greco}
\altaffiliation{NSF Astronomy \& Astrophysics Postdoctoral Fellow}
\affiliation{Center for Cosmology and AstroParticle Physics (CCAPP), The Ohio State University, Columbus, OH 43210, USA}

\begin{abstract}
Low-mass satellites around Milky Way (MW)-like galaxies are important probes of small scale structure and galaxy formation. However, confirmation of satellite candidates with distance measurements remains a key barrier to fast progress in the Local Volume (LV). We measure the surface brightness fluctuation (SBF) distances to recently cataloged candidate dwarf satellites around 10 massive hosts within $D<12$ Mpc to confirm association. The satellite systems of these hosts are complete and mostly cleaned of contaminants down to $M_g{\sim}-9$ to $-10$, within the area of the search footprints. Joining this sample with hosts surveyed to comparable or better completeness in the literature, we explore how well cosmological simulations combined with common stellar to halo mass relations (SHMR) match observed satellite luminosity functions in the classical satellite luminosity regime. Adopting a SHMR that matches hydrodynamic simulations, the predicted overall satellite abundance agrees well with the observations. The MW is remarkably typical in its luminosity function amongst LV hosts. Contrary to recent results, we find that the host-to-host scatter predicted by the model is in close agreement with the scatter between the observed systems, once the different masses of the observed systems are taken into account. However, we find significant evidence that the observed systems have more bright and fewer faint satellites than the SHMR model predicts, necessitating a higher normalization of the SHMR around halo masses of $10^{11}$ \msun\ than present in common SHMRs. These results demonstrate the utility of nearby satellite systems in inferring the galaxy-subhalo connection in the low-mass regime.
\end{abstract}

\section{Introduction}
For over two decades, the satellites of the Milky Way (MW) have been an important testing ground for the $\Lambda$CDM model of structure formation. Within the last few years, hydrodynamic simulations have achieved the resolution required to resolve the formation of the bright MW `classical' ($M_*\gtrsim10^5$~\msun) satellites. Results from the APOSTLE \citep{sawala2016}, FIRE \citep{wetzel2016, gk2019}, and NIHAO \citep{buck2019} projects have demonstrated that the inclusion of baryonic physics leads to simulated satellite systems that have similar satellite numbers and internal kinematics as observed satellites of the MW and M31. Together, this ensemble of results suggests baryonic resolutions of the long-standing `Missing Satellites' \citep{klypin1999, moore1999} and `Too Big to Fail' Problems \citep{bk2011, bk2012} that are associated with dissipationless dark matter only (DMO) simulations of structure formation.

In recent years, solutions to the `Missing Satellites' problem has shifted to determining what stellar to halo mass relation (SHMR) can reproduce the observed abundance of dwarf satellites of the MW and comparing that to the SHMR predicted from hydrodynamic simulations of galaxy formation. The SHMR is an important observational benchmark and can help refine the importance of the physical processes involved in dwarf galaxy formation \citep[e.g.][]{agertz2020}. While SHMRs proposed in the literature reproduce the abundance of MW satellites and also appear to broadly agree with predictions from suites of hydrodynamic simulations \citep[e.g.][]{elvis, gk_2017}, there is still significant uncertainty on the details of the SHMR over the mass regime of the MW classical satellites \citep{gk_2017, read2017, wheeler2019}.

However, by only considering the dwarf satellites of the MW (and sometimes M31), there is the risk of {\it over-tuning} the models to reproduce the abundance and properties of the MW satellites. There is still no consensus on what a `normal' satellite system is and, thus, no way of ascertaining if the MW satellite system is abnormal. Therefore, there is a strong motivation to study the satellites of host galaxies other than the MW (and M31). We would be able to define, for the first time, what a `normal' satellite system is for MW-like galaxies. By probing MW-analogs, the host-to-host scatter in the satellite systems can be quantified. The host-to-host scatter in observed satellite systems is sensitive to both the statistics of the DM subhalo populations around MW-like hosts and also to the stochasticity of galaxy formation on these small scales. Additionally, by broadening the range of properties spanned by the set of MW analogs, both in terms of mass and host environment, the effect of these on the satellite systems can be explored.

Despite considerable investment by a number of groups, the challenge of both identifying and confirming the low-mass companions of $L^\star$ hosts has limited such study. To date, only a handful of galaxies have been surveyed at a level comparable to the classical satellites of the MW. These include M31 \citep{mcconnachie2009,martin2016, mcconnachie2018}, M81 \citep{chiboucas2009, chiboucas2013}, Centaurus A \citep{crnojevic2014, crnojevic2019, mullerCenA, muller2015, muller2019}, M94 \citep{smercina2018}, and M101 \citep{danieli101, bennet2017, sbf_m101, bennet2019, bennet2020}. 

Complementary to large-area searches of the nearest galaxies, the Satellites Around Galactic Analogs Survey \citep[SAGA;][]{geha2017} characterized the bright ($M_r<-12.3$) satellites of 8 MW-analogs at larger distances in the range $20<D<40$ Mpc (i.e. beyond the LV). At these distances, SAGA is only sensitive to roughly the brighter half of the classical satellite regime. However, the full survey will include $\sim3$ times more hosts than are available out to 10 Mpc. 

Despite the challenges facing these projects, their early observational results suggest that the host-to-host scatter between satellite systems of nearby MW-analogs is larger than anticipated by DMO $\Lambda$CDM simulations. More specifically, \citet{geha2017} noted that the scatter in satellite richness between hosts appeared to be larger than that predicted from abundance matching (AM) applied to DMO simulations. In a focused study, \citet{smercina2018} found only two satellites with $M_V<-9$ in the inner projected 150 kpc volume around the MW-analog M94 (compared to seven in this range found around the MW). They argue that common AM relations applied to the DMO results from the EAGLE Project \citep{schaye2015} have far too little scatter to explain M94's anemic satellite population. \citeauthor{smercina2018} suggest that significantly increasing the scatter in the stellar-halo mass relation (SHMR) could explain M94's satellite system. However, \citeauthor{smercina2018} considered all observed satellite systems together as `MW-analogs' whereas the different host stellar masses (and presumably halo masses) amongst the surveyed hosts will contribute to the observed scatter in satellite abundances. Thus, isolating the true host-to-host scatter requires careful controlling for the observed host mass when comparing to simulations, which is one goal of the current paper.

Measuring distances to individual candidates is a major challenge in this work. Many more hosts have been surveyed for candidates \citep[e.g.][]{mullerLeo, park2017, kim2011, park2019, byun2020} than have distance-confirmed satellite populations. For the systems with distance-confirmed satellites, the contamination from unrelated background galaxies can be quite high. \citet{sbf_m101} and \citet{bennet2019} found that the contamination fraction of the candidate satellite catalog of \citet{bennet2017} for M101 was ${\sim}80$\%. These contaminants will obfuscate the interpretation of host-to-host scatter in satellite number. While some science questions can be overcome by careful statistical subtraction of a background luminosity function (LF) \citep[e.g.][]{nierenberg2016}, it is the goal of our work to study satellite systems that are fully confirmed with distance measurements.

In this paper, we use surface brightness fluctuations (SBF) to confirm candidate satellites recently uncovered around nearby hosts in the LV in \citet{LV_cat}. SBF has been shown to be a very efficient distance measure for low surface brightness (LSB) dwarfs \citep[e.g.][]{jerjen_sculpt,jerjen_cenA,jerjen_field,jerjen_fornax,jerjen_virgo,mieske_fornax,mieske_calib, sbf_calib}. \citet{sbf_calib} determined that the SBF-based distances reproduced the tip of the red giant (TRGB) distances to dwarfs with ${\sim}15$\% accuracy, even for $\mu_0{\sim}26$ mag arcsec$^{-2}$ dwarfs. This precision is sufficient, in almost all cases, to distinguish a candidate as a real satellite or a background galaxy. SBF measurements can be performed using the same ground based data that was used to discover the candidate satellites, obviating the need for expensive follow-up (either \emph{HST} or spectroscopic).

Using this much expanded sample of cleaned satellite systems around the hosts in \citeauthor{LV_cat}, we produce the luminosity functions for low mass satellites. We compare these LFs to each other as well as drawing conclusions on the ensemble. We explore how well stellar to halo mass relations applied to modern cosmological simulations reproduce the observed LFs for a total of 12 systems, a sample large enough to examine the host-to-host scatter.

This paper is structured as follows: in \S\ref{sec:data} we describe the candidate sample and data reduction, in \S\ref{sec:sbf_methods} we outline the SBF methodology used for our study, in \S\ref{sec:sbf_r_calib} we derive an absolute SBF calibration for the $r$ band, and in \S\ref{sec:distances} we present our distance results. In \S\ref{sec:lfs}, we collate all of the satellite systems currently surveyed in the LV. In \S\ref{sec:models} we introduce the simulations and models that we use to compare with the data, in \S\ref{sec:results} we discuss the results of the comparison, and, finally, we conclude in \S\ref{sec:concl}. Readers interested primarily in the analysis of the satellite systems and comparison with models can skip to \S\ref{sec:lfs}.

\section{Data}
\label{sec:data}
The foundation of this paper is the catalog of candidate satellites from \citet{LV_cat}. \citet{LV_cat} searched for candidates satellites around 10 massive primaries in the LV using wide-field deep archival CFHT/MegaCam imaging. The surveyed hosts are: NGC 1023, NGC 1156, NGC 2903, NGC 4258, NGC 4565, NGC 4631, NGC 5023, M51, M64, and M104 (see Table 1 of \citeauthor{LV_cat} for characteristics of these hosts). The area and surface brightness completeness were heterogeneous but several of the hosts were nearly completely surveyed within a projected radius of 150 kpc. Through careful mock recovery tests, we determine that we are complete at $\gtrsim90$\% for satellites down to $\mu_{r,0}{\sim}26-26.5$ mag arcsec$^{-2}$.

For the SBF measurements in this paper, we use the same archival CFHT/MegaCam \citep{megacam} imaging data as used by \citet{LV_cat}. Either $g$ and $r$ or $g$ and $i$ band imaging is used, depending on the availability in the CFHT archive. Exposure times are characteristically ${\sim}1$ hour in each of the bands. The data reduction follows that in \citet{LV_cat} and we refer the reader to that paper for details. 

\citet{LV_cat} used the object detection algorithm of \citet{johnny2}, which is specifically optimized for low surface brightness (LSB) galaxies, to detect 155 candidate satellite galaxies around these 10 hosts. While the detection algorithm focused on LSB galaxies, \citet{LV_cat} also cataloged many high surface brightness (HSB) candidates. We use the catalogs of \citet{LV_cat} as the basis for the SBF analysis presented here. While most of the cataloged galaxies have no prior distance information, some have redshifts and some even have TRGB distances. Where possible, we take these into account when determining the nature of a candidate. We refer the reader to \citet{LV_cat} for the full catalogs of candidates.

\section{SBF Methodology}
\label{sec:sbf_methods}
In this section, we describe the methodology we use in the SBF analysis. We follow the procedure detailed in \citet{sbf_calib}, which largely follows the usual SBF measurement process \citep[e.g.][]{blakeslee2009,cantiello2018}. We briefly outline the important steps here. The analysis starts with modelling the smooth surface brightness profile for each candidate. Then the amount of fluctuation in the brightness profile relative to the smooth profile is quantified. This quantity is expressed in terms of the apparent SBF magnitude. The absolute SBF magnitude for a certain stellar population is defined as:
\beq
\bar{M} = -2.5 \log\left(\ddfrac{\sum_i n_i L_i^2}{\sum_i n_i L_i}\right) + \mathrm{z.p.}
\label{eq:sbf_def}
\eeq
\noindent where $n_i$ is the number of stars with luminosity $L_i$ in the stellar population and $\mathrm{z.p.}$ is the zero-point of the photometry. To determine this quantity for a given candidate, we use the calibration of \citet{sbf_calib} that relates the absolute SBF magnitude to the broad band color of the stellar population. Bluer stellar populations have brighter SBF magnitudes as those populations have, on average, brighter stars. With an apparent and absolute SBF magnitude in hand, we determine the distance modulus to the candidate.

We use the S\'{e}rsic profile \citep{sersic} fits reported by \citet{LV_cat} as the model for the smooth surface brightness profile. While the light profiles are often more complex (e.g. lopsided and/or twisted) than is captured by a single S\'{e}rsic, the candidates are generally too small and faint to use non-parametric modeling as a function of radius. Using a S\'{e}rsic profile as a model for the smooth underlying profile where, in reality, the profile is more complicated can lead to spurious fluctuation power in the SBF measurement that can bias the distance significantly. To overcome this, for a sub-sample of the galaxies, we produce new S\'{e}rsic fits that are restricted to the outer regions of the galaxies, which are often much smoother and more amenable to SBF than the inner structured regions. A small sub-sample ($\sim$10\%) of the galaxies are too irregular to attempt an SBF measurement in any form. For these galaxies, we use other distance measures (TRGB and redshift) where possible or just leave the candidate as a `possible/unconfirmed' satellite, as described more in \S\ref{sec:distances}.

\begin{figure*}
\includegraphics[width=0.95\textwidth]{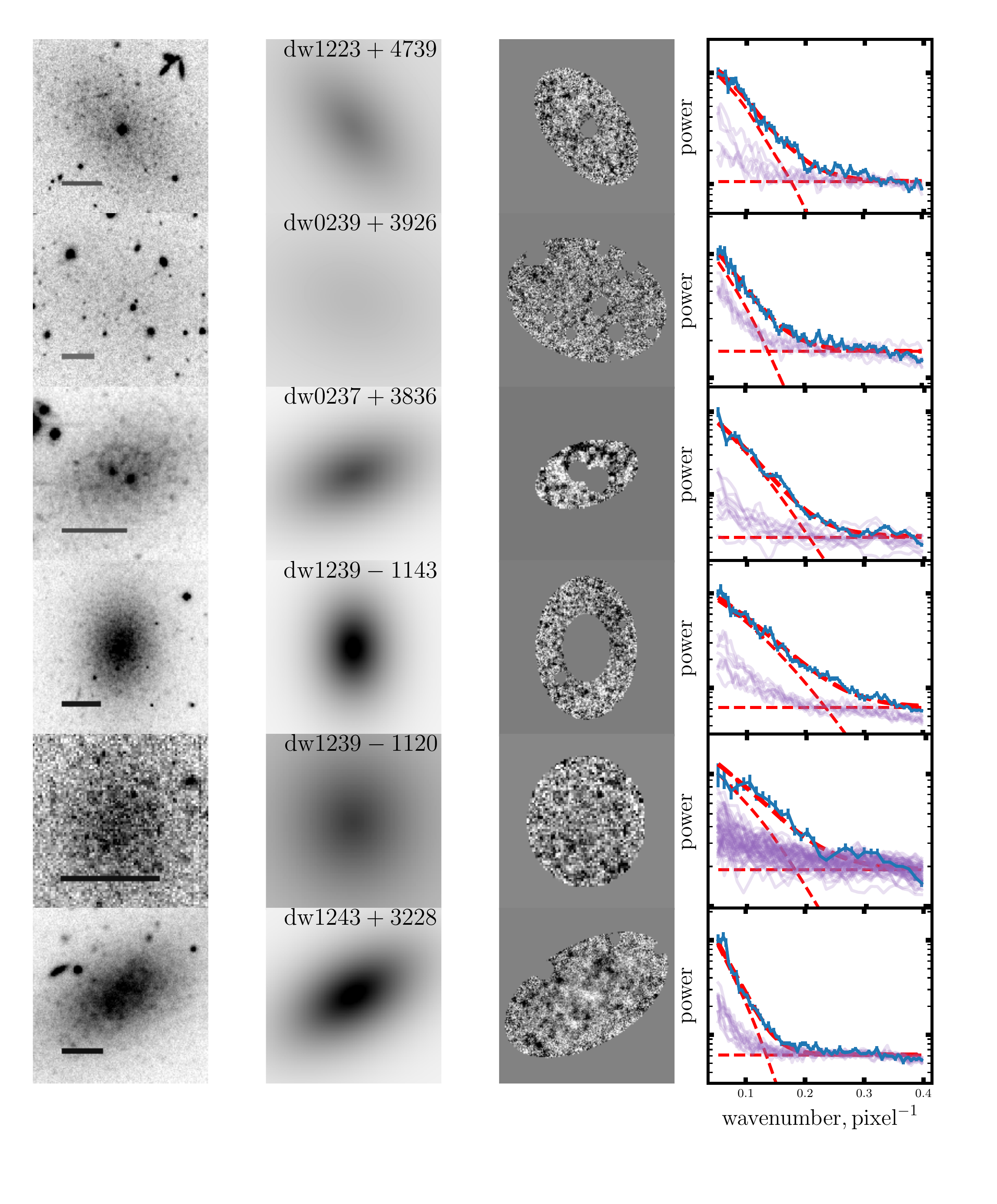}
\caption{Demonstration of the SBF measurement process adopted in this work. The stacked $r$ or $i-$band images of the dwarfs are shown in the left column. The black bars in each image indicate 10$''$. The S\'{e}rsic fit used to model the smooth galaxy profile is shown in the second column. This smooth model is subtracted from the galaxy and used to normalize the galaxy. Any contaminating point sources are masked and an annulus is chosen within which to measure the SBF. This result is shown in the third column. The azimuthally averaged (and normalized) power spectrum of the image is shown in the right column along with the fitted combination of PSF power spectrum and white noise. The faint purple lines are the power spectrum measured in nearby background fields. The fluctuation power measured in these fields is subtracted from that measured from the galaxy. Note that even though dw0239+3926 (second from top) is very low surface brightness, a high S/N${\sim}15$ measure of the SBF is possible.}
\label{fig:sbf_ex}
\end{figure*}

Using the fits for the smooth brightness profile, the fluctuation power is measured in the usual Fourier way described in detail in \citet{sbf_calib}. The main steps in the SBF measurement are shown in Figure \ref{fig:sbf_ex} for six example candidates in our catalog. Each of these dwarfs is confirmed to be at the distance of their host. Many of the dwarfs are LSB with $\mu_{0,g}{\sim}26$ mag arcsec$^{-2}$ but high S/N SBF measurements are still possible with the depth of the archival imaging.

To turn the SBF measurement into a distance constraint, we use the empirical calibration of \citet{sbf_calib}. This calibration accounts for the dependence of SBF on stellar population via the integrated $g-i$ color of a galaxy and provides the absolute SBF magnitude in the $i$-band. However, for seven of our hosts, our imaging data is in the $g$ and $r$-bands, not $g$ and $i$. In \S\ref{sec:sbf_r_calib}, we extend the calibration of \citet{sbf_calib} into the $r$-band using simple stellar population isochrone models and the subsample of calibrator galaxies of \citet{sbf_calib} that also have $r$-band imaging data. This calibration produces the absolute $r$-band SBF magnitude as a function of integrated $g-r$ color.

Using either calibration, we follow the same procedure to turn the SBF measurement into a distance constraint. Here, we use a Monte Carlo approach. For each of 10,000 iterations, we sample a color from a Gaussian with mean equal to the measured color of the galaxy and standard deviation equal to the estimated uncertainty in the color. With this color, we use the SBF calibration to derive an absolute SBF magnitude. To account for the uncertainties in the calibration, in each iteration, we sample the calibration (slope and $y$-intercept) parameters from the Markov Chain Monte Carlo chains produced in the calibration fit of \citet{sbf_calib}. This step accounts for the strong covariance between the slope and $y$-intercept in the calibration formula. We are left with a distribution of distances that are consistent with the measured SBF and color for a galaxy. From this distribution, we calculate a median distance and $\pm1\sigma$ and $\pm2\sigma$ distance bounds.

For much of the candidate sample, the measured SBF level is very low and it is possible to show that these must be background galaxies. Stated differently, a dwarf satellite at the distance of the host should show a certain level of SBF, the lack of which provides a meaningful constraint. Following \citet{sbf_m101}, we consider any dwarf whose 2$\sigma$ distance lower bound is beyond the distance of the host to be in the background of the host. It is important to emphasize that the lack of detected SBF is not due to limited S/N. Rather the S/N is sufficient to firmly establish a lack of fluctuation at the level expected. Additionally, classifying these candidates as background is not simply due to the galaxies being too faint to measure SBF. The uncertainty of the SBF measurement accounts for the faintness and is thus included in the distance constraint. \citet{sbf_m101} concluded that many candidate satellites of M101 were background, and this has since been confirmed by HST imaging \citep{bennet2019}, demonstrating that SBF distance lower bounds set in this way are reliable. 

Some of the galaxies that we confirm to be background appear to have relatively strong fluctuation signals. This signal is more often than not coming from residuals between assuming a S\'{e}rsic profile and the galaxy having, in reality, a more complicated galaxy profile; thus the signal is not real SBF. The conclusion that these galaxies are background is, however, reliable because even with this added fluctuation power, the galaxies do not show the fluctuations that would be expected for a galaxy at the distance of the host.

Examples of galaxies that we conclude to be background along with examples of galaxies that we conclude to be real satellites from the same host are shown in Appendix \ref{app:sbf_details}.

\section{$\MakeLowercase{r-}$band SBF Calibration}
\label{sec:sbf_r_calib}
\citet{sbf_calib} provides a calibration for $\bar{M}_i$ as a function of $g-i$ color. However, many of the host galaxies in this work only have imaging in $r$, or the $r$ coverage is substantially deeper than the $i$-band. Therefore, in this section we derive an absolute SBF calibration for the $r$-band. While SBF is less prominent in the $r$-band and the seeing is generally worse than $i$-band \citep{scott_psf}, robust SBF distances are still possible in the $r$-band. In this section, we extend the work of \citet{sbf_calib} and provide a calibration for $\bar{M}_r$ as a function of $g-r$ color. Twelve of the galaxies used in the calibration of \citet{sbf_calib} have $r$-band data and we measure the $r$-band SBF magnitudes for these galaxies. We supplement this sample with two additional dwarf satellites in the M81 group that have CFHT $g$ and $r-$band imaging and \emph{HST} TRGB distances \citep{chiboucas2009, chiboucas2013}. These 14 galaxies are listed in Table \ref{tab:rband_sample} (we refer the reader to \citet{sbf_calib} for more details on the sample).

\begin{deluxetable}{cc}
\tablecaption{Galaxies used in the $r$-band calibration \label{tab:rband_sample}}
\tablehead{\colhead{Name} & \colhead{TRGB Distance (Mpc)}}
\startdata
FM1 & 3.78\\
KDG 061 & 3.66\\
BK5N & 3.7\\
UGCA 365 & 5.42\\
DDO 044 & 3.21\\
d0939+71 & 3.7\\
d0944+71 & 3.4 \\
LVJ1218+4655 & 8.28\\
NGC 4258-DF6 & 7.3\\
KDG 101 & 7.28\\
M101-DF1 & 6.37\\
M101-DF2 & 6.87\\
M101-DF3 & 6.52\\
UGC 9405 & 6.3\\
\enddata
\end{deluxetable}

Unfortunately, there are significantly fewer calibration galaxies available for the $r$-band than the $i$-band. Therefore, we do not simply fit a $\bar{M}_r$ vs $g-r$ calibration but instead convert the $\bar{M}_i$ vs $g-i$ calibration into the $r$ band using theoretical isochrones. We show that the calibration is consistent with the SBF observations of the galaxies in Table \ref{tab:rband_sample}. The uncertainties associated with the filter transform are smaller than the uncertainties that will come from fitting the limited sample of calibrator galaxies. \citet{sbf_calib} found good agreement with the theoretical $\bar{M}_i$ vs $g-i$ relation predicted by either the MIST \citep{mist_models} or PADOVA \citep{parsec, parsec2} isochrone models for colors $g-i\gtrsim0.5$. In that work, it was unclear whether to attribute the disagreement at bluer colors to the isochrone models or the SBF measurements. However, recently, \citet{greco2020} demonstrated good agreement between that calibration and MIST models at bluer colors if instead of assuming a single stellar population, a double burst star formation history is adopted for the bluest galaxies. Either way, we are not using the isochrones to provide an independent, absolute $r$-band calibration but rather to convert the existing, empirical $i$-band calibration into the $r$-band, and this is more reliable. 

To do the filter conversion, we transform $\bar{M}_i$ to $\bar{M}_r$ and $g-i$ to $g-r$ using SSP models from the MIST project with ages between 3 and 10 Gyr and metallicities in the range $-2<\mathrm{[Fe/H]}<0$. Both conversions are fitted by linear functions in the $g-r$ color. These conversions are shown in Appendix \ref{app:rband}. The $\bar{M}_i$ to $\bar{M}_r$ conversion is fit only in the color range $g-r<0.6$, which is the range appropriate for the low-mass galaxies studied here.

With filter conversion functions of the form:
\beq
\bar{M}_i-\bar{M}_r = a (g-r) + b\\
(g-r)-(g-i) = a_2(g-r)+b_2\\
\eeq
and the $i$-band calibration of the form:
\beq
\bar{M}_i = \alpha (g-i)+ \beta ,
\eeq
the $r$-band calibration can be written as:
\beq
\bar{M}_r = (\alpha-a-\alpha a_2)(g-r) - b - \alpha b_2 + \beta.\\
\label{eq:mrbar_th}
\eeq
Performing the fits, we find $a=-0.92$, $b=-0.243$, $a_2=-0.530$, and $b_2=0.0319$ to determine a  final calibration (using $\alpha/\beta$ from \citet{sbf_calib}):
\beq
\bar{M}_r = 4.21(g-r) - 3.00.\\
\label{eq:mrbar_num}
\eeq
To calculate distance uncertainties resulting from this calibration, we sample parameters from the chains in the MCMC fit of \citet{sbf_calib} and convert those into uncertainties in $\bar{M}_r$ using Equation \ref{eq:mrbar_th} above. Using the chains is crucial to capture the covariance between the slope and y-intercept in the calibration. As shown in Appendix \ref{app:rband}, the uncertainty stemming from the filter transforms is $\lesssim0.1$ mag (5\% in distance) and is sufficiently subdominant to other sources of error that we do not include it.

\begin{figure}
\includegraphics[width=0.48\textwidth]{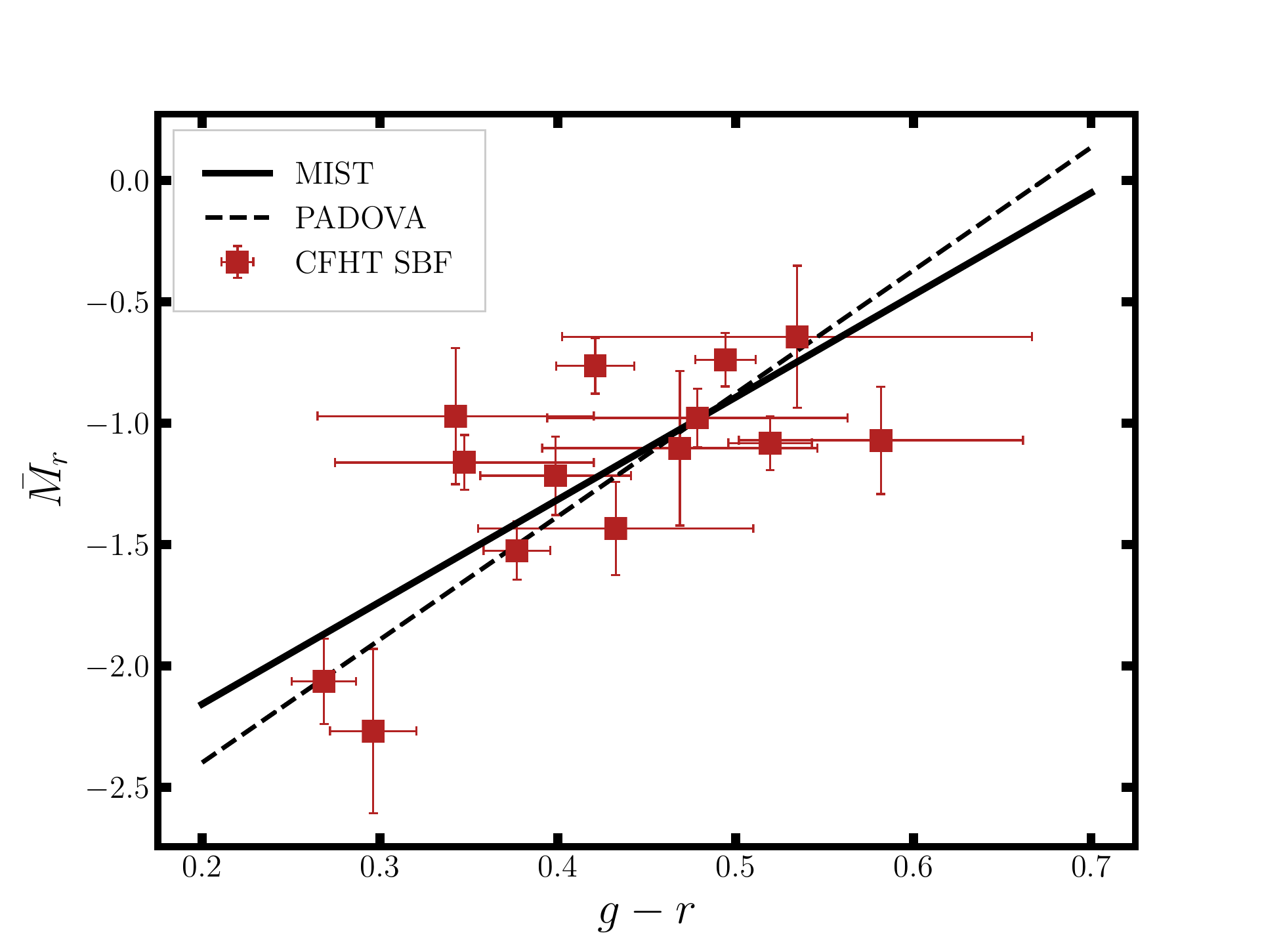}
\caption{The SBF $r$-band calibration determined and used in this work. The black lines show the $i$-band calibration of \citet{sbf_calib} transformed into the $r$-band using theoretical isochrones as described in \S\ref{sec:sbf_r_calib}. The points are CFHT SBF measurements of galaxies with known TRGB distances as listed in \autoref{tab:rband_sample}.}
\label{fig:rband}
\end{figure}

The calibration given in Equation \ref{eq:mrbar_num} is shown in Figure \ref{fig:rband} along with the 14 calibrator galaxies. The agreement in the color range $0.3<g-r<0.6$ is good between the observations and the converted $i$-band calibration, particularly in the $y$-intercept. Without the two bluest data points, it is unclear how well the slope of the data points matches that of the converted calibration. We calculate a reduced $\chi^2$ (e.g., Eq 5 of \citet{sbf_calib}) of the data points relative to the MIST line of $\chi^2_\mathrm{red}$=2.0 (including the whole color range), indicating the agreement is acceptable. We take this as evidence that the systematic uncertainties involved in the filter transform are minimal. Also shown in the dashed line is the calibration that results from using PADOVA isochrones instead of MIST isochrones. We see that the difference is minimal in the color range $0.3<g-r<0.6$ that describes the majority of the galaxies in this paper.

\section{SBF Distances to Local Volume Satellites}
\label{sec:distances}

In this section, we measure SBF and apply the calibration in \citet{sbf_calib} to determine an SBF-based distance for each candidate. Based on these distances, we classify each candidate into one of three categories: confirmed physical satellites, confirmed background contaminants, or galaxies where no SBF constraint is possible that we will refer to as `unconfirmed' or `possible' satellites. This last category is generally composed of galaxies that were so faint that the uncertainty in the SBF measurement is large. Additionally, some galaxies that were markedly non-S\'{e}rsic or had other problems (for instance, being behind a saturation spike) making the SBF measurement impossible are conservatively put into this category. We label a dwarf to be a confirmed satellite if the SBF is measured at a S/N $>5$ and the distance is within $\sim2\sigma$ of the host's distance. We define the SBF S/N as simply the measured SBF variance level divided by its estimated uncertainty\footnote{For very noisy measurements, this quantity can be negative if the inferred SBF variance is negative.}. 

A summary of the SBF-based candidate classifications is given in Table \ref{tab:sbf_overview}. We list the number of confirmed satellites, confirmed background galaxies, and unconfirmed galaxies. In this table, we give the number of candidates confirmed via any method (including TRGB and/or redshift), although the vast majority are confirmed via SBF. Details for each host (including what outside information is used in the confirmation of satellites) is given in Appendix \ref{app:sbf_details}. As discussed in Appendix \ref{app:sbf_details}, when TRGB and SBF distances exist for the same dwarf, the SBF distances agree very well with the TRGB distances.

Overall 52 of the 155 candidates of \citet{LV_cat} are confirmed as physical satellites while 55 are constrained to be background. We confirm 41 candidates as real satellites via SBF. A further 11 are confirmed via other distance measures available in the literature, particularly TRGB and redshift. The SBF results constrain 49 candidates to be background, and other distance measurements from the literature constrain a further 6 to be background. The remaining 48 candidates are still unconstrained. Only 25 of these are above our fiducial completeness limit of $M_V<-9$, assuming they are at the distance of the hosts. Deeper imaging or (most likely) \emph{HST} will be required to ascertain the distances to these candidates.  Our results broadly demonstrate the power of SBF in mapping and characterizing the dwarf galaxy population in the Local Volume. 


For NGC 4631, extremely deep archival HSC imaging exists for several of the candidate satellites. We acquired and reduced these data (described in detail in Appendix \ref{app:sbf_details}) and used it to analyze the SBF of dwarfs around this host, as a check for the CFHT data. As detailed in the Appendix, we find very close agreement with the shallower CFHT data. Additionally, we are able to constrain an additional candidate to be background that was ambiguous from the CFHT data.

\subsection{Classification Details}

Here we discuss more details of the satellite confirmation process.

\subsubsection{Setting the SBF S/N Threshold}
We determine the SBF S/N threshold using image simulations performed by injecting dwarfs with SBF into the CFHT imaging; we refer the reader to \citet{sbf_calib} for details of the SBF image simulations. We find that dwarfs with $M_g\sim-9$ mag and moderate surface brightness ($\mu\sim25$ mag arcsec$^{-2}$) have SBF measurable with S/N$\sim$5 with the depth of the CFHT data. The calibration galaxies of \citet[][all of which had TRGB distances]{sbf_calib} almost all had S/N $>$ 5 and spanned the same luminosity, surface brightness, and distance range as the candidates satellites in the current sample. Galaxies whose SBF distance result is consistent with the host's distance but the SBF is of low significance (S/N $<$ 5) are placed in the `unconfirmed' category. 

Using our prior work on M101, the S/N threshold we use here is conservative enough to prevent false positive satellite confirmation. \citet{sbf_m101} confirmed two satellites around M101 using SBF. These satellites had SBF S/N $\geq7$, which means they would be confirmed by the threshold used here. Both of these have been confirmed by the \emph{HST} imaging of \citet{bennet2019}. \citet{sbf_m101} also highlighted 2 other candidates as promising follow-up targets that had reasonably strong signal with S/N$\sim2-3$. The \emph{HST} imaging of \citet{bennet2019} showed that this signal was not from SBF and the galaxies were background contaminants. The signal instead appeared to be coming from unmasked background galaxies. Using the threshold adopted here, these two candidates would be conservatively included in the `unconfirmed' candidate category.

\subsubsection{Visual Inspection} 
We also carefully visually inspect each candidate to make sure that the SBF signal is coming from the bulk stellar population of the galaxy and not from twists or other irregularities in the light profile. This visual inspection check is an important step to prevent false positives, particularly for the smaller candidates. From our experience with the calibration sample \citep{sbf_calib}, SBF should be clearly visible in dwarfs of the luminosity, surface brightness, and distance as the current candidates. We emphasize that we do not discard galaxies or conclude galaxies are background on the visual check alone. In a handful of cases ($\lesssim5\%$), we conservatively move a candidate from the `confirmed' bin into the `unconfirmed' bin if its visual appearance generates concern that the fluctuation signal is not actually coming from SBF. 

\subsubsection{False Negatives}
Additionally, it is possible to have false negatives in the SBF analysis. The most likely cause is if the color of the candidate is measured incorrectly.  We estimate the error in the galaxy colors using image simulations that should, in principle, capture the systematic uncertainty associated with the sky subtraction. However, it is possible that significant systematic errors in the sky subtraction linger. If the candidate was measured to be bluer than it actually is, the SBF distance can be greatly overestimated and vice versa. For our analysis, the most likely impact of this failure mode is for galaxies that are too faint for a meaningful SBF distance constraint. If one of these galaxies is measured to be significantly bluer than it actually is, we could falsely conclude it must be background because it lacks the strong SBF expected at that blue color. Note that erroneously measuring one of these galaxies to be too red would not have the same effect and would not change the categorization of this galaxy from being an `unconfirmed' candidate.

\begin{deluxetable*}{cccccc}
\tablecaption{Overview of the SBF results for each host.\label{tab:sbf_overview}}
\tablehead{\colhead{Host Name} & \colhead{Host Distance (Mpc)} &\colhead{Host $M_K$} &\colhead{\# Confirmed} &\colhead{\# Possible} & \colhead{\# Background}}
\startdata
NGC 1023 & 10.4 & -23.9 & 15 & 6 & 10 \\
NGC 1156 & 7.6 & -19.9 & 0 & 2 & 1 \\
NGC 2903 & 8.0 & -23.5 & 2 & 2 & 0 \\
NGC 4258 & 7.2 & -23.8 & 7 & 4 & 22 \\
NGC 4565 & 11.9 & -24.3 & 4 & 15 & 2 \\
NGC 4631 & 7.4 & -22.9 & 10 & 0 & 7 \\
NGC 5023 & 6.5 & -19.3 & 0 & 1 & 1 \\
M51 & 8.6 & -24.2 & 2 & 6 & 8 \\
M64 & 5.3 & -23.3 & 0 & 0 & 1 \\
M104 & 9.55 & -24.9 & 12 & 12 & 3 \\
\enddata
\end{deluxetable*}

\subsection{Completeness of Satellite Systems}
The completeness of the catalogs of candidate satellites is quantified in detail in \citet{LV_cat}, but we give some overview here. In that work, we conducted extensive mock injection tests to quantify the detection-efficiency as a function of dwarf luminosity and surface brightness. Most of the hosts had fairly similar completeness levels. Completeness was generally $\gtrsim90$\% for dwarfs up to central surface brightness of $\mu_{0,V}\sim26.5$ mag arcsec$^{-2}$ and for sizes greater than $r_e\gtrsim4$ \arcs. Thus, the catalogs are likely complete down to luminosities of $M_V\sim-9$ at the distances of these hosts over the survey footprints. We note that roughly half of the unconstrained/inconclusive satellite candidates are actually below this fiducial completeness limit.  For the discussion below, we assume these hosts are 100\% complete to $M_V\sim-9$ and to $\sim\mu_{0,V}\sim26.5$ mag arcsec$^{-2}$ over the survey footprint. The survey footprints are given in \citet{LV_cat} and cover roughly the inner 150 kpc projected area for the six best surveyed hosts (NGC 1023, NGC 4258, NGC 4565, NGC 4631, M51, M104), which are the focus for  the rest of the paper.

\subsection{Structural Parameters of Confirmed Satellites}

Tables giving the properties of the confirmed and possible satellites, including physical sizes and absolute luminosities are given in Appendix \ref{app:sat_systems_tables}. For the physical quantities, we assume the confirmed and possible satellites are at the distance of the host, instead of using the individual SBF distances, to prevent artificially inflating the scatter of these quantities due to the lower precision of the SBF distances. This implicitly assumes that the confirmed satellites are likely within the virial radius of the host along the line-of-sight and, hence, at about the same distance\footnote{While we do expect some of the confirmed satellites to be nearby field objects ($\sim500-1000$ kpc in front of or behind the host), the dSph morphology of the majority of the confirmed satellites strongly implies that the majority are bona fide virialized satellites of their hosts. From our experience with the simulations (see below), we expect this population of nearby field dwarfs to constitute $\sim10-15$\% of the confirmed dwarfs.}.

\begin{figure*}
\includegraphics[width=0.98\textwidth]{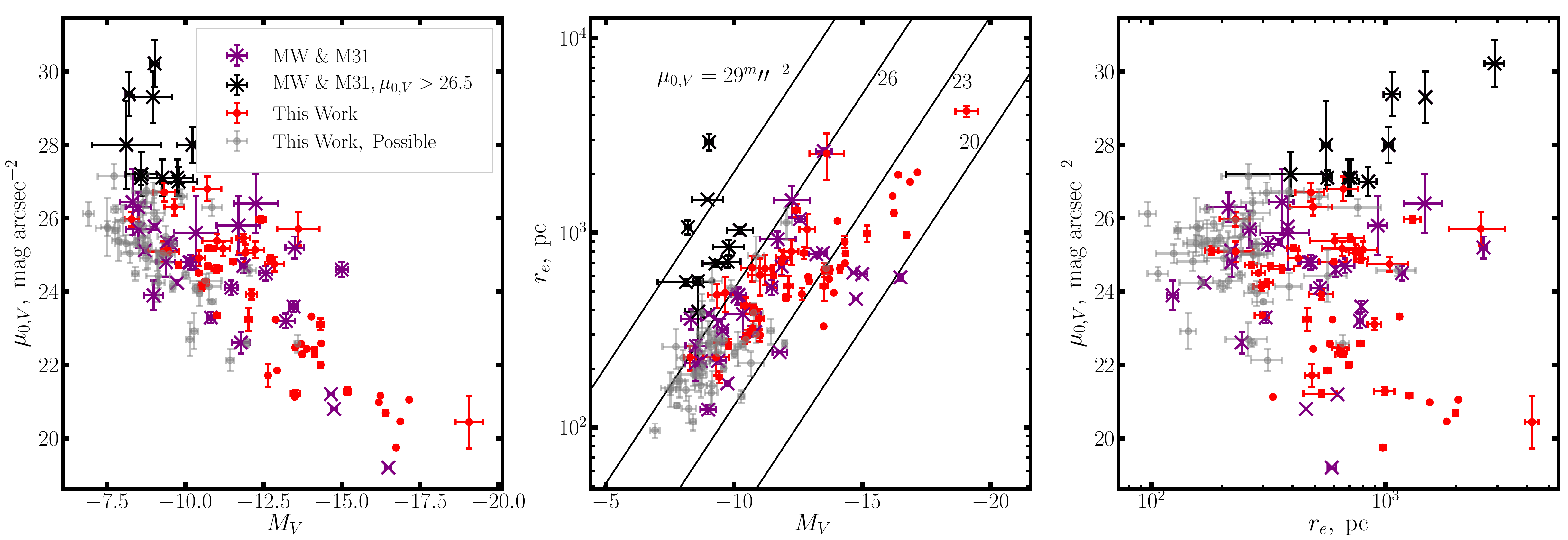}
\caption{Structural parameters of the confirmed satellites (red). Shown in purple are the classical ($M_V<-8$) MW and M31 satellites from \citet{mcconnachie2012}, with black indicating the satellites with $\mu_{0,V}>26.5$ mag arcsec$^{-2}$, which we consider the surface brightness limit of our survey. The gray points show the possible/unconfirmed satellites from the current work. Generally these are smaller and have lower surface brightness than the confirmed satellites.}
\label{fig:struct}
\end{figure*}

Figure \ref{fig:struct} shows various structural parameters for the confirmed satellites of the 10 hosts. They show close agreement with the scaling relations of the MW and M31 classical satellites. The confirmed satellites show better agreement with the LG dwarfs than the entire sample shown in Figure 6 of \citet{LV_cat}. Many of the objects in \citet{LV_cat} were smaller than the LG dwarfs at fixed luminosity, indicating they were likely background. As seen in all three panels of Figure \ref{fig:struct}, the surface brightness completeness of \citet{LV_cat} is $\mu_{0,V}\sim26.5$ mag arcsec$^{-2}$.

\subsection{The Importance of Distances}
It is worth discussing why it is important that we consider only satellite systems with full (or nearly full) distance constraints on all candidate satellites. We could consider many more surveyed satellite systems if we relaxed this requirement and used a statistical background subtraction \citep[e.g.][]{wenting2012} to remove background contaminants. However, as shown in Table \ref{tab:sbf_overview}, a majority of candidate satellites often turn out to be background. Furthermore, the scatter between hosts in the amount of background contamination is immense, due to differing amount of structure along the line of sight. Including the six systems NGC 1023, NGC 2903, NGC 4258, NGC 4631, M51, and M104 that had good SBF results (with relatively few inconclusive candidates) and M101 from \citet{sbf_m101}, the rms scatter in confirmed background contaminants is $\sim11$ per host (over roughly the inner 150kpc projected area). This is \emph{significantly} more than the rms scatter in confirmed satellites of $\sim5$ per host over the same area. Therefore, the scatter introduced by any statistical background subtraction will overwhelm the true, intrinsic host-to-host scatter in satellite abundance, and $\sim5$ times the host sample size would be required to get a similar constraint on the average number of satellites. To do a detailed analysis of the satellite abundances in nearby systems, distance constraints for the majority of candidate satellites are crucial, either from SBF/TRGB or redshift \citep[e.g.][]{sales2013,geha2017}.

\section{Satellite Systems from the Literature}
\label{sec:lfs}

In this section, we assemble the information on satellite systems of the nearby hosts that have been previously surveyed in the literature and combine these with the new hosts from the current work. For the rest of the paper, we only consider the six best-surveyed hosts from the current work (those whose surveys cover roughly the inner 150 kpc): NGC 1023, NGC 4258, NGC 4565, NGC 4631, M51, and M104. As discussed in the Introduction, there are six other nearby systems that have been well searched for satellites previously.  We give an overview of the literature that gives the satellite properties, along with estimates of the completeness for each system. All of the satellites in these systems have been confirmed with distance measurements and the surveys are complete down to at least $M_V\sim-10$ to $-9$ over a large fraction of the host's virial volume. For reference, we list all of the satellite properties for each host in tables in Appendix \ref{app:sat_systems_tables}.

\subsection{Previously Surveyed Systems}

Positions for the MW classical satellites are taken from \citet{mcconnachie2012}. Luminosities are taken from \citet{munoz2018}, where available, and \citet{mcconnachie2012} otherwise. The distances are taken from the compilation of \citet{fritz2018} and individual references are given in the Appendix. We assume that the census of MW classical satellites is complete throughout the virial volume. For the luminosity function, we assume the MW has an absolute magnitude of $M_V=-21.4$ \citep{bland2016}.

We take the sample of M31 satellites from \citet{martin2016} and \citet{mcconnachie2018}. The luminosities come from \citet{mcconnachie2018}, altered to account for updated distances. The distances, themselves, come from a variety of sources (references are provided in the Appendix), prioritizing \emph{HST} distances over ground-based and variable star over TRGB, where possible. Due to the faintness of these satellites, the RGB is often not well populated leading to relatively uncertain TRGB distances. The PandAS survey is sensitive to ultra-faint satellites of M31 with $M_V\lesssim-6$, but their imaging covers the inner projected 150 kpc volume. However, with Pan-STARRS the census of M31 satellites is likely complete through the virial volume down to $M_V\sim-9$ \citep[e.g.][]{martin2013a, martin2013b}.  We assume an absolute magnitude of M31 of $M_V=-22$ \citep{walterbos, geha2017}.

The satellites of Centaurus A come from \citet{crnojevic2019} and \citet{muller2019}. \citet{crnojevic2019} estimate their completeness at 90\% for dwarfs brighter than $M_V\sim-9$ over their Magellan/Megacam survey footprint which roughly covers the inner projected 150 kpc. Similarly, \citet{muller2019} estimate that they are complete down to $M_V\sim-10$ over the inner projected 200 kpc. 

The list of satellites of M81 comes from \citet{chiboucas2013} and \citet{chiboucas2009}. The photometry for NGC 3077, M81, M82, NGC 2976, IC 2574, and DDO 82 come from \citet{galex2007}. The photometry for IKN, BK5N, KDG061, and KDG064 come from the recent HSC imaging of \citet{okamoto2019}. The rest come from \citet{chiboucas2013}. We convert the $r$ magnitudes reported in \citet{chiboucas2013} into $V$ magnitudes assuming $M_V\sim M_r+0.4$ \citep{crnojevic2019}. The TRGB distances come from \citet{chiboucas2013} and \citet{karachentsev}. We do not include any of the dwarfs that \citet{chiboucas2013} consider to be tidal dwarf galaxies. We assume that the census of satellites of M81 is complete for all `classical'-like satellites ($M_V\lesssim -8$) throughout the inner projected 250 kpc volume. 

The satellite system of M101 comes from \citet{t15}, \citet{danieli101}, \citet{sbf_m101}, and \citet{bennet2019}. The photometry for M101 uses the updated distance of \citet{beaton2019}. To convert from the $B$ magnitudes reported by \citet{t15}, we assume $M_V= M_B-0.3$. We use the \emph{HST} photometry of \citet{bennet2019} for dwA and dw9. We use the \emph{HST} photometry for DF1, DF2, and DF3 (S. Danieli, priv. comm.). We note that the magnitudes we take for these objects are significantly ($\sim1-2$ mag) brighter than those listed by \citet{bennet2019}. This is likely due to the aggressive sky subtraction used in the CFHT Legacy Survey data \citep{gwyn2012} used in \citet{bennet2017}. We also include UGC 8882 among the M101 satellites. \citet{sbf_m101} gives an SBF distance to UGC 8882 of $8.5\pm1.0$ Mpc which is marginally consistent with the distance of M101 ($D=6.52$ Mpc). The \citeauthor{sbf_m101} SBF distance agrees well with that of \citet{jerjen_field2}, $D=8.3$ Mpc, who use a completely different (albeit somewhat outdated) calibration. To investigate this dwarf more closely, we measured its SBF distance in the $r$ band of the CFHT Legacy Survey which interestingly gives a somewhat smaller distance of $D=7$ Mpc, close to that of M101. We also measured its SBF with completely independent data using a color from DECaLS \citep{decals} and SBF magnitude from archival HSC $r$ band imaging which agreed with the CFHT $r$ band distance. Thus, we tentatively include this object as a satellite of M101. Its extremely regular, quenched \citep{huchtmeier2009} dSph morphology supports this association. We assume that the satellite system of M101 is complete down to $M_V\sim-8.5$ within the inner projected 200 kpc \citep[see Fig 1 of ][ for the different search footprints covering M101]{sbf_m101}.

The properties of the satellites of M94 come from \citet{smercina2018}. We assume the census is complete to $M_V\sim -9$ throughput the inner projected 150 kpc volume. 

For these previously surveyed hosts, we do not include satellites that have lower surface brightness than $\mu_{0,V}=26.5$ mag arcsec$^{-2}$. As discussed above and in \citet{LV_cat}, this is the surface brightness limit of the satellite systems surveyed in this work. Satellites with significantly fainter surface brightness are detectable around the MW and M31 (and, to a lesser extent, M81 and CenA) from resolved stars, so to compare all systems on equal footing, these satellites are not included in the following. The tables in Appendix \ref{app:sat_systems_tables} indicate which satellites satisfy this criterion. Additionally, several of the extremely low surface brightness satellites (e.g. Crater 2 and AndXIX) are clearly the result of tidal stripping, and it is unclear if these satellites are appropriate to include in the comparison with simulations below. The subhalos hosting such stripped systems might not be recognized by the halo finders used in the simulations, as discussed more below.

We note that two of the confirmed satellites from the hosts in the current work are below the $\mu_{0,V}=26.5$ mag arcsec$^{-2}$ limit. They are also excluded in the LF comparisons below and are marked in the Tables in Appendix \ref{app:sat_systems_tables}. However, we note that our conclusions do not qualitatively change if we keep this population of very low surface brightness satellites.

\subsection{MW-Analogs vs. Small Group Hosts}

For the rest of this paper, we consider these six previously surveyed systems from the literature along with the six best surveyed hosts from the current work in more detail. In some of the comparisons below, we do not consider all 12 LV hosts together but instead roughly split them into hosts we argue are MW-like in halo mass and hosts that are more massive, which we term `small group'. It is important to recognize that several of the surveyed LV hosts are significantly more massive than the MW, and should not be directly compared to the MW. The specific mass bins that we choose are somewhat arbitrary, but it is important to look at trends in matched halo mass bins, and this is a crude way to do that. The MW-sized halos are those with halo mass roughly in the range $0.8-3\times10^{12}$ \msun, and the small groups have halo mass in the range $3-8\times10^{12}$\msun. These rough limits come from dynamical estimates of the total mass of these hosts from the literature. Table \ref{tab:mh_estimates} lists these estimates where available for our sample of hosts. For the halo estimates of NGC 4258, M94, and M104, the estimates from \citet{karachentsev2014_masses} are likely overestimated. The estimates comes from the dynamics of nearby group members, but considering that some of the group members included are likely not actual group members (many do not have redshift-independent distances), the dynamical mass is probably overestimated. While the estimated mass of M104 is ostensibly above the upper end of our `small-group' mass range, we include M104 in the small-group category because this mass is likely overestimated, but we note that it might be more massive.  We do not list dynamical mass estimates of NGC 4565, NGC 4631, and M51. Based on their stellar mass and peak rotation speed, we put NGC 4565, NGC 4631, and M51 into the MW-like group \citep[see][for these quantities]{LV_cat}.

\begin{deluxetable}{ccc}
\tablecaption{Rough dynamical halo mass estimates for the LV hosts \label{tab:mh_estimates}}
\tablehead{\colhead{Name} & \colhead{$M_\mathrm{halo}$ ($\times10^{12}$ \msun)} & \colhead{Source}}
\startdata
\hline
\multicolumn{3}{c}{MW-like Hosts}\\
MW & $\sim1$ & 1,2 \\ 
M31 & $\sim1.5$ & 3,4,5  \\
M101 & $1.5\pm1$ &  6,7 \\ 
M94  &  $2.7\pm0.9$ & 7 \\ 
NGC 4258 & $3\pm1$  & 7\\ 
M51 & -- & \\
NGC 4631 & -- &  \\
NGC 4565 & --  & \\
\hline
\multicolumn{3}{c}{Small Group Hosts}\\
M81 & $5\pm1$ & 7 \\
Cen A & $7\pm2$ & 7,8 \\
M104 & $30\pm20$ & 7\\
NGC 1023 & $\sim6$  & 9 \\
\enddata
\tablecomments{Sources: 1-\citet{callingham2019}, 2-\citet{watkins2019}, 3-\citet{watkins2010}, 4-\citet{gonzalez2014}, 5-\citet{penarrubia2014}, 6-\citet{t15}, 7-\citet{karachentsev2014_masses}, 8-\citet{woodley2006}, 9-\citet{trentham2009}  }
\end{deluxetable}

As we will see below, this distinction by halo mass of the hosts is also reflected in the LFs. The small-group hosts have significantly richer satellite systems than the MW-analogs. We note that the small-group hosts are different from the MW-analogs in other ways as well. The small-group hosts include the only two ellipticals in the whole sample (M104 and CenA) and the only S0. M81 is also unique in including two central late type galaxies of similar stellar mass (M81 and M82 with $M_\star=5\times10^{10}$~\msun\ and $M_\star=3\times10^{10}$~\msun, respectively).

\section{Theoretical Models}
\label{sec:models}
In this section, we introduce the theoretical model that we compare against and use to interpret our observed satellite systems. We primarily compare the observed satellite systems with those predicted from dark matter only (DMO) simulations combined with a stellar-to-halo mass relation (SHMR). We could have alternatively used hydrodynamic simulations (obviating the need to use a SHMR) or a semi-analytic model (SAM) combined with a cosmological simulation. The public hydrodynamic simulations (e.g. Illustris and EAGLE) do not have sufficient baryonic resolution to comfortably resolve satellites of the luminosity we probe ($M_V<-9$). However, we can still make meaningful comparisons with hydrodynamic simulations for the brighter ($M_V<-16$) satellites. While SAMs could be used to explore the properties of satellites of virtually any mass, their added complication over a simple SHMR makes extracting physical interpretations more complicated. Therefore, as our primary point of comparison, we use halo catalogs from DMO simulations combined with a SHMR to populate the halos, but we also compare the bright satellite populations with those predicted from public hydrodynamic simulations.

\subsection{Simulation Suites}
For the DMO simulations, we use the halo catalogs from the Illustris-TNG100 project \citep{tng1, tng2, tng3, tng4, tng5, tng6} and the high-resolution ELVIS zoom simulations \citep{elvis}. Each simulation suite has its strength. TNG has a better constraint on the host-to-host scatter because of the large number of MW-like hosts ($>1000$) within the simulated volume, more than the 48 simulated hosts in the ELVIS project. On the other hand, the higher resolution of ELVIS allows us to consider the effect of resolution on the halo catalogs. 

While the baryonic results of TNG will not resolve all of the satellites we are interested in (baryonic particle mass $\sim10^6$ \msun), the DM particle mass of $7.5\times10^6$ \msun\ implies that DM subhalos hosting the satellites of interest ($M_{\mathrm{vir}}\sim5\times10^9$ \msun, see below) will be resolved. Note that we do not use the explicit DMO TNG simulation; instead, we use the dark matter halo catalog from the full baryonic run. This will capture any effect that the baryons might have on the halo abundances. In particular, this accounts for the enhanced destruction of subhalos by the baryonic disk which has been shown to have a dramatic impact on subhalo abundance, particularly near the host galaxy \citep[e.g.][]{donghia2010, gk_lumpy, errani18, kelley2019}. 

For the TNG simulation, we select host halos from the friends-of-friends group catalog provided in the TNG public data release. To avoid any problems with the periodic boundary conditions, we only select halos that are more than 1.5 Mpc from a simulation box edge. We employ several different selection criteria for the hosts, as described below. When selecting on halo mass, we use the $M_{200}$ value given in the friends-of-friends group catalog. When selecting on stellar mass, we use the stellar mass (from the hydro run) of the most massive subhalo in each friends-of-friends group. This corresponds to the stellar mass of the central host (i.e. the MW). We use the SubFind catalog of subhalos to procure a list of subhalos in each FoF group.

The ELVIS suite consists of 24 isolated MW sized hosts and 12 pairs of hosts in a Local Group (LG)-like configuration. We treat all 48 of these hosts in the same way. The ELVIS hosts range in mass fairly uniformly between 1 and 3$\times 10^{12}$ \msun. While this does cover the range we expect for the MW-sized observed hosts, due to the halo mass function \citep[e.g.][]{tinker2008}, it is more likely that an observed host occupies a $10^{12}$~\msun\ halo than a $3\times10^{12}$~\msun\ halo. Therefore, we expect that the ELVIS hosts to be, in general, more rich in subhalos than the corresponding MW-like hosts from TNG. 

In some comparisons, we also make use of the hydrodynamic results of IllustrisTNG. With baryonic particle mass $1.4\times10^6$ \msun, the simulations will resolve a $M_V=-16$ dwarf with roughly 100-200 stellar particles. Thus, we can meaningfully compare the full hydrodynamic results to the bright $M_V<-16$ end of the satellite LFs. The halos and subhalos are selected as described above, and we directly use the $M_V$ quantities reported in the TNG subhalo catalogs.

\subsection{Stellar-Halo Mass Relation}
With a catalog of subhalos in hand, we populate the halos with luminous galaxies using a SHMR. We use the peak virial mass of each subhalo, $M_{\mathrm{peak}}$, to determine the stellar mass of the galaxy. This is important to account for the effect of tidal stripping once a halo becomes a subhalo of a more massive galaxy. To determine $M_{\mathrm{peak}}$, we use the TNG merger trees and record the peak virial mass that each subhalo attains along its main progenitor branch. The ELVIS halo catalogs list $M_{\mathrm{peak}}$ directly.

The well-known SHMRs from abundance matching \citep[e.g.][]{behroozi2013, moster2013} are only valid for $M_*\gtrsim10^8$ \msun, which is larger than the stellar masses of many satellites in our sample. It is possible to extrapolate these relations down, but it is known that the SHMR of \citet{behroozi2013} will over-predict the luminosity function of MW and M31 satellites \citep{elvis}. A steeper relation between stellar mass and halo mass is needed\footnote{We note that the more recent SHMR of \citet{behroozi2019} does show a steeper slope.}. 

We take as our fiducial SHMR the relation from \citet{elvis} and \citet{gk_2017}. This relation has the same functional form as the \citet{behroozi2013} SHMR but uses a steeper power law slope at the low mass end. \citet{elvis} used the  GAMA stellar mass function \citep{baldry2012} to infer a power law slope of 1.92 ($M_*\propto M_{\rm halo}^{1.92}$), as opposed to the slope of 1.412 inferred in \citet{behroozi2013} using an SDSS-derived stellar mass function. \citet{elvis} showed that this SHMR could reproduce the stellar mass function of LG dwarfs down to $M_*\sim5\times10^5$ \msun. \citet{gk_2017} found that a slightly shallower slope of 1.8 fits the LG dwarf stellar mass functions a little better. The SHMR of \citet{gk_2017} is a popular relation often referenced in the literature.

Thus, we use the functional form of the \citet{behroozi2013} SHMR but modified to have a power law slope of 1.8 at the low mass end ($M_{\rm halo}\lesssim 10^{11.5}$ \msun). All of the parameters other than the low-mass slope are taken from \citet{behroozi2013}. For the fiducial model, we assume a fixed lognormal scatter of 0.2 dex about this relation. While, the scatter in the SHMR will likely increase for lower halo masses \citep[e.g.][]{munshi2017}, there is no current understanding (observational or theoretical) of specifically what the scatter should be. Thus we assume the scatter is the same as it is constrained to be at higher masses \citep[e.g.][]{behroozi2013}.

The SHMR is then used to assign a stellar mass to each subhalo. We assume a fixed mass-to-light ratio of $M_*/L_V=1.2$ to convert this stellar mass into a $V$ band magnitude. This mass-to-light ratio is roughly the average ratio inferred for the MW satellites \citep{woo2008}. We note that our sample of satellites do not exhibit a noticeable color-luminosity trend and, thus, a constant mass-to-light ratio for all satellite luminosities is justified. 

We do not attempt to assign a size and/or surface brightness to the model satellite galaxies. We simply assume that all galaxies above our fiducial luminosity limit of $M_V\sim-9$ would be detectable. Relaxing this assumption will be an important step in future work. 

One additional consideration to note is that we do not account for the possibility of dark subhalos. Presumably, some low-mass subhalos exist that do not contain a luminous galaxy as the UV background associated with cosmic reionization completely suppressed star formation in those halos. The halo mass scale at which this process becomes important is often estimated as a few $\times10^9$ \msun~ \citep[e.g.][]{okamoto2008, okamoto2009, sawala2016b, okvirk2016}, however recent work is pushing this scale down to smaller masses \citep[e.g.][]{kim2018, wheeler2019, graus2019}. These masses are at the low end of (or well below) the halo masses expected for classical-sized satellites so we do not expect this to be a relevant physical process for the type of satellites we consider here.

\section{Dwarf Satellite System Luminosity Functions}
\label{sec:results}
In this section, we show the results of comparing the observed satellite systems to the ones predicted from the SHMR model described above. We show four main comparisons. First, we simply compare the observed systems with each other. We clearly see that observed hosts with higher inferred halo mass have richer satellite systems. Second, we compare the luminosity functions for each observed host to those predicted from the models for each host. Third, we explore the number of satellites as a function of host stellar mass. This comparison demonstrates that the scatter between observed satellite systems closely matches that predicted by the simulations, once the mass of the host is accounted for. Finally, we look more closely at the average shape of the LFs by comparing the combined LF of all observed systems to the simulated systems to show that, while the total number of satellites agrees between observations and simulations, the observed hosts have more bright satellites and fewer faint systems than the SHMR model predicts.

\subsection{Observed Luminosity Functions}
\label{sec:lf_obs}
\begin{figure*}
\includegraphics[width=\textwidth]{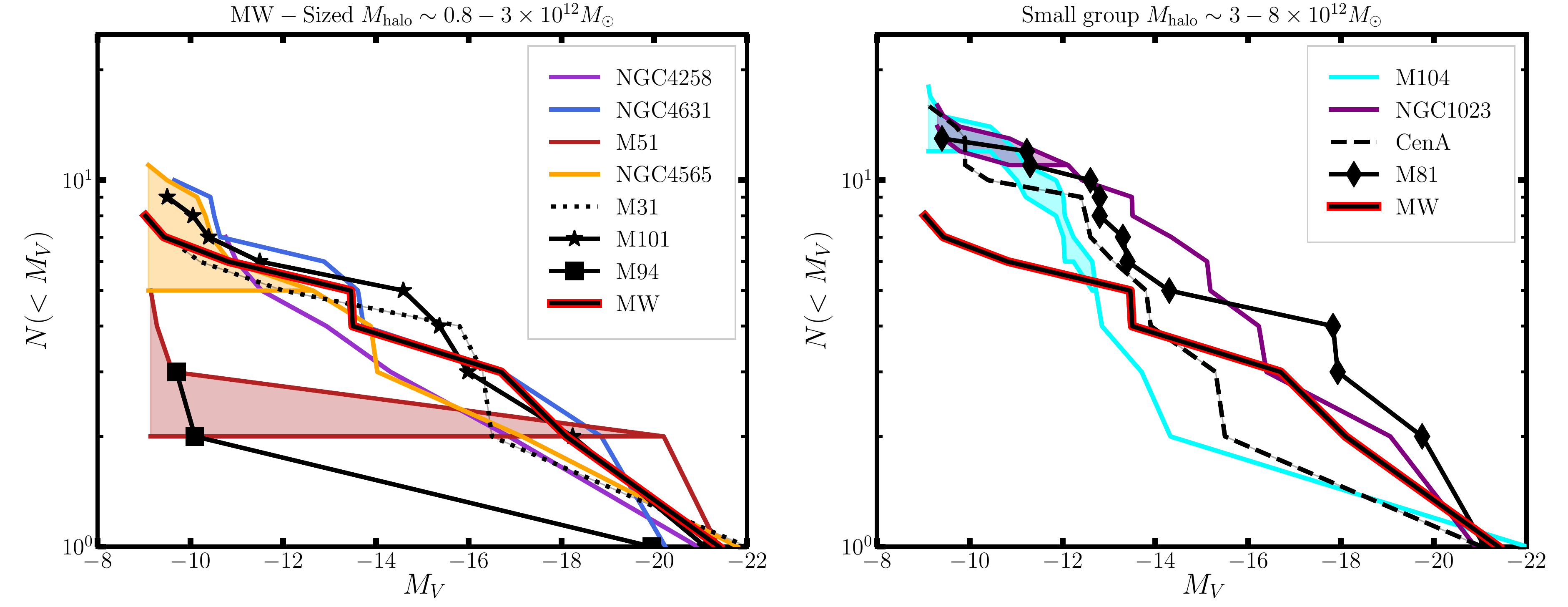}
\caption{The cumulative luminosity functions for the 12 LV hosts with well-measured luminosity functions. The two panels split the hosts roughly by halo mass, see text for details. The MW is only shown in the right panel for comparison. The spread in some of the LFs indicates the membership uncertainty for the candidates where the SBF analysis was ambiguous. Both a lower and upper bound for the luminosity function is given. All hosts are restricted to the inner 150 kpc radius, but no further area corrections are applied.  }
\label{fig:lfs}
\end{figure*}

In this section, we directly compare the observed satellite systems with each other. Figure \ref{fig:lfs} shows the cumulative luminosity functions for the hosts considered here split into the two groups (MW-like vs small group, see \S\ref{sec:lfs}) by halo mass. To address the very different survey footprints for the different hosts, only satellites within 150 kpc (3D distance for the MW and M31, projected for the other hosts) are included, but further area correction is not performed.  We note that 150 kpc is roughly half the virial radius for the MW-like hosts, but less for the more massive hosts.

We highlight a few interesting things in Figure \ref{fig:lfs}. First, the MW appears to have a typical satellite LF compared to the other MW-like hosts. Second, there is large scatter in the luminosity functions within each host class. The scatter would be even more if we compared the two mass bins together, emphasizing the importance of considering them separately. We will quantify the host-to-host scatter in \S\ref{sec:n150} below.

There are some very large separations between satellite magnitudes in the LF. In particular, M94, CenA, and M104 show large gaps between the largest and second largest member in each satellite system. Interestingly, CenA and M104 are the only two ellipticals in the sample, and their merging history might be reflected in these magnitude gaps. To interpret these luminosity functions further, we need to compare with predictions from theoretical models, which is what we turn to next.

\begin{figure*}
\includegraphics[width=\textwidth]{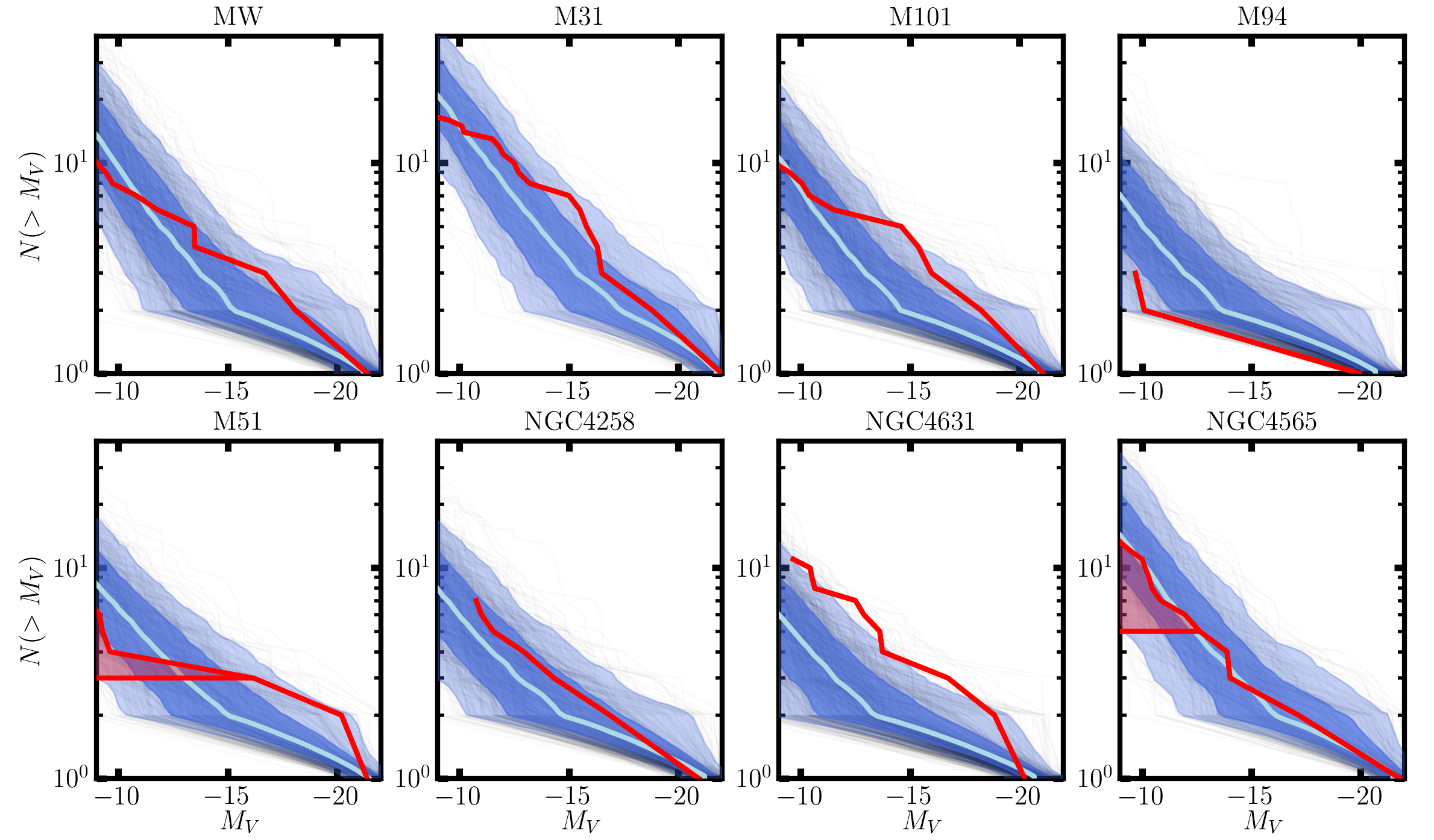}
\caption{The cumulative luminosity functions for the 8 `MW-sized' hosts in our sample (red). The thin black lines show the predicted LFs from the abundance matching model described in the text. The simulation hosts have been selected to have roughly the same stellar mass as the corresponding observed host. The blue regions show the $\pm 1,2 \sigma$ spread in the models. The luminosity completeness is different for each host but is  $M_V\sim-9$ in all cases. For each host, the model satellite systems have been forward-modeled considering the survey area selection function for each specific host. For the hosts that had inconclusive results from the SBF distances, a spread of possible LFs is shown, accounting for uncertain membership.}
\label{fig:mw_lf_comp}
\end{figure*}

\subsection{Individual Luminosity Functions}
\label{sec:ind_lf}
In this section, we compare the individual observed luminosity functions with those of matched simulated hosts from the IllustrisTNG simulations. In Figure \ref{fig:mw_lf_comp}, we compare the observed LFs with those predicted from the models for the 8 `MW-sized' hosts. The IllustrisTNG hosts are selected based on their stellar mass from the hydro results to roughly match the stellar mass of each observed host. In particular, each TNG host is given a probability to be included given by a Gaussian distribution in log stellar mass centered on the stellar mass of the observed host with spread 0.1 dex. We assume 0.1 dex is an appropriate estimate of the error in determining the stellar mass of nearby massive galaxies \citep[e.g.][]{leroy2019}. Thus the distribution of the stellar mass of selected TNG hosts is peaked at the observed stellar mass of the LV host but allows some spread due to possible measurement error. There are roughly 500 TNG hosts selected for each observed host following this prescription. 

To account for the different survey area coverage between the observed hosts, for each observed host, the models are forward modeled through the survey area selection function for that specific host. For comparison with the MW and M31, all model satellites within 300 kpc of the host are included. For the hosts surveyed in the current work, the area coverage of each host is taken from the survey footprints shown in Figure 1 of \citet{LV_cat}. For each observed host (other than the MW and M31), the model hosts are mock observed from a random direction at the distance of the real host and satellite galaxies are selected that project into the survey footprint. For the non-circular footprints shown in Figure 1 of \citet{LV_cat}, a random direction is taken to be North. 

To account for uncertainties in the distances to the dwarf satellites, model satellites are selected within 500 kpc of the host along the line of sight. This will include some splash-back satellites and field dwarfs that have not yet fallen into their host, but presumably the observed satellite systems include a few of these dwarfs as well. The 500 kpc limit is chosen as a compromise between the hosts that have had their satellites confirmed with TRGB and those that have had their satellites confirmed with SBF. \emph{HST} TRGB can yield distances accurate to 5\% which at $D=7$ Mpc is $\sim300$ kpc, whereas SBF, as applied here, can yield distances accurate to 15\% which at $D=7$ Mpc is $\sim1$ Mpc. Our results are qualitatively unchanged if a larger (1 Mpc) line-of-sight limit is used instead.

For the systems that had inconclusive SBF distance constraints for some of their candidate satellites, Figure \ref{fig:mw_lf_comp} shows the spread of possible LFs, given the uncertain/possible members.

\begin{figure*}
\includegraphics[width=\textwidth]{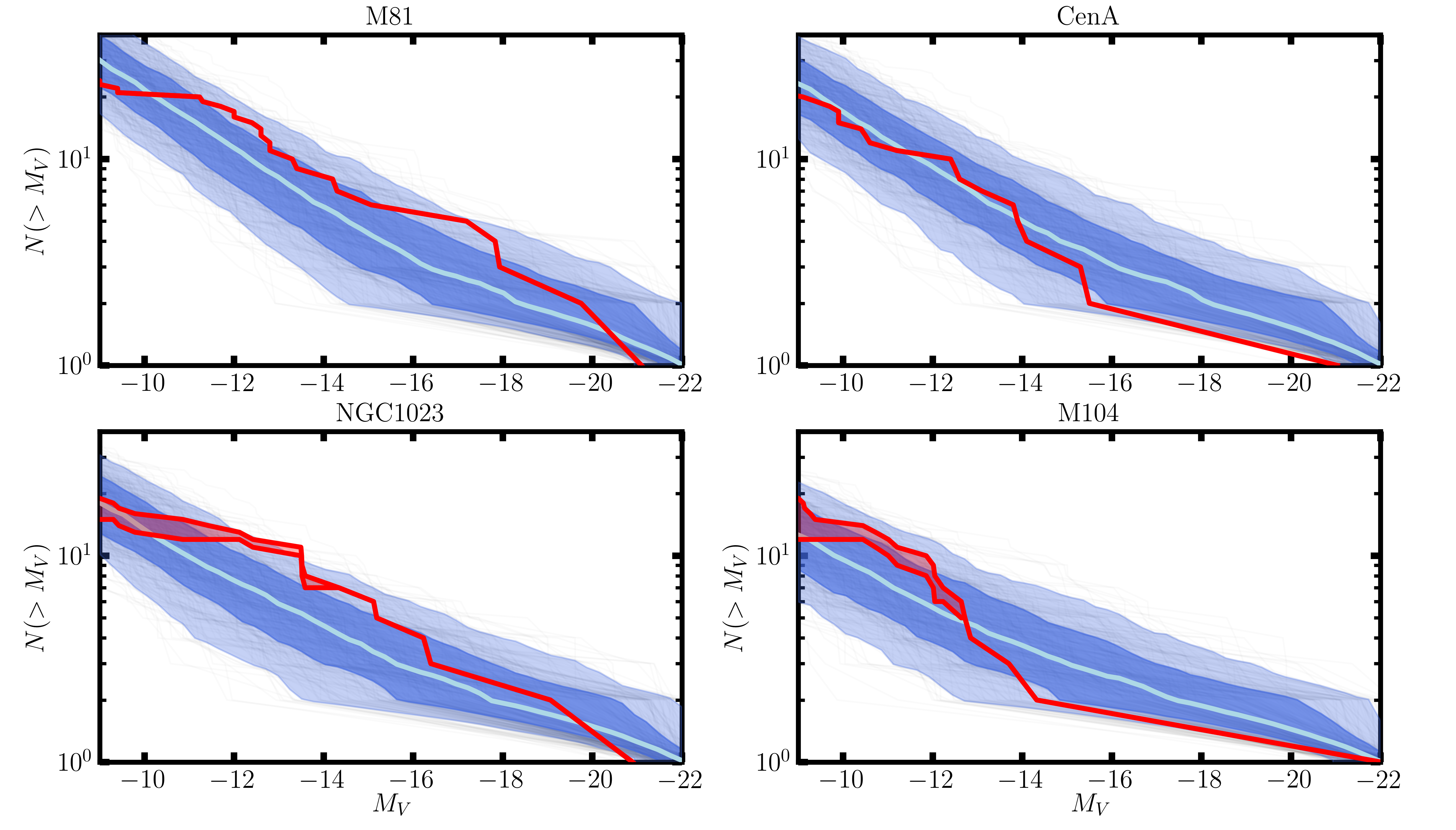}
\caption{The cumulative luminosity functions for the 4 `small-group' hosts that have been well surveyed for satellites (red). The thin black lines show the predicted LFs from the abundance matching model described in the text. The simulation hosts have been selected to have roughly the same stellar mass as the corresponding observed host. The blue regions show the $\pm 1,2 \sigma$ spread in the models. The luminosity completeness is different for each host but is  $M_V\sim-9$ in all cases. For each host, the model satellite systems have been forward modeled considering the survey area selection function for that specific host. For the hosts that had inconclusive results from the SBF distances of their satellite candidates, a spread of possible LFs is shown, accounting for uncertain membership.}
\label{fig:group_lf_comp}
\end{figure*}

In Appendix \ref{app:elvis}, we show the comparison between the observed LFs and those predicted using the ELVIS high-resolution zoom DMO simulations. The results are similar to Figure \ref{fig:mw_lf_comp}, demonstrating that the resolution of IllustrisTNG does not appear to be affecting our results. The ELVIS-predicted LFs are noticeably richer than the TNG LFs because the ELVIS host halos are more massive, on average, than the TNG `MW-like' hosts, as discussed above. Recall that the TNG hosts are selected to have stellar masses similar to the observed hosts (most often with halo masses around $\sim1\times10^{12}$ \msun), but the ELVIS hosts have halo mass fairly evenly distributed between $1-3\times10^{12}$ \msun. Another reason is that the ELVIS subhalos do not experience the enhanced tidal disruption of the central disk, but the TNG subhalos do. The disk can reduce subhalo counts within $r\sim100$ kpc (3D radius) by roughly a factor of two \citep{gk_lumpy}. The reduction of subhalo numbers is less in projection.

There are several interesting things to note from Figure \ref{fig:mw_lf_comp}. The first is that the overall abundance of satellites is well-matched by the SHMR model. This confirms the result of \citet{gk_2017} that this SHMR can reproduce the stellar mass function of MW satellites. However, the observations seem to all fall above the model for bright ($M_V\lesssim-15$ mag) satellites. We will investigate this more closely in \S\ref{sec:average_lf}. These bright satellites appear to be rare in the model hosts. The second is that while there is large spread between the observed systems, the luminosity functions all fall within the $\pm2\sigma$ spread of the models. While M94 is clearly a deficient satellite system \citep{smercina2018}, it is still within the $\pm2\sigma$ spread of the models. We explore the scatter between systems in more detail in the next section.

Figure \ref{fig:group_lf_comp} shows the analogous results for the more massive (`small-group') hosts. The increased richness of these satellite systems is well reproduced in the SHMR model. There are fewer hosts of this mass in the TNG-100 volume, and this is reflected in the smaller number of model lines in Figure \ref{fig:group_lf_comp}. While still within the scatter of the models, M104 and CenA show a larger than typical magnitude gap between first and second brightest group member. It is possible that this is related to the elliptical morphology of these two galaxies, but we leave further exploration of this to future work.

\subsection{Satellite Richness versus Stellar Mass}
\label{sec:n150}
The main goal of this section is to quantify the host-to-host scatter in the observations and compare with that of the models. More massive halos are expected to host more subhalos. While we have rough halo mass estimates for each host in our sample (see above) these estimates are not accurate enough to explore how satellite richness depends on halo mass. Instead, in this section we explore how satellite richness depends on stellar mass, which we use as a proxy for halo mass.

\begin{figure*}
\includegraphics[width=\textwidth]{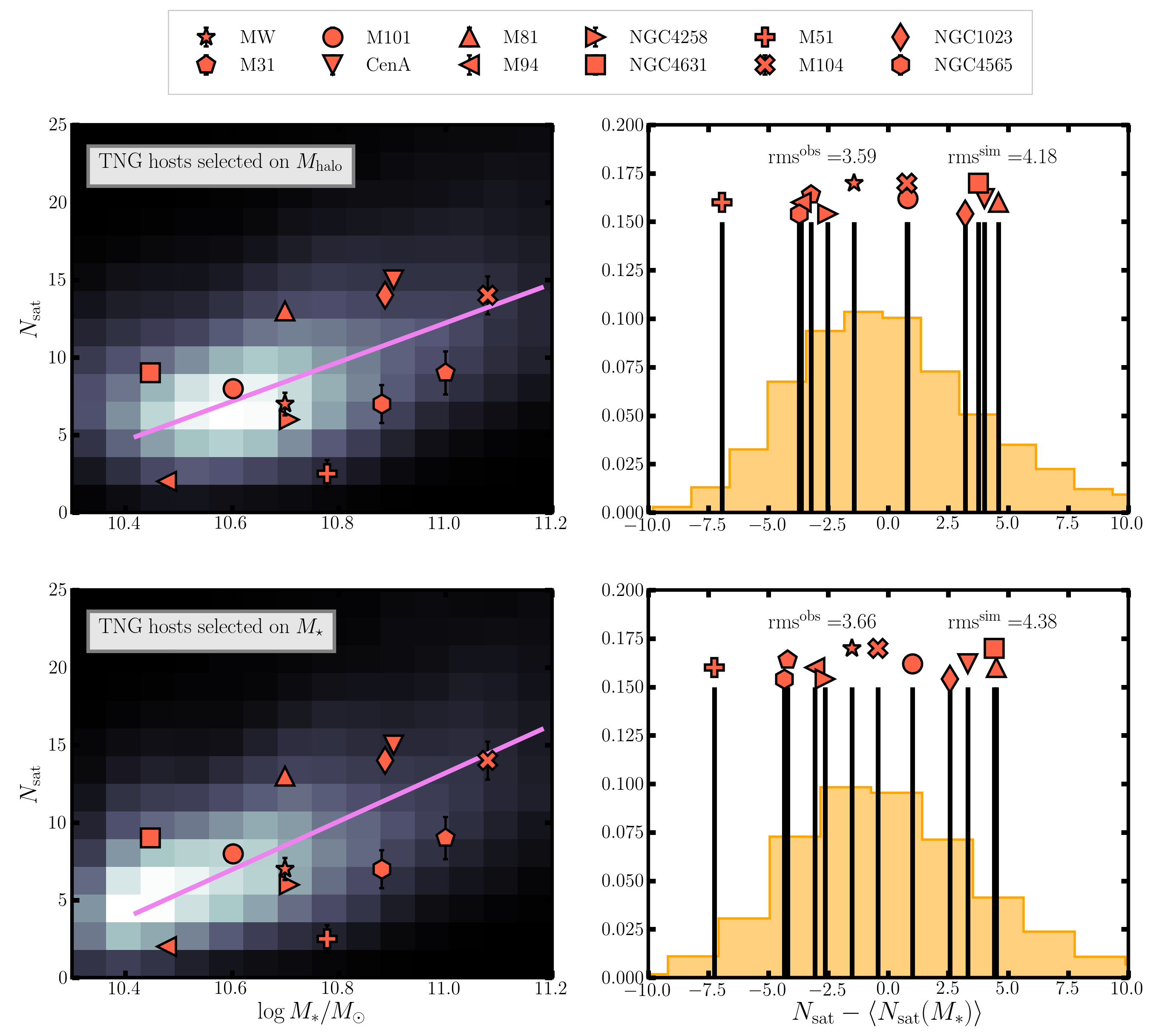}
\caption{Left column: the number of satellites $M_V<-9$ within a projected radius of 150 kpc for each observed host (indicated by the symbols) as a function of host stellar mass. The background color-map shows the results for the simulated hosts. The simulated hosts are hosts drawn from Illustris-TNG100 combined with the stellar halo mass relation (SHMR) of \citet{gk_2017}. The purple line shows the average relation for the simulated hosts. Right column: the residual in the number of satellites, corrected for the average relation of the simulated hosts. The rms scatter agrees fairly well between the simulated and observed hosts. The top panels use TNG hosts selected on halo mass while the bottom panels use a stellar mass cut on TNG hosts.}
\label{fig:n150}
\end{figure*}

Figure \ref{fig:n150} shows the relation between satellite richness and stellar mass for both the observed hosts and the simulated hosts. We show the comparison when we select TNG hosts based on halo mass in the range $0.8\times10^{12}<M_{200}<8\times10^{12}$ \msun\ and also on stellar mass in the range $10^{10.3}<M_\star<10^{11.2}$ \msun. This shows that the results are largely unaffected by the specifics of how we select the simulated hosts.

To account for the different area coverage of the different hosts, only the satellites within 150 projected kpc are included, and we make the assumption that each observed host is complete to this radius.  For the MW and M31, for which we have detailed 3D locations of the satellites, the observed satellite systems are mock observed at a distance of 7 Mpc (which is roughly the average distance of the LV hosts). The errorbars show the spread ($\pm1\sigma$) in the satellite number for many different viewing directions. For the systems with inconclusive SBF results, the errorbars show the spread ($\pm1\sigma$) in possible satellite richness accounting for this uncertainty\footnote{Specifically, each uncertain member is given a 50-50 chance of being a real satellite.}. The simulated systems are mock observed at a distance of 7 Mpc. Only satellites brighter than $M_V<-9$ are included. For the simulations, the stellar mass used for each host is the actual stellar mass for that host predicted by the hydrodynamic component of IllustrisTNG, not a stellar mass from abundance matching (the results are unchanged if we use a stellar mass predicted from the SHMR).

There is clearly a positive relation between host stellar mass and satellite richness, in both the observed hosts and the simulated hosts. The steepness of the relation between host stellar mass and satellite richness appears to agree quite well between the observations and the model predictions. The purple line shows the average trend of the simulated hosts. The right panels of Figure \ref{fig:n150} shows the number of satellites corrected for the general trend of the models with host stellar mass. Both the models and observations are symmetric around zero, indicating that the SHMR we use accurately reproduces the normalization of the satellite luminosity functions. Also shown in the plot is the rms scatter of the observed systems and the simulated systems. They agree well, indicating that the host-to-host scatter in the observed systems is quantitatively what one would expect from the models, once variations in the host mass are accounted for. The scatter in the observed hosts is actually somewhat below the scatter in the simulated hosts. Framed this way, M51 is even more deficient in satellites than the M94 system, considering its higher stellar mass.

This result shows that the observed host-to-host scatter in satellite richness amongst nearby MW-like systems is comparable to that predicted by $\Lambda$CDM simulations without the need of greatly increased scatter in the SHMR. Thus, we do not confirm the conclusions of \citet{smercina2018} who argued that significantly increased scatter is required to explain M94's satellite system. We come to a different conclusion for a few reasons. First, we show that much of the observed scatter between hosts is due to the difference in host halo masses (as proxied by stellar mass). This is important to take into account when inferring host-to-host scatter. M94 has a low abundance of satellites largely because M94 has a relatively small stellar mass amongst `MW analogs'. Second, our sample of 12 systems offers much improved statistics over the five considered by \citet{smercina2018}. Finally, the average number of satellites that our model predicts seems to be somewhat lower than that of \citet{smercina2018}. M94 appears to be a $>3\sigma$ outlier from their simulated LFs while is is only $\sim1-2\sigma$ for ours (see Figure \ref{fig:mw_lf_comp}). It is unclear where this discrepancy originates since we use a very similar SHMR as that used by \citet{smercina2018}. However, we note that our model reproduces the mean observed satellite abundance well (Figure \ref{fig:n150}), while the model of \citet{smercina2018} produces too many satellites compared to all 5 observed hosts they compare with. With this said, it certainly is possible that a SHMR with large scatter could also reproduce the host-to-host scatter, but we have shown that it is not needed.

\begin{figure*}
\includegraphics[width=\textwidth]{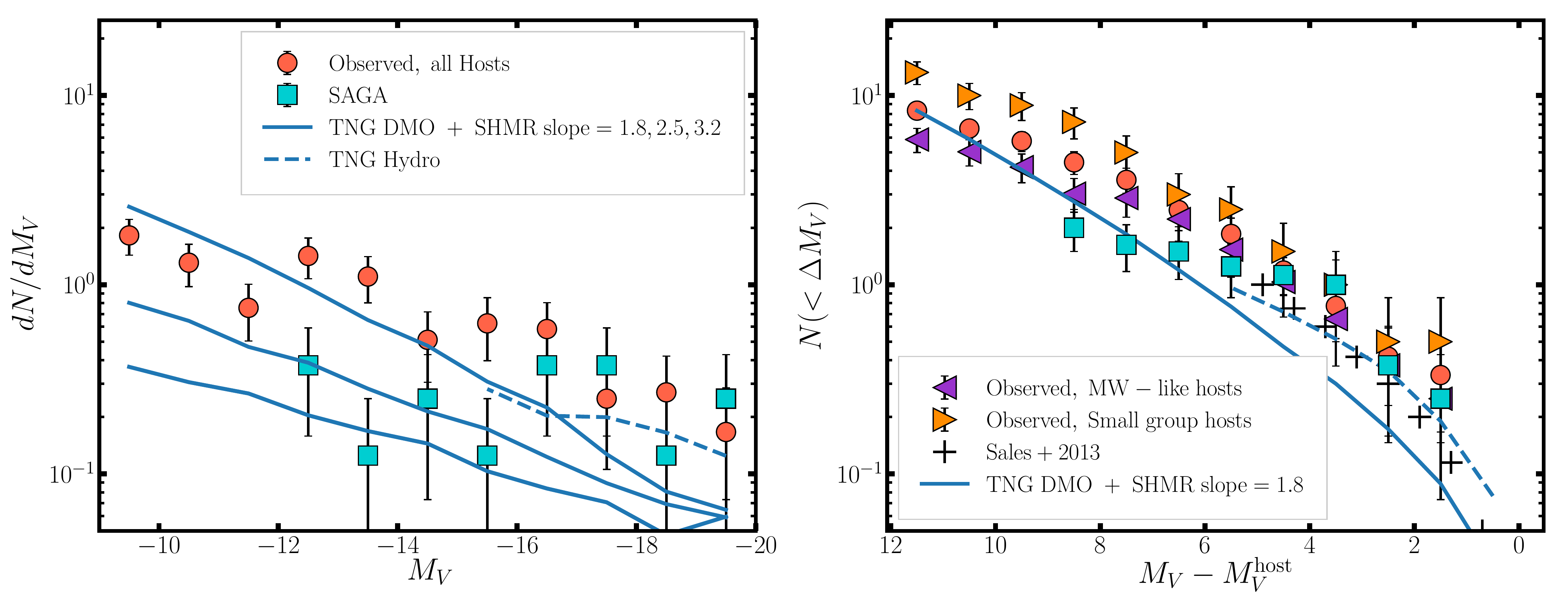}
\caption{\textit{Left}: The average differential LF of all 12 observed hosts we consider for satellites within 150 kpc projected of the host.  The average simulated LFs are shown in blue.  Curves corresponding to three different values of the low mass slope of the SHMR of \citet{gk_2017} are shown. The smallest slope value corresponds to the highest (most rich) LF. Note that no matter the slope used in the SHMR, the simulated LF cannot match the shape of the observed LF. There are too many observed bright satellites. \textit{Right}: The same observational samples are shown as a cumulative LF where the satellite luminosities have been scaled by the luminosity of the host. The LV host sample are shown all together and also split into the MW-like hosts and `small-group' hosts (see \S\ref{sec:lfs} for discussion and definition). The average SAGA \citep{geha2017} LF is shown in turquoise. The SAGA results agree well with the Local Volume hosts for the bright satellites ($M_V<-14$), but include fewer faint satellites. The satellite LF function results from the SDSS analysis of \citet{sales2013} for hosts in the mass range $10.5<\log{M_\star}<11$ are also shown. Errorbars show the Poisson scatter in the total number of satellites in each magnitude bin.}
\label{fig:average_lf}
\end{figure*}

\subsection{Average LF Shape}
\label{sec:average_lf}
In this section, we explore the shape of the LFs in detail. The shape of the LF is a sensitive probe of the low-mass slope of the SHMR. To make the comparison, we construct the average differential luminosity function of the 12 observed hosts by considering the total number of satellites in different magnitude bins and compare with the average LF of the simulated hosts. Figure \ref{fig:average_lf} (left panel) shows this comparison for satellites with $M_V<-9$ (and also $\mu_{0,V}<26.5$ mag arcsec$^{-2}$) within a projected separation of 150 kpc of their host,  assuming all 12 observed hosts are complete at this level. Therefore, the average LF will be a lower bound to the true LF since our hosts are not quite complete to 150 kpc. To account for the effect of projection angle on the satellite systems of the MW and M31, we average over many different projection angles. The uncertain membership of some satellites is also accounted for by averaging over all possible combinations of the uncertain satellites being members or not.

We compare with the TNG simulation results. Subhalos around the simulated hosts are selected in the same way as in the previous section. The TNG hosts are selected based on halo mass in the range $0.8\times10^{12}<M_{200}<8\times10^{12}$\msun. We compare the observed average LF to the average simulated LF using the fiducial SHMR (with low mass slope of 1.8) along with the result of using a slope of 2.5 and 3.2. The left panel of Figure \ref{fig:average_lf} shows that no matter the low mass slope used in the SHMR of \citet{gk_2017}, the average simulated LF will not quite match the shape of the observed LF. The observed LF has too many bright satellites and an overall flatter LF. 

\begin{figure}
\includegraphics[width=0.5\textwidth]{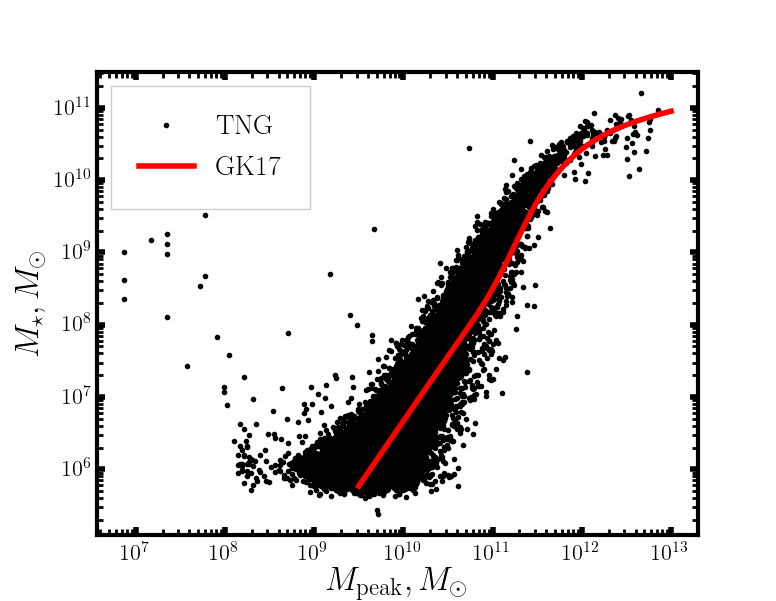}
\caption{The inferred SHMR for satellites of hosts in the stellar mass range $10^{10.3}<M_\star<10^{11.2}$ \msun~ in the hydrodynamic results of the IllustrisTNG simulation. Satellites are selected as subhalos within 300 kpc of their hosts. Note the higher normalization of the hydro SHMR compared to that of \citet{gk_2017} around $M_\mathrm{peak}\sim10^{11}$ \msun. }
\label{fig:shmr_hydro}
\end{figure}

Here, we also make use of the full hydrodynamic results of the TNG simulation and compare with the bright ($M_V<-16$) end of the observed satellite LF. In this case, the agreement with the observations is much better, although they are still lower than the observed LF. Part of this might be due to the set of observed hosts having more numerous massive `small-group' hosts than the set of simulated halos. We estimate four out of twelve of the observed hosts are in this more massive category while ~1/5 of the simulated halos are in that mass range. It appears that the TNG hydrodynamic results are in better agreement with the observations because halos with $M_h\sim10^{11}$ \msun\ end up with more stars than predicted with the SHMR. The stellar and halo mass of galaxies in the hydrodynamic TNG results are shown in Figure \ref{fig:shmr_hydro} compared to the SHMR of \citet{gk_2017}. The hydrodynamic results have a noticeably higher normalization at $M_h\sim10^{11}$ \msun\ which effectively leads to a higher abundance of bright satellites.

To explore whether the spread in host luminosities is affecting the average LF shape, we normalize by the host luminosity in the right panel of Figure \ref{fig:average_lf}. This approach has been used to look at the shape of the average LF when including hosts of different luminosity (and mass) before \citep[e.g.][]{sales2013, nierenberg2016}. Framed this way, the observed satellite systems still have more bright satellites than the SHMR model predicts. The hosts all have $M_V$'s within roughly 1-2 mag of each other, so this normalization does not greatly affect the shape of the LF.

We also compare with the results from SDSS reported in \citet{sales2013}. The results of \citet{sales2013} come from spectroscopically confirmed satellites in SDSS. \citet{sales2013} report $\Delta M_r$ which we assume is roughly equivalent to $\Delta M_V$. Additionally, their results include all satellites in the virial radius of the hosts ($\sim300$ kpc). To roughly compare with our satellite LFs that only include satellites in the inner 150 kpc, we simply divide their satellite counts by two. This is roughly the fraction of satellites that the SAGA Survey \citet{geha2017} find within 150 kpc of the host compared to within 300 kpc. We find fair agreement with their satellite counts in the region of overlap. Their results extend only to $\Delta M=5$ whereas our results extend 7 magnitudes fainter in satellite luminosity.

In this figure, we compare with the average observed LF of the 8 MW analogs of the first SAGA release \citep{geha2017}. We estimate that 8 out of our 12 hosts would qualify as `MW-analogs' according to the criteria of \citet{geha2017}, including some of the `small-group' hosts, so this is a fairly reasonable comparison. Only one host (M104) is above their sample range of $M_K$ ($-23<M_K<-24.6$), and one (NGC 4631) is actually below this range. Two others (M51 and M81) would not qualify due to the presence of a bright nearby companion (NGC 5195 and M82, respectively)\footnote{It is unclear how the second environmental cut that SAGA uses which removes hosts that are within two virial radii of a $5\times10^{12}$ \msun\ host from the 2MASS group catalog would restrict the LV sample.}. In comparing with the SAGA results, we assume $M_V\sim M_r+0.2$ and only take satellites within 150 kpc of their host. These hosts also show a surplus of bright satellites and a flatter LF slope. This was noticed by \citet{geha2017} and \citet{zhang2019}. The SAGA hosts appear to have fewer satellites in the range $-14<M_V<-12$ than our observed hosts.  Even conservatively limiting to our `MW-like' hosts, the LV hosts are richer at these magnitudes. This is possibly indicating some incompleteness in the SDSS catalogs used in SAGA, although a more detailed comparison of the host samples is merited to understand this difference.

The observed systems are the most discrepant from the SHMR predictions around $M_V\sim-16$ to $-17$, but they are in surplus at even brighter, LMC-like magnitudes as well. This holds for both the `MW-like' and `small-group' hosts. Several observational results have argued that Magellanic Cloud (MC) analogs are fairly rare around MW-analogs \citep[e.g.][]{tollerud2011, liu2011}. \citet{liu2011} find that 81\% of MW-analogs in SDSS do not have any MC-like satellite within a projected 150 kpc, with 11\% having one and 4\% having two. They define a MC-analog as a satellite between 2 and 4 magnitudes fainter than the host. We note that 5 (NGC 4631, MW, M31\footnote{Depending on whether M33 is projected within 150 kpc.}, M81, and M101) of the 12 LV hosts would qualify as having one or more MC-analogs according to this definition. Similarly 6 out of 8 of the SAGA hosts have at least one satellite within 2 and 4 magnitudes fainter than the host and within 150 kpc projected. \citet{liu2011} define a MW-analog as having $-21.4<M_r<-21.0$. Most of the SAGA hosts are within this range while some of our hosts are above and some are below.

We note that we are assuming a constant $M/L_V$ ratio in the abundance matching model for all of the simulated galaxies. It does seem feasible that the discrepancy in the LF could be due to changing $M/L_V$ ratio for different luminosity satellites due to different star formation histories. Brighter satellites will likely continue to form stars longer after infall than very faint satellites \citep{fillingham2015}. However, as mentioned above, the satellites do not exhibit a strong color vs luminosity trend.

\subsection{Summary and Implications for the SHMR}
\label{sec:shmr}

In \S\ref{sec:results}, we have used a SHMR combined with DMO simulations to generate model satellite systems and compared them with the observed systems. In particular, we used the SHMR of \citet{gk_2017}.  This SHMR is well reproduced in several high-resolution zoom hydrodynamic simulations. Both the FIRE \citep{fitts2017} and NIHAO \citep{buck2019} projects produce galaxies that fall on or near this relation (see Figure 6 of \citet{gk_2017} for a detailed comparison with simulation results). In particular, \citet{buck2019} find a SHMR with a slope of 1.89 for their simulated dwarf satellites, very similar to the slope we adopt here. We note, however, that there is still significant scatter in the predicted SHMR among different simulation projects \citep[see e.g.][]{agertz2020}. 

We found that the overall number of satellites and host-to-host scatter of the observations was closely matched by this SHMR. However, we find that the SHMR of \citet{gk_2017} does not quite match the observed shape of the composite satellite LF. This is independent of the assumed slope of the relation in the low-mass regime, suggesting that the problem is in the normalization of the standard \citet{behroozi2013} SHMR around $\sim10^{11}$ \msun. We find a similar result if we use other popular SHMRs from the literature, including that of \citet{rp17} and \citet{moster2018}. Both of these SHMRs have the same or lower normalizations than \citet{behroozi2013}, causing a lack of bright satellites. The SHMR of \citet{brook2014}, which comes directly from abundance matching the LG satellites, has a higher normalization at $\sim10^{11}$ \msun, and does more closely match the average observed LF. 

The natural next question is: What SHMR does reproduce the observed LFs? Fitting for a SHMR is beyond the scope of the current paper but will be an important avenue for future work. It will also be important to explore how a different SHMR might change our conclusion that the host-to-host scatter in satellite abundance for simulated galaxies closely matches that of observed galaxies. On a broader note, the LFs shown in Figure \ref{fig:average_lf} show that the statistics are already sufficient with the current sample of LV satellites systems to place powerful constraints on what the SHMR can be in the low-mass regime.

\section{Conclusions}
\label{sec:concl}

The dwarf satellites of the MW are a premier probe of small-scale structure formation and the properties of dark matter. Dwarf galaxies are also important probes of galaxy formation, particularly of the effect of stellar feedback, and dwarf satellites, in particular, are sensitive to the effect of quenching by a massive host. The effectiveness of these processes may differ on a host-to-host or even satellite-to-satellite basis. To get a full picture of small scale structure, satellite systems beyond the MW must be studied to comparable levels of detail. We do not yet have a sense of what a `normal' satellite system is and, thus, no way of knowing if the MW satellites (in properties or abundance) are ``typical''. In studying the low-mass satellites around hosts other than the MW, the limiting step is usually the difficulty in getting distances to candidate satellites to confirm their association with a host. In this paper we measure the distance to candidate satellites around a large number of hosts and perform an in-depth analysis of their satellite luminosity functions to investigate the stellar to halo mass relation of low-mass dwarf galaxies.

We used surface brightness fluctuation measurements to confirm satellite candidates identified in \citet{LV_cat} around several hosts in the Local Volume. The SBF analysis cleans the satellite systems of background contaminants, allowing for an in-depth analysis of the satellite abundance and properties, previously only possible for a handful of very nearby systems. There were six hosts (NGC 1023, NGC 4258, NGC 4565, NGC 4631, M51, and M104) whose survey footprints were a significant portion of the host's virial volume and had usable SBF results. The remaining four (NGC 1156, NGC 2903, NGC 5023, and M64) either had ambiguous SBF results with most candidates remaining unconstrained or had very limited survey area coverage.

The systems with nearly complete distance constraints show significant scatter in the amount of background contamination present in each field. This scatter completely overwhelms the true host-to-host scatter in the abundance of satellites, highlighting the importance of getting distances to candidate satellites discovered around nearby galaxies.

For the group of well-surveyed systems, we explore the luminosity functions of these satellite systems in more detail. We combine this sample of six with a sample of six nearby hosts that have been previously well-surveyed for satellites. This is by far the largest sample of nearby roughly MW-sized hosts whose satellite systems have been surveyed down to approximately the faintest classical satellites. Instead of considering all of these systems together, we separately consider the hosts that are the most MW-like (NGC 4258, NGC 4565, NGC 4631, M51, MW, M31, M94, and M101) and the hosts that are somewhat more massive (NGC 1023, M104, CenA, M81), which we refer to as `small-group' hosts. The more massive systems have clearly more rich satellite systems than the MW-like hosts, and we see a clear correlation between satellite abundance and host stellar mass (cf. Figures \ref{fig:lfs} and \ref{fig:n150}). We find that the LF of MW satellites is remarkably typical compared to the other MW-like hosts.

To further interpret the luminosity functions of the observed satellite systems, we develop a simple model based on $N$-body cosmological simulations coupled with a stellar-to-halo mass relation (SHMR). Luminous galaxies are painted onto the DMO results with a SHMR. The fiducial SHMR we use is known to reproduce the normalization of the luminosity function of the MW and agrees fairly well with the results of high resolution hydrodynamic simulations from multiple projects. The predicted satellite systems from this model are able to well reproduce both the normalization and spread of the observed satellite systems, for both the `MW-like' hosts and the `small-group' hosts (cf. Figures \ref{fig:mw_lf_comp} and \ref{fig:group_lf_comp}). 

We consider the satellite richness as a function of the host stellar mass, which we use as a rough proxy for the host halo mass. Both the observed systems and simulated systems show a similar positive relation between satellite number and host stellar mass. Using this relation, we quantitatively show, for the first time, that the observed systems exhibit the same host-to-host scatter as the simulated systems once host mass is accounted for, without the need to invoke increased scatter in the SHMR (cf. Figure \ref{fig:n150}). Thus, we do not confirm previous results that conclude the observed scatter is more than expected from simulations \citep[e.g.][]{smercina2018, geha2017}. This difference is due to a combination of our use of a larger sample of observed hosts and also carefully accounting for the fact that the observed hosts have different masses (stellar and halo).

Finally, we consider the average shape of the observed LF and compare with the average simulated LF. We find that while the simulations and SHMR can produce the right total number of satellites, the simulations seem to under-produce bright satellites and over-produce faint ones (cf. Figure \ref{fig:average_lf}). This appears to be independent of the power law slope of the SHMR in the low mass regime, as long as the SHMR is fixed to the relation of \citet{behroozi2013} at higher masses (halo mass of $\sim10^{11}$ \msun). The hydrodynamic results of IllustrisTNG over the range in satellite luminosities that are resolved in the hydrodynamic simulation ($M_V<-16$) seem to show better agreement with the observations. Our observations seem to require a higher normalization of the SHMR around a halo mass of $\sim10^{11}$ \msun\ in order to match the observed abundance of massive satellites.

We find that our average LF agrees quite well with the initial SAGA Survey \citep{geha2017} results at the bright end which show a similar surplus of bright satellites. The LV systems do show significantly more faint ($M_V\sim-13$) satellites than the SAGA results, however. 

The true SHMR valid for this low-mass regime remains a significant open question in the field of dwarf galaxy formation and near-field cosmology. In the future, the observed satellite systems around the MW and similar nearby hosts will continue to play a significant role in constraining the SHMR. With the Vera Rubin Observatory, the observational sample has the potential to grow tremendously with SBF playing a facilitating role in providing distances to low mass satellites in systems out to 20 Mpc \citep[e.g.][]{greco2020}.

\section*{Acknowledgements}
Support for this work was provided by NASA through Hubble Fellowship grant \#51386.01 awarded to R.L.B.by the Space Telescope Science Institute, which is operated by the Association of  Universities for Research in Astronomy, Inc., for NASA, under contract NAS 5-26555. J.P.G. is supported by an NSF Astronomy and Astrophysics Postdoctoral Fellowship under award AST-1801921. J.E.G. is partially supported by the National Science Foundation grant AST-1713828. S.G.C acknowledges support by the National Science Foundation Graduate Research Fellowship Program under Grant No. \#DGE-1656466. AHGP is supported by National Science Foundation Grant Numbers AST-1615838 and AST-1813628.

Based on observations obtained with MegaPrime/MegaCam, a joint project of CFHT and CEA/IRFU, at the Canada-France-Hawaii Telescope (CFHT) which is operated by the National Research Council (NRC) of Canada, the Institut National des Science de l'Univers of the Centre National de la Recherche Scientifique (CNRS) of France, and the University of Hawaii.

\software{\texttt{SExtractor} \citep{SExtractor}, \texttt{sep} \citep{sep}, \texttt{Scamp} \citep{scamp}, \texttt{SWarp} \citep{swarp}, \texttt{astropy} \citep{astropy}, \texttt{imfit} \citep{imfit}}

\bibliographystyle{aasjournal}
\bibliography{calib}

\appendix

\section{Details on the SBF Results}
\label{app:sbf_details}
The main results of the SBF analysis are given in Tables \ref{tab:ngc1023_sbf}-\ref{tab:m104_sbf}. In these tables, we only list the galaxy name and the SBF results. To remind the reader, we split the dwarfs into three categories: confirmed physical satellites, confirmed background contaminants, or galaxies where no SBF constraint is possible that we refer to as `unconfirmed' or `possible' satellites. More information, including photometry can be found in \citet{LV_cat}. For convenience, we include the photometry for the confirmed and possible satellites in Appendix \ref{app:sat_systems_tables}. Physical sizes and absolute magnitudes are included in those tables. In the following sub-sections, we go through each host and discuss the SBF results in detail, focusing on the dwarfs that are exceptions to our general classification guidelines of confirmed/background/unconstrained. We also describe any auxiliary distance information used in confirming or discarding satellites.

\begin{figure*}
\includegraphics[width=\textwidth]{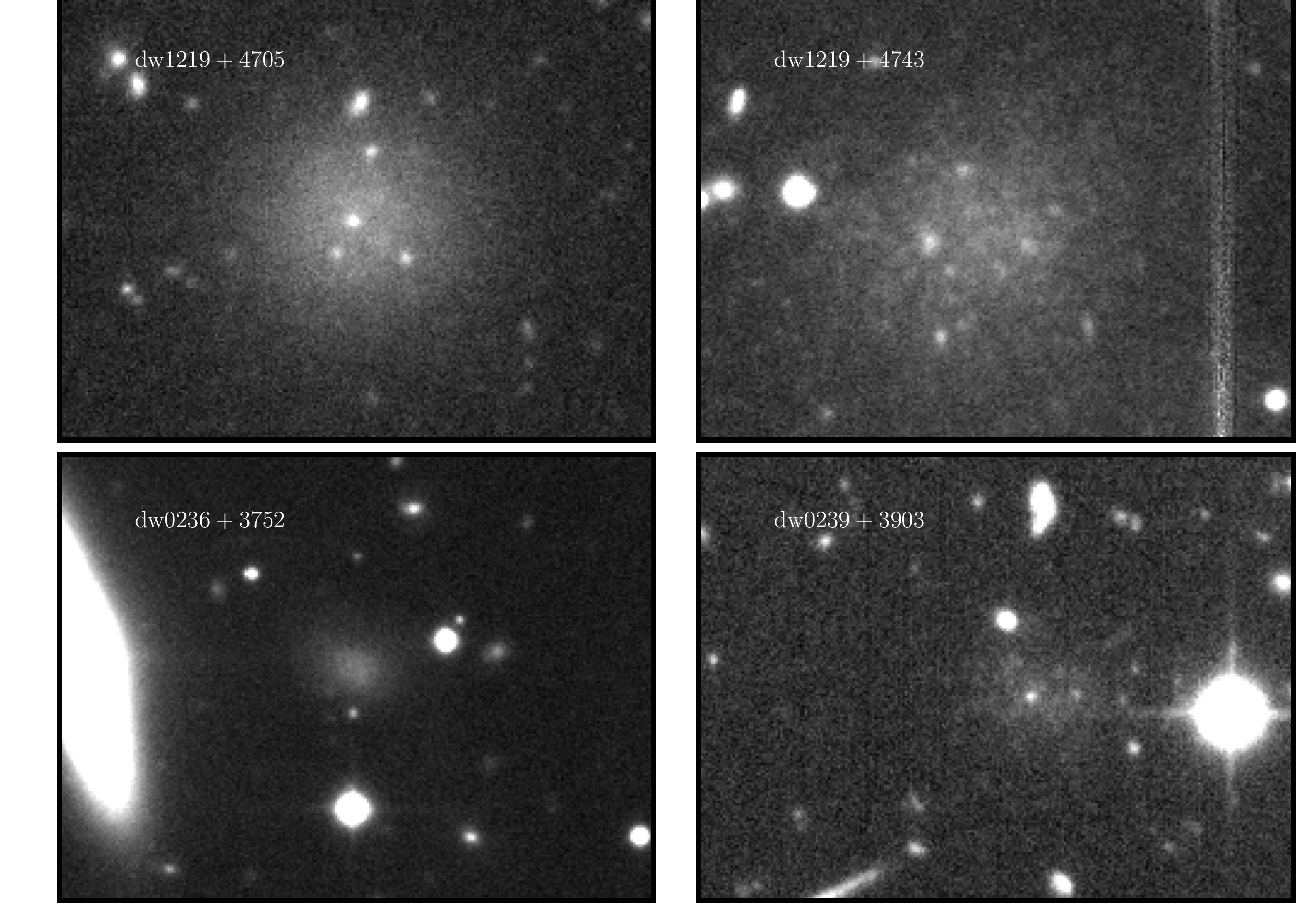}
\caption{The left column shows examples of galaxies that we constrain to be background in the SBF analysis and the right column shows examples of confirmed satellites. The top row galaxies are from the NGC 4258 ($D=7.2$ Mpc) region and the bottom row are from NGC 1023 ($D=10.4$ Mpc). The real satellites exhibit clearly visible SBF while the background galaxies are nearly perfectly smooth. The pairs of galaxies from each region are roughly matched in surface brightness, size, and color. Each image is 45\arcs~ wide. Top row is $r$ band and the bottom row is $i$.}
\label{fig:bkgd_ex}
\end{figure*}

To show examples of these three categories, Figure \ref{fig:bkgd_ex} shows examples of galaxies that we conclude to be background along with examples of galaxies that we conclude to be real satellites from the same host. The galaxies that we constrain to be background are roughly the same surface brightness as the confirmed satellites but show visibly smoother surface brightness profiles without any SBF. The example background galaxy from the NGC 4258 region (dw1219+4705) was confirmed to be background by \citet{cohen2018} as well.  

\begin{figure}
\includegraphics[width=0.45\textwidth]{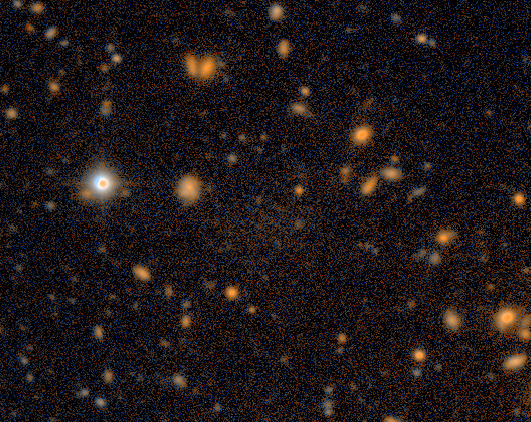}
\caption{An example of a dwarf (dw1218+4623) that was too low surface brightness to be either confirmed as a satellite or constrained to be background. }
\label{fig:unconf_ex}
\end{figure}

Figure \ref{fig:unconf_ex} shows an example dwarf that is too low surface brightness to be confirmed as either a satellite or background contaminant with the current data.

\subsection{NGC 1023}
\label{sec:sbf_ngc1023}

\begin{deluxetable*}{ccc|ccc|ccc}
\tablecaption{NGC 1023 SBF Results\label{tab:ngc1023_sbf}}

\tablewidth{\textwidth}

\tablehead{\colhead{} & \colhead{Confirmed} &\colhead{} & \colhead{} & \colhead{Possible} &\colhead{} & \colhead{} & \colhead{Background} &\colhead{} \\ 
\colhead{Name} & \colhead{SBF S/N} & \colhead{Dist (Mpc)}  & \colhead{Name} & \colhead{SBF S/N} & \colhead{Dist (Mpc)} & \colhead{Name} & \colhead{SBF S/N} & \colhead{Dist (Mpc)}  } 

\startdata
dw0233+3852 & 5.7 & $12.5^{+2.0,4.5}_{-1.7,3.1}$ & dw0238+3805 & -- & --  & dw0234+3800 & 4.8 & $>17.5$ \\ 
dw0235+3850 & 11.3 & $11.7^{+0.9,1.9}_{-0.9,1.8}$ & dw0239+3910 & 2.2 & $11.8^{+4.4,\infty}_{-2.6,4.4}$ & dw0236+3752 & 5.5 & $>11.9$ \\ 
IC 239 & -- & --  & dw0241+3852 & 2.1 & $15.3^{+6.3,\infty}_{-3.4,5.6}$ & dw0236+3925 & 1.5 & $>14.1$ \\ 
dw0237+3855* & 37.5 & $7.1^{+0.5,1.0}_{-0.5,1.0}$ & dw0241+3829 & 1.9 & $12.7^{+5.7,\infty}_{-2.5,3.9}$ & dw0237+3903 & 1.9 & $>13.1$ \\ 
dw0237+3836 & 20.8 & $10.6^{+0.8,1.6}_{-0.8,1.7}$ & dw0242+3757 & 0.4 & $16.0^{+\infty,\infty}_{-7.5,9.8}$ & dw0238+3808 & -0.7 & $>16.8$ \\ 
dw0239+3926 & 14.4 & $10.8^{+0.7,1.4}_{-0.7,1.5}$ & dw0243+3915 & 0.9 & $14.0^{+\infty,\infty}_{-4.3,6.2}$ & dw0239+3824 & 1.3 & $>13.3$ \\ 
dw0239+3903 & 8.2 & $8.6^{+1.8,3.9}_{-1.6,3.0}$ &  & & & dw0240+3844 & 7.0 & $>16.7$ \\ 
dw0239+3902 & 7.0 & $11.5^{+1.2,2.7}_{-1.1,2.1}$ &  & & & dw0240+3829 & -0.2 & $>21.6$ \\ 
UGC 2157 & -- & --  &  & & & dw0241+3923 & 0.4 & $>15.2$ \\ 
dw0240+3854 & 19.1 & $11.2^{+0.5,1.0}_{-0.4,0.8}$ &  & & & dw0241+3934 & 1.6 & $>13.5$ \\ 
dw0240+3903 & -- & --  &  & & &  & & \\ 
dw0240+3922 & 6.3 & $11.6^{+1.2,2.8}_{-1.0,1.9}$ &  & & &  & & \\ 
dw0241+3904* & 22.2 & $12.4^{+0.5,1.1}_{-0.5,0.9}$ &  & & &  & & \\ 
UGC 2165 & 47.5 & $10.7^{+0.8,1.5}_{-0.8,1.8}$ &  & & &  & & \\ 
dw0242+3838 & 6.2 & $9.9^{+1.2,2.6}_{-0.9,1.7}$ &  & & &  & & \\ 
\enddata
\tablecomments{SBF results for candidates around NGC 1023 ($D=10.4$ Mpc). Objects are ordered as confirmed satellites, then possible (still unconfirmed) satellites, and then confirmed background contaminants. The SBF distances give $+1\sigma, +2\sigma$ errors in superscipt and $-1\sigma, -2\sigma$ errors in subscript. Lower distance limits (2$\sigma$) are given for the background objects. Objects with dashes through the measurements were too irregular and no SBF measurement was attempted. The objects that are confirmed without SBF measurements have redshifts. Objects with asterisks (*) are exceptions to the confirmation criteria outlined in \S\ref{sec:distances}, see text for details.}

\end{deluxetable*}
Table \ref{tab:ngc1023_sbf} gives the SBF results for candidate satellites in the field of NGC 1023. Several of the dwarfs had very strong SBF signals that put them at the distance of NGC 1023 ($D=10.4$ Mpc). For four of the candidates, we did not attempt an SBF measurement, either because the candidate was too irregular or because there was too much scattered light from a nearby star. For three of these IC 239, UGC 2157, and dw0240+3903, the candidates have redshifts from \citet{trentham2009} that are within $\pm300$ km/s of NGC 1023. \citet{trentham2009} consider these three to be high confidence members for the NGC 1023 group, and we consider these to be confirmed members as well. All three have visible SBF that looks similar to other confirmed members of similar color. There were two objects dw0237p3855 and dw0241p3904 which had very strong (S/N$\gtrsim20$) SBF signals but distances that were slightly inconsistent with NGC 1023.  As discussed in \citet{LV_cat}, NGC 1023 has contamination from scattered light from bright stars due to its low galactic latitude. Both of these objects were heavily contaminated by scattered light, which is likely causing the discrepant distances. Both objects had visually similar SBF to other confirmed objects, and we consider it very likely that both are genuine members of the group. We note that there are a few candidates that we consider to be background with strong fluctuation signal and $2\sigma$ distance lower bounds only slightly beyond NGC 1023 (e.g. dw0236p3752). These galaxies do not suffer from the same amount of contamination as the confirmed candidates with discrepant distances, and there is no reason to believe the SBF distance is biased in these cases.

\subsection{NGC 1156}
\label{sec:sbf_ngc1156}
\begin{deluxetable*}{ccc|ccc|ccc}
\tablecaption{NGC 1156 SBF Results\label{tab:ngc1156_sbf}}

\tablewidth{\textwidth}

\tablehead{\colhead{} & \colhead{Confirmed} &\colhead{} & \colhead{} & \colhead{Possible} &\colhead{} & \colhead{} & \colhead{Background} &\colhead{} \\ 
\colhead{Name} & \colhead{SBF S/N} & \colhead{Dist (Mpc)}  & \colhead{Name} & \colhead{SBF S/N} & \colhead{Dist (Mpc)} & \colhead{Name} & \colhead{SBF S/N} & \colhead{Dist (Mpc)}  } 

\startdata
 & & & dw0300+2514* & 5.4 & $6.4^{+1.7,3.8}_{-1.4,2.5}$ & dw0300+2518 & 0.7 & $>7.8$ \\ 
 & & & dw0301+2446 & 4.5 & $3.3^{+1.7,4.4}_{-1.2,2.0}$ &  & & \\ 
\enddata
\tablecomments{Same as Table \ref{tab:ngc1023_sbf} for NGC 1156 ($D=7.6$ Mpc). Objects marked with an asterisk (*) are discussed in detail in the text.}

\end{deluxetable*}
The SBF results for NGC 1156 are shown in Table \ref{tab:ngc1156_sbf}.  We do not confirm any of the candidates to be genuine satellites. One of the candidates is likely background, and the other two are possible satellites. One of the possible satellites (dw0300p2514) was above our fiducial S/N$>5$ threshold and had distance consistent with NGC 1156, but due to the galactic cirrus, we could not visually confirm this signal was actual SBF. Thus, we conservatively include this galaxy into the unconfirmed/possible satellite category. dw0300p2514 and dw0301p2446 are the two objects cataloged by \citet{karachentsev2015} and are both promising targets for follow-up.

\subsection{NGC 2903}
\label{sec:sbf_ngc2903}
\begin{deluxetable*}{ccc|ccc|ccc}
\tablecaption{NGC 2903 SBF Results\label{tab:ngc2903_sbf}}

\tablewidth{\textwidth}

\tablehead{\colhead{} & \colhead{Confirmed} &\colhead{} & \colhead{} & \colhead{Possible} &\colhead{} & \colhead{} & \colhead{Background} &\colhead{} \\ 
\colhead{Name} & \colhead{SBF S/N} & \colhead{Dist (Mpc)}  & \colhead{Name} & \colhead{SBF S/N} & \colhead{Dist (Mpc)} & \colhead{Name} & \colhead{SBF S/N} & \colhead{Dist (Mpc)}  } 

\startdata
dw0930+2143 & 5.8 & $8.4^{+1.0,2.3}_{-0.8,1.5}$ & dw0933+2114 & 1.3 & $6.1^{+6.1,\infty}_{-1.9,3.0}$ &  & & \\ 
UGC 5086 & 10.1 & $8.7^{+0.9,1.9}_{-0.9,1.7}$ & dw0934+2204 & 2.9 & $7.8^{+2.3,6.6}_{-1.6,2.8}$ &  & & \\ 
\enddata
\tablecomments{Same as Table \ref{tab:ngc1023_sbf} for NGC 2903 ($D=8.0$ Mpc).}

\end{deluxetable*}
Table \ref{tab:ngc2903_sbf} lists the SBF results for NGC 2903. We confirm two candidates as satellites and consider two more as possible/unconfirmed satellites. Our SBF distance of UGC 5086 is trustworthy given the smooth, round morphology of that galaxy. The other confirmed satellite, dw0930+2143, is bluer, more irregular, and H\textsc{I}-rich \citep{irwin2009} so the SBF distance is less certain. \citet{irwin2009} measured a redshift for this dwarf via H\textsc{I} observations that is close to that of NGC 2903 ($\Delta cz\sim30$ km/s). Given the redshift and the fact that the SBF distance is at least consistent with that of NGC 2903, we consider this dwarf a confirmed satellite.

\subsection{NGC 4258}
\label{sec:sbf_ngc4258}
\begin{deluxetable*}{ccc|ccc|ccc}
\tablecaption{NGC 4258 SBF Results\label{tab:ngc4258_sbf}}

\tablewidth{\textwidth}

\tablehead{\colhead{} & \colhead{Confirmed} &\colhead{} & \colhead{} & \colhead{Possible} &\colhead{} & \colhead{} & \colhead{Background} &\colhead{} \\ 
\colhead{Name} & \colhead{SBF S/N} & \colhead{Dist (Mpc)}  & \colhead{Name} & \colhead{SBF S/N} & \colhead{Dist (Mpc)} & \colhead{Name} & \colhead{SBF S/N} & \colhead{Dist (Mpc)}  } 

\startdata
NGC 4248 & 49.5 & $7.6^{+0.4,0.9}_{-0.5,1.0}$ & dw1218+4623 & 3.7 & $5.4^{+1.9,4.4}_{-1.4,2.4}$ & dw1214+4726 & 0.1 & $>12.3$ \\ 
LVJ1218+4655 & 7.6 & $7.4^{+0.7,1.4}_{-0.6,1.0}$ & dw1220+4922 & 4.4 & $6.7^{+1.2,2.8}_{-1.0,1.8}$ & dw1214+4621 & 4.5 & $>12.2$ \\ 
dw1219+4743 & 8.6 & $7.6^{+0.8,1.8}_{-0.8,1.5}$ & dw1220+4748 & 2.1 & $11.3^{+5.9,\infty}_{-3.3,5.4}$ & dw1214+4743 & -0.3 & $>9.5$ \\ 
UGC 7356 & 22.8 & $6.3^{+0.7,1.4}_{-0.7,1.4}$ & dw1223+4848 & 1.3 & $10.0^{+9.7,\infty}_{-2.7,4.3}$ & dw1216+4709 & -0.5 & $>9.4$ \\ 
dw1220+4729 & 5.2 & $9.2^{+2.6,6.1}_{-2.0,3.5}$ &  & & & dw1217+4639 & 0.3 & $>20.2$ \\ 
dw1220+4649 & 9.2 & $7.9^{+1.1,2.3}_{-1.0,1.9}$ &  & & & dw1217+4703* & 1.8 & $>3.8$ \\ 
dw1223+4739 & 8.9 & $7.4^{+0.8,1.8}_{-0.8,1.5}$ &  & & & dw1217+4759 & 13.8 & $>8.3$ \\ 
 & & &  & & & dw1217+4747 & 2.4 & $>9.8$ \\ 
 & & &  & & & dw1217+4656 & 9.4 & $>10.6$ \\ 
 & & &  & & & dw1218+4748 & 2.0 & $>8.1$ \\ 
 & & &  & & & dw1218+4801 & 0.0 & $>7.4$ \\ 
 & & &  & & & dw1219+4921 & 1.6 & $>8.5$ \\ 
 & & &  & & & dw1219+4718 & 3.8 & $>10.1$ \\ 
 & & &  & & & dw1219+4727 & 4.4 & $>12.7$ \\ 
 & & &  & & & dw1219+4705 & 1.5 & $>13.0$ \\ 
 & & &  & & & dw1219+4939 & -0.8 & $>18.1$ \\ 
 & & &  & & & dw1220+4919 & 3.7 & $>7.9$ \\ 
 & & &  & & & UGC 7392 & 16.5 & $>12.7$ \\ 
 & & &  & & & dw1220+4700 & 3.6 & $>11.5$ \\ 
 & & &  & & & UGC 7401 & 7.8 & $>12.7$ \\ 
 & & &  & & & dw1222+4755 & 2.8 & $>15.3$ \\ 
 & & &  & & & dw1223+4920 & 2.0 & $>12.4$ \\ 
\enddata
\tablecomments{Same as Table \ref{tab:ngc1023_sbf} for NGC 4258 ($D=7.2$ Mpc). dw1217+4703 is constrained to be background by \emph{HST} imaging, see text for details. Objects marked with an asterisk (*) are discussed in detail in the text.}

\end{deluxetable*}
The results for NGC 4258 are shown in Table \ref{tab:ngc4258_sbf}. Many candidates are shown to be background while only a few were inconclusive. Seven satellites are confirmed with the SBF. Four of these have TRGB distances that put them at the distance of NGC 4258: NGC 4248 \citep{sabbi2018}, dw1219p4743 \citep{cohen2018}, UGC 7356, and LVJ1218+4655 \citep{karachentsev}. The SBF distances agree well in these cases. We confirm another three that had no prior distance information. Several of the confirmed background galaxies are worth discussing in detail. dw1217p4703 only has a distance lower bound of $\sim4$ Mpc. However, \citet{cohen2018} used \emph{HST} imaging to show that this galaxy is in the background of NGC 4258. Furthermore, \citet{cohen2018} showed that dw1219p4705 and dw1220p4700 are also background, which agrees with the SBF results.

Also classified as background are two candidates that \citet{spencer2014} considered to be confirmed satellites via their redshifts. These two are dw1214+4621 and dw1217+4759. They have fairly strong fluctuation signals, but it is visually clear that this signal is coming from their irregular morphology, not true SBF. Even with this added power, the analysis indicated they are background galaxies. Neither show any visible SBF, which should be quite apparent given their blue colors ($g-r\sim0.3$). Both of these had redshifts within 250 km/s of NGC 4258. These results highlight the dangers of confirming satellites with only redshifts, especially if there are multiple groups at different distances projected onto the same area of sky, as is the case for NGC 4258. 

Two candidate dwarfs around NGC 4258 have imaging in the \emph{HST} archive: dw1219+4718 and dw1219+4727. In the \emph{HST} images, these galaxies do not resolve into stars, which is expected at $D\sim7$ Mpc, indicating they are background. This is in line with the SBF results for these two candidates. \citet{spencer2014} considered these both to be confirmed satellites due to TRGB distances from \citet{munshi2007}\footnote{We note that this reference is a AAS abstract, and no further details on the TRGB distances can be found.}. It is unclear how a TRGB distance is possible since the dwarf galaxies are not resolved in the \emph{HST} imaging. 

The possible satellite dw1218+4623 is significantly fainter in Table \ref{tab:ngc4258_sats} than in \citet{LV_cat} due to a different sky subtraction procedure we used here which should be more accurate for this extremely low surface brightness dwarf.

\subsection{NGC 4565}
\label{sec:sbf_ngc4565}
\begin{deluxetable*}{ccc|ccc|ccc}
\tablecaption{NGC 4565 SBF Results\label{tab:ngc4565_sbf}}

\tablewidth{\textwidth}

\tablehead{\colhead{} & \colhead{Confirmed} &\colhead{} & \colhead{} & \colhead{Possible} &\colhead{} & \colhead{} & \colhead{Background} &\colhead{} \\ 
\colhead{Name} & \colhead{SBF S/N} & \colhead{Dist (Mpc)}  & \colhead{Name} & \colhead{SBF S/N} & \colhead{Dist (Mpc)} & \colhead{Name} & \colhead{SBF S/N} & \colhead{Dist (Mpc)}  } 

\startdata
dw1234+2531 & 9.4 & $11.9^{+0.9,2.0}_{-0.9,1.8}$ & dw1233+2535 & 2.1 & $9.7^{+3.8,\infty}_{-1.8,2.8}$ & dw1235+2606* & 3.7 & $>6.9$ \\ 
NGC 4562* & 13.3 & $10.1^{+0.6,1.2}_{-0.6,1.2}$ & dw1233+2543 & 2.4 & $13.1^{+4.0,\infty}_{-2.3,3.8}$ & dw1238+2536 & -1.0 & $>16.5$ \\ 
IC 3571 & -- & --  & dw1234+2627 & 1.1 & $11.5^{+\infty,\infty}_{-3.7,5.6}$ &  & & \\ 
dw1237+2602 & 8.9 & $11.0^{+0.8,1.6}_{-0.7,1.4}$ & dw1234+2618 & 2.2 & $7.1^{+2.6,15.6}_{-1.3,2.1}$ &  & & \\ 
 & & & dw1235+2616 & 1.6 & $13.2^{+8.8,\infty}_{-3.4,5.6}$ &  & & \\ 
 & & & dw1235+2534 & 0.2 & $>10.9$ &  & & \\ 
 & & & dw1235+2637 & 0.6 & $17.8^{+\infty,\infty}_{-9.8,13.5}$ &  & & \\ 
 & & & dw1235+2609 & 1.3 & $12.1^{+12.6,\infty}_{-3.5,5.4}$ &  & & \\ 
 & & & dw1236+2616 & 2.7 & $6.8^{+2.0,6.4}_{-1.2,2.2}$ &  & & \\ 
 & & & dw1236+2603 & 1.3 & $15.8^{+\infty,\infty}_{-4.5,7.0}$ &  & & \\ 
 & & & dw1236+2634 & 0.3 & $>11.3$ &  & & \\ 
 & & & dw1237+2605 & 4.6 & $7.7^{+1.6,3.7}_{-1.3,2.3}$ &  & & \\ 
 & & & dw1237+2637 & 0.4 & $16.2^{+\infty,\infty}_{-7.8,10.0}$ &  & & \\ 
 & & & dw1237+2631 & 1.5 & $7.4^{+5.6,\infty}_{-2.1,3.4}$ &  & & \\ 
 & & & dw1238+2610 & 0.9 & $10.5^{+\infty,\infty}_{-3.7,5.5}$ &  & & \\ 
\enddata
\tablecomments{Same as Table \ref{tab:ngc1023_sbf} for NGC 4565 ($D=11.9$ Mpc). Objects marked with an asterisk (*) are discussed in detail in the text.}

\end{deluxetable*}
The SBF results for NGC 4565 were inconclusive, as shown in Table \ref{tab:ngc4565_sbf}. Only a few galaxies could be confirmed as either satellites or background due to NGC 4565's larger distance of $D=11.9$ Mpc and the poor seeing in the CFHT data. NGC 4562 is irregular, such that the SBF distance is likely underestimated. Given that the redshift is within 100 km/s of NGC 4565, this galaxy is likely a companion of NGC 4565. The candidate dw1234p2531 has an SDSS redshift which is 600 km/s less than NGC 4565. This candidate has a very regular, nucleated dSph morphology which means the SBF will be trustworthy. The signal is certainly coming from the SBF of the bulk stellar population. We therefore consider this candidate as a confirmed satellite and note that the SDSS redshift might be inaccurate. Looking at the SDSS spectrum, we believe it is likely that the SDSS pipeline erroneously identified an artifact as H$\alpha$ emission from the galaxy, leading to a spurious redshift. IC 3571 was too irregular to attempt an SBF measurement but has a redshift consistent with NGC 4565 so we consider it a likely satellite. \citet{zschaechner2012} noted a bridge in H\textsc{I} between this galaxy and NGC 4565, in line with this conclusion. The candidate dw1235+2606 is located directly in the middle of the H\textsc{I} warp on the northwest edge of the disk of NGC 4565. \citet{radburn2014} used HST observations to show that there is a clump of young ($\sim600$ Myr) stars located in the warp which is likely what our detection algorithm identified as a candidate satellite. They argue that these stars formed in-situ in the warp. In this case, this candidate should not be considered a real satellite, and we include it in the `background' category. We note that \citet{gilhuly2019} interpret this candidate as the core of an accreted satellite whose disruption produced other LSB structures seen in their data. However, since \citet{radburn2014} did not find an old stellar population along with the young, the in-situ star formation scenario seems more likely.

\subsection{NGC 4631}
\label{sec:sbf_ngc4631}
\begin{deluxetable*}{ccc|ccc|ccc}
\tablecaption{NGC 4631 SBF Results\label{tab:ngc4631_sbf}}

\tablewidth{\textwidth}

\tablehead{\colhead{} & \colhead{Confirmed} &\colhead{} & \colhead{} & \colhead{Possible} &\colhead{} & \colhead{} & \colhead{Background} &\colhead{} \\ 
\colhead{Name} & \colhead{SBF S/N} & \colhead{Dist (Mpc)}  & \colhead{Name} & \colhead{SBF S/N} & \colhead{Dist (Mpc)} & \colhead{Name} & \colhead{SBF S/N} & \colhead{Dist (Mpc)}  } 

\startdata
NGC 4656 & -- & --  &  & & & UGCA 292 & -- & --  \\ 
dw1239+3230 & 7.3 & $7.3^{+0.8,1.7}_{-0.7,1.2}$ &  & & & dw1240+3239 & 9.2 & $>10.2$ \\ 
dw1239+3251 & 7.0 & $5.9^{+1.5,3.2}_{-1.2,2.3}$ &  & & & dw1242+3224 & 9.0 & $>9.6$ \\ 
dw1240+3216 & 12.0 & $6.7^{+0.8,1.6}_{-0.7,1.4}$ &  & & & dw1242+3231* & 9.2 & $>6.0$ \\ 
dw1240+3247 & 9.5 & $7.2^{+6.4,17.5}_{-3.4,5.3}$ &  & & & dw1242+3227* & 3.5 & $>4.8$ \\ 
dw1241+3251* & 15.0 & $6.2^{+0.5,1.0}_{-0.4,0.8}$ &  & & & dw1243+3229 & 17.9 & $>8.6$ \\ 
NGC 4627 & -- & --  &  & & & dw1243+3232 & 3.6 & $>11.5$ \\ 
dw1242+3237 & 8.8 & $7.2^{+2.9,7.1}_{-2.1,3.7}$ &  & & &  & & \\ 
dw1242+3158 & 6.9 & $7.1^{+0.9,2.0}_{-0.8,1.6}$ &  & & &  & & \\ 
dw1243+3228 & 18.3 & $8.1^{+0.4,0.9}_{-0.4,0.9}$ &  & & &  & & \\ 
\enddata
\tablecomments{Same as Table \ref{tab:ngc1023_sbf} for NGC 4631 ($D=7.4$ Mpc). Note that UGCA 292 is not background but significantly in the foreground of NGC 4631. Objects marked with an asterisk (*) are discussed in detail in the text.}

\end{deluxetable*}
Table \ref{tab:ngc4631_sbf} shows the SBF analysis results for NGC 4631. There are 10 confirmed satellites, and no candidates that are possible/unconfirmed.  dw1242p3231 is an exception to our usual criteria. It has strong SBF signal that is consistent with being at the distance of NGC 4631. However, it is a small compact system that is projected onto the outskirts of the disk of NGC 4631. The SBF signal appears to be coming from the outer disk stars of NGC 4631. This dwarf has archival \emph{HST} imaging in which it does not resolve into stars, strongly suggesting that it is background. Thus we include it in the `background' category. The confirmed candidate dw1241p3251 is barely inconsistent with the distance of NGC 4631 within $2\sigma$. This galaxy is somewhat non-S\'{e}rsic, and so the distance is likely underestimated. This galaxy also has a redshift consistent with NGC 4631 ($\Delta cz\sim60$ km/s). We do not attempt an SBF measurement for NGC 4627, but it is clear this galaxy is physically associated to NGC 4631 both from redshift and ongoing tidal disruption. The SBF distance errorbars for dw1240p3247 are large ($\pm4$ Mpc), even though the SBF signal is strong. This is driven by the large error on the measured color of this galaxy. This galaxy is the progenitor of a large tidal stream around NGC 4631 and is clearly physically associated \citep{delgado2015}. UGCA 292 is a foreground dwarf galaxy as evidenced by both a TRGB distance \citep{dalcanton2009} and the SBF distance. Several of the background galaxies are surprising given their LSB, spheroidal morphology and no clear massive host in the background of NGC 4631. We note that the apparent SBF signal coming from the confirmed background galaxy dw1243+3229 is from its irregular morphology and not real SBF. dw1243+3229 has a redshift that is $>250$ km/s larger than that of NGC 4631 and is almost certainly background. dw1242p3227 was too faint to have a robust distance constraint from the CFHT data alone.

To confirm our SBF results for many of the candidates found around NGC 4631, we used the much deeper HSC data of \citet{tanaka2017}. The CFHT/Megacam data we used for this region had $\sim 1$ hour exposure times for most of the field. The HSC data, on the other hand, has $\sim 10$ hour exposure times (on a telescope with twice the aperture of CFHT). The CFHT data has a wider field, however, so we identify a few candidates that were outside of the footprint of \citet{tanaka2017}. 

To use the HSC data, we downloaded the raw data from the Subaru archive\footnote{\url{https://smoka.nao.ac.jp/fssearch.jsp}} and reduced it using version 4 of the HSC pipeline \citep{bosch2018}. For the sake of saving computing time, we only downloaded and stacked $\sim3$ hours of $g$ and $i$ band data each. We then did an SBF analysis on cutouts for several of the candidates that are in the HSC footprint. Because the HSC pipeline does a local (128$\times$128 pixel grid) background estimation and subtraction, we did not attempt an SBF analysis for dw1240+3247 or dw1242+3237. These two dwarfs are large and low surface brightness (and also near NGC 4631 in projection), and the pipeline sky subtraction was clearly over-subtracting some diffuse light from these galaxies. This over-subtraction can have a significant effect on the SBF results, so we did not look at these dwarfs. To turn the measured SBF magnitudes into distances, we used the $i$ band calibration of \citet{sbf_calib}. As our goal is mostly just to confirm the CFHT results, we do not bother with any filter conversions to convert the CFHT/Megacam calibration into the HSC filter system. Both filter systems are based on SDSS filters so they should not differ by much. We assume a 0.1 mag uncertainty in the $g-i$ color of each galaxy.

\begin{deluxetable}{ccc}
\tablecaption{NGC 4631 SBF Results using HSC\label{tab:hsc_4631}}

\tablewidth{\textwidth}

\tablehead{\colhead{Name} & \colhead{SBF S/N} & \colhead{Dist (Mpc)} } 

\startdata
dw1242+3227, HSC-1 & 3.7 & $>10.4$ \\
dw1243+3232, HSC-5 & 10 & $>11.0$ \\
dw1243+3228, HSC-6 & 18 & $6.6\pm0.8$ \\
dw1240+3239, HSC-7 & 21 & $>8.6$ \\ 
dw1241+3251, HSC-8 & 57 & $7.0\pm0.8$ \\
dw1240+3216, HSC-9 & 21 & $7.4\pm0.9$ \\
dw1242+3158, HSC-10 & 17 & $7.0\pm0.8$ \\
\enddata
\tablecomments{SBF results for candidates around NGC 4631 ($D=7.4$ Mpc), using the extremely deep HSC data of \citet{tanaka2017}. For the distance lower bounds, $2\sigma$ lower bounds are given. For the distances, $\pm1\sigma$ uncertainties are given.}

\end{deluxetable}

Table \ref{tab:hsc_4631} gives the SBF results for the HSC data. The results are remarkably consistent with what we found with the CFHT data. The HSC data confirms the surprising result above that several of the dSphs are actually background. These objects are prototypical dSphs and were clustered around NGC 4631. There does not appear to be an obvious possible massive host for these objects behind NGC 4631. The HSC data confirms that dw1243+3228, dw1240+3239, and dw1242+3227 are all background. In the CFHT data, the SBF results were ambiguous for dw1242+3227 given its extreme faintness, but with the HSC data, it is clearly background. For the remaining objects, the HSC data confirms them to be at the distance of NGC 4631 with $\textit{much}$ higher S/N than was possible with the CFHT data. 

The consistency between the HSC results and the shallower CFHT results gives us confidence in the distance constraints we derive for candidates around other hosts in the CFHT data.

\subsection{NGC 5023}
\label{sec:sbf_ngc5023}
\begin{deluxetable*}{ccc|ccc|ccc}
\tablecaption{NGC 5023 SBF Results\label{tab:ngc5023_sbf}}

\tablewidth{\textwidth}

\tablehead{\colhead{} & \colhead{Confirmed} &\colhead{} & \colhead{} & \colhead{Possible} &\colhead{} & \colhead{} & \colhead{Background} &\colhead{} \\ 
\colhead{Name} & \colhead{SBF S/N} & \colhead{Dist (Mpc)}  & \colhead{Name} & \colhead{SBF S/N} & \colhead{Dist (Mpc)} & \colhead{Name} & \colhead{SBF S/N} & \colhead{Dist (Mpc)}  } 

\startdata
 & & & dw1314+4420 & 0.2 & $21.0^{+\infty,\infty}_{-12.3,15.1}$ & dw1310+4358 & 0.2 & $>7.6$ \\
\enddata
\tablecomments{Same as Table \ref{tab:ngc1023_sbf} for NGC 5023 ($D=6.5$ Mpc).}

\end{deluxetable*}
Due to the shallower data for NGC 5023, we do not confirm any candidates as satellites. As listed in Table \ref{tab:ngc5023_sbf}, one candidate is likely background while the other is unconstrained from the SBF analysis.

\subsection{M51}
\label{sec:sbf_m51}
\begin{deluxetable*}{ccc|ccc|ccc}
\tablecaption{M51 SBF Results\label{tab:m51_sbf}}

\tablewidth{\textwidth}

\tablehead{\colhead{} & \colhead{Confirmed} &\colhead{} & \colhead{} & \colhead{Possible} &\colhead{} & \colhead{} & \colhead{Background} &\colhead{} \\ 
\colhead{Name} & \colhead{SBF S/N} & \colhead{Dist (Mpc)}  & \colhead{Name} & \colhead{SBF S/N} & \colhead{Dist (Mpc)} & \colhead{Name} & \colhead{SBF S/N} & \colhead{Dist (Mpc)}  } 

\startdata
NGC 5195 & -- & --  & dw1327+4637 & -0.6 & $>4.8$ & dw1327+4654 & 1.7 & $>9.1$ \\ 
NGC 5229* & 29.3 & $7.6^{+0.3,0.6}_{-0.3,0.6}$ & dw1327+4626 & 1.4 & $9.9^{+8.8,\infty}_{-2.7,4.2}$ & dw1328+4718 & 6.9 & $>11.8$ \\ 
 & & & dw1328+4703 & 2.7 & $6.8^{+2.0,6.9}_{-1.2,2.1}$ & dw1329+4634 & 1.1 & $>13.2$ \\ 
 & & & dw1330+4731 & 1.6 & $12.1^{+8.8,\infty}_{-3.6,5.7}$ & dw1329+4622* & 2.6 & $>6.4$ \\ 
 & & & dw1331+4654 & -0.1 & $>5.6$ & dw1330+4708* & 1.6 & $>7.0$ \\ 
 & & & dw1331+4648 & 0.3 & $19.1^{+\infty,\infty}_{-10.3,12.7}$ & dw1330+4720 & 2.2 & $>15.0$ \\ 
 & & &  & & & dw1332+4703 & 0.5 & $>9.7$ \\ 
 & & &  & & & dw1333+4725 & 3.0 & $>13.1$ \\ 
\enddata
\tablecomments{Same as Table \ref{tab:ngc1023_sbf} for M51 ($D=8.6$ Mpc). Objects marked with an asterisk (*) are discussed in detail in the text.}

\end{deluxetable*}
Table \ref{tab:m51_sbf} gives the SBF results for the candidates found around M51. NGC 5229 shows strong SBF and is clearly not far in the background. \citet{sharina1999} give a brightest-stars distance of 5.1 Mpc which would put it significantly in the foreground of M51. However, they quote a $\pm2$ Mpc uncertainty in that distance. Due to NGC 5229's disky morphology, the SBF results are not trustworthy. Still, we find an SBF distance of $\sim7.7$ Mpc which is likely somewhat underestimated and, therefore, suggestive of association with M51. The redshift of this galaxy is also consistent with being bound to M51 ($\Delta cz\sim$100 km/s). Thus, we tentatively include this galaxy as a confirmed satellite, but we note that a firm confirmation will likely require an \emph{HST} TRGB distance. There is one candidate, dw1329+4622, that had inconclusive SBF results but had a redshift that indicates it is background from \citet{dalcanton1997}. dw1330+4708 also had inconclusive SBF results but has archival \emph{HST} imaging in which it is not resolved, strongly suggesting it is background.

\subsection{M64}
\label{sec:sbf_m64}
\begin{deluxetable*}{ccc|ccc|ccc}
\tablecaption{M64 SBF Results\label{tab:m64_sbf}}

\tablewidth{\textwidth}

\tablehead{\colhead{} & \colhead{Confirmed} &\colhead{} & \colhead{} & \colhead{Possible} &\colhead{} & \colhead{} & \colhead{Background} &\colhead{} \\ 
\colhead{Name} & \colhead{SBF S/N} & \colhead{Dist (Mpc)}  & \colhead{Name} & \colhead{SBF S/N} & \colhead{Dist (Mpc)} & \colhead{Name} & \colhead{SBF S/N} & \colhead{Dist (Mpc)}  } 

\startdata
 & & &  & & & dw1255+2130 & 3.8 & $>5.4$ \\ 
\enddata
\tablecomments{Same as Table \ref{tab:ngc1023_sbf} for M64 ($D=5.3$ Mpc).}

\end{deluxetable*}
\citet{LV_cat} only found one candidate satellite in the vicinity of M64, at least partly due to the small survey footprint and shallow data. As shown in Table \ref{tab:m64_sbf}, the SBF analysis indicates that this dwarf is background.

\subsection{M104}
\label{sec:sbf_m104}
\begin{deluxetable*}{ccc|ccc|ccc}
\tablecaption{M104 SBF Results\label{tab:m104_sbf}}

\tablewidth{\textwidth}

\tablehead{\colhead{} & \colhead{Confirmed} &\colhead{} & \colhead{} & \colhead{Possible} &\colhead{} & \colhead{} & \colhead{Background} &\colhead{} \\ 
\colhead{Name} & \colhead{SBF S/N} & \colhead{Dist (Mpc)}  & \colhead{Name} & \colhead{SBF S/N} & \colhead{Dist (Mpc)} & \colhead{Name} & \colhead{SBF S/N} & \colhead{Dist (Mpc)}  } 

\startdata
dw1237-1125 & 6.4 & $7.5^{+2.0,4.6}_{-1.8,3.2}$ & dw1238-1208 & 0.7 & $11.6^{+\infty,\infty}_{-5.5,7.9}$ & dw1237-1110 & 1.8 & $>10.6$ \\ 
dw1239-1152 & 6.9 & $8.2^{+1.1,2.4}_{-0.9,1.8}$ & dw1238-1116 & 4.8 & $9.2^{+2.4,5.7}_{-1.8,3.3}$ & dw1240-1155 & 15.0 & $>16.9$ \\ 
dw1239-1159 & 6.2 & $11.3^{+1.4,3.2}_{-1.2,2.3}$ & dw1238-1122 & -- & --  & dw1241-1210 & 3.1 & $>12.5$ \\ 
dw1239-1143 & 21.5 & $9.4^{+0.6,1.2}_{-0.6,1.3}$ & dw1238-1102 & 2.5 & $8.5^{+3.0,11.0}_{-1.8,3.1}$ &  & & \\ 
dw1239-1113 & 12.8 & $7.9^{+1.2,2.4}_{-1.2,2.3}$ & dw1239-1154 & 2.9 & $7.3^{+2.4,6.7}_{-1.6,2.8}$ &  & & \\ 
dw1239-1120 & 11.7 & $9.7^{+0.6,1.3}_{-0.6,1.2}$ & dw1239-1118 & 1.7 & $11.2^{+6.1,\infty}_{-2.5,4.1}$ &  & & \\ 
dw1239-1144 & 9.6 & $9.0^{+1.5,3.1}_{-1.3,2.5}$ & dw1239-1106 & -0.1 & $>8.3$ &  & & \\ 
dw1240-1118 & 35.4 & $8.8^{+0.5,1.0}_{-0.6,1.2}$ & dw1241-1123 & 0.0 & $>8.0$ &  & & \\ 
dw1240-1140* & 5.8 & $4.2^{+0.6,1.3}_{-0.5,1.1}$ & dw1241-1105 & 2.4 & $8.4^{+2.7,10.7}_{-1.6,2.7}$ &  & & \\ 
dw1241-1131 & 6.1 & $7.2^{+1.2,2.5}_{-1.0,2.0}$ & dw1242-1116* & 2.0 & $14.5^{+5.9,\infty}_{-2.9,4.7}$ &  & & \\ 
dw1241-1153 & 6.1 & $11.2^{+1.4,2.9}_{-1.2,2.2}$ & dw1242-1129 & 1.6 & $15.3^{+9.7,\infty}_{-3.8,6.2}$ &  & & \\ 
dw1241-1155 & 16.3 & $9.0^{+0.8,1.5}_{-0.8,1.6}$ & dw1243-1137 & 2.5 & $5.3^{+1.7,6.7}_{-1.0,1.8}$ &  & & \\ 
\enddata
\tablecomments{Same as Table \ref{tab:ngc1023_sbf} for M104 ($D=9.55$ Mpc). Objects marked with an asterisk (*) are discussed in detail in the text.}

\end{deluxetable*}
The SBF results for M104 are shown in Table \ref{tab:m104_sbf}. Due to the good seeing of the data and brightness of SBF in the $i$ band, we were able to confirm a large number of the candidates to be at the distance of M104 ($D=9.55$ Mpc). A few of the dwarfs were exceptions to our usual classification criteria. dw1240m1140 showed a strong SBF signal that put it significantly in the foreground. However, this dwarf is located very close to M104 in projection, and the halo of M104 could be adding signal to the SBF measurement causing the distance to be underestimated. Considering its dSph morphology, proximity to M104, and SBF, we suspect this dwarf is physically associated with M104, and include it in the `confirmed' category. The SBF measurement of dw1242m1116 indicated that it is background, but this dwarf only partially fell on a chip in the MegaCam data which might make the measurement unreliable. Thus, we include this dwarf in the unconfirmed/possible category. dw1238m1122 was contaminated by a large saturation spike in the MegaCam data and so we did not attempt an SBF measurement of this galaxy.

\section{$\MakeLowercase{r}$-band SBF Calibration}
\label{app:rband}

\begin{figure*}
\includegraphics[width=\textwidth]{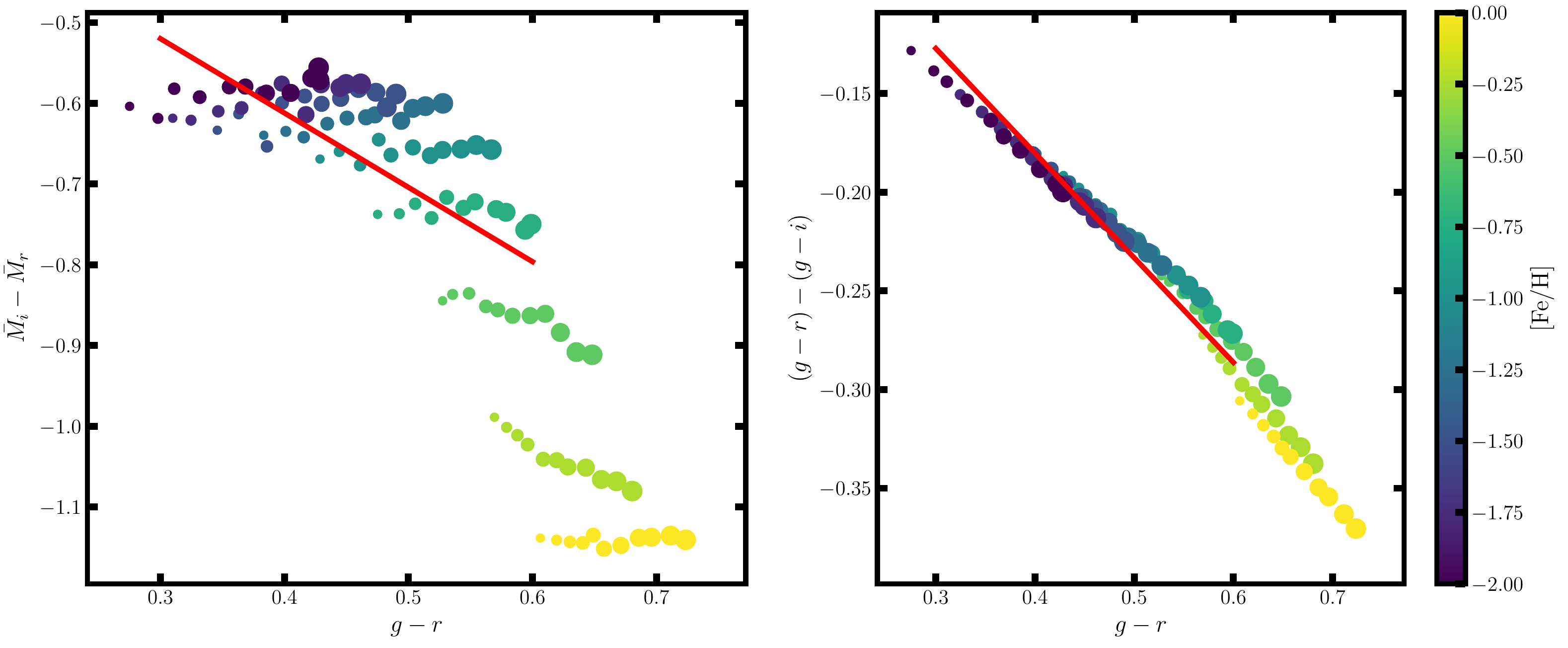}
\caption{The filter conversions used in deriving the $r$-band calibration. The points show the SSP models where point size represents age ($3<$ age $<10$ Gyr) with biggest sizes indicating oldest ages. The red lines show the fits that are used in the conversion.}
\label{fig:mr_2_mi}
\end{figure*}

In Figure \ref{fig:mr_2_mi} we show the conversions between $\bar{M}_r$ and $\bar{M}_r$ and $g-i$ and $g-r$ that we use to derive the $r$-band calibration used in this work. The color-color transformation has quite low scatter. The SBF magnitude transformation looks significantly worse but we note that the galaxies we analyze in this paper all have $g-r\lesssim0.6$ where the scatter is $\sim0.1$ mag. An error of 0.1 mag in the conversion between $i$ and $r$ band SBF magnitudes will only introduce a 5\% error in distance which is less than the usual distance uncertainties we find in the SBF analysis. While a quadratic looks to be more appropriate for the SBF magnitude conversion, using a linear fit has the attractive property that the $\bar{M}_r$ versus $g-r$ relation will be linear as well.

\section{LV Satellite Systems}
\label{app:sat_systems_tables}

\subsection{Systems Surveyed In the Current Work}
Tables \ref{tab:ngc1023_sats} - \ref{tab:m104_sats} list the properties of the confirmed and possible satellites in the systems that we surveyed in \S\ref{sec:distances} of this work. Position, magnitudes, and sizes are given for all system members, including the hosts. The host photometry comes from \citet{galex2007}. For the satellites, the host distance is used to calculate absolute magnitudes and physical sizes and not the individual SBF distances. We are not able to resolve the 3D structure of these groups with the precision of SBF, and using the SBF distances would simply increase the scatter in size and magnitude.

In \citet{LV_cat}, $R$-band photometry from \citet{trentham2009} was used for some of the largest candidates around NGC 1023. We convert from $R$ into $V$ using $V\approx R+0.56$ \citep{fukugita1995}. The rest of the dwarf photometry comes from \citet{LV_cat}. Errors are estimated from injecting mock galaxies into the data and measuring the spread in the recovered photometry.

\begin{deluxetable*}{ccccccc}
\tablecaption{NGC 1023 Satellites\label{tab:ngc1023_sats}}

\tablewidth{\textwidth}

\tablehead{ 
\colhead{Name} & \colhead{RA} & \colhead{Dec}  & \colhead{$M_g$} & \colhead{$M_V$} & \colhead{$g-i$} & \colhead{$r_e$} \\
\colhead{} & \colhead{deg} &\colhead{deg} & \colhead{} & \colhead{} & \colhead{} & \colhead{pc} } 

\startdata
NGC 1023 & 40.1000 & 39.0633 & -- & -20.9 & -- & -- \\
\hline
\multicolumn{7}{c}{Confirmed Satellites}\\
dw0233+3852 & 38.4286 & 38.8716 & -11.7$\pm$0.27 & -11.92$\pm$0.27 & 0.52$\pm$0.11 & 745.3$\pm$49.7  \\
dw0235+3850 & 38.9763 & 38.8361 & -13.27$\pm$0.15 & -13.52$\pm$0.15 & 0.6$\pm$0.01 & 532.0$\pm$86.4  \\
IC 239 & 39.1171 & 38.969 & -18.8$^{**}$ & -19.1$^{**}$ & 0.73$^{**}$ & 4216.1$^{**}$  \\
dw0237+3855 & 39.3284 & 38.9328 & -14.9$\pm$0.11 & -15.19$\pm$0.11 & 0.72$\pm$0.01 & 992.1$\pm$98.3  \\
dw0237+3836 & 39.414 & 38.6004 & -11.86$\pm$0.17 & -12.12$\pm$0.17 & 0.65$\pm$0.07 & 533.0$\pm$63.0  \\
dw0239+3926 & 39.8318 & 39.435 & -12.12$\pm$0.14 & -12.42$\pm$0.14 & 0.75$\pm$0.02 & 1305.3$\pm$103.8  \\
dw0239+3903 & 39.8451 & 39.0559 & -9.02$\pm$0.49 & -9.3$\pm$0.49 & 0.7$\pm$0.18 & 228.8$\pm$44.0  \\
dw0239+3902 & 39.9476 & 39.0484 & -9.46$\pm$0.09 & -9.79$\pm$0.09 & 0.82$\pm$0.02 & 267.3$\pm$15.8  \\
UGC 2157 & 40.1046 & 38.563 & -16.1$^{**}$ & -16.4$^{**}$ & 0.66$^{**}$ & 1986.6$^{**}$  \\
dw0240+3854 & 40.1376 & 38.9004 & -13.32$\pm$0.03 & -13.49$\pm$0.03 & 0.42$\pm$0.01 & 330.2$\pm$3.3  \\
dw0240+3903 & 40.1555 & 39.0554 & -- & -15.1 & -- & --  \\
dw0240+3922 & 40.165 & 39.3791 & -13.4$\pm$0.07 & -13.51$\pm$0.07 & 0.26$\pm$0.01 & 644.2$\pm$53.4  \\
dw0241+3904 & 40.2514 & 39.0721 & -14.16$\pm$0.03 & -14.34$\pm$0.03 & 0.42$\pm$0.01 & 781.8$\pm$25.9  \\
UGC 2165 & 40.3148 & 38.7438 & -15.88$\pm$0.04 & -16.23$\pm$0.04 & 0.89$\pm$0.01 & 1261.0$\pm$40.3  \\
dw0242+3838 & 40.6023 & 38.635 & -9.24$\pm$0.11 & -9.43$\pm$0.11 & 0.45$\pm$0.06 & 181.0$\pm$12.3  \\
\hline
\multicolumn{7}{c}{Possible Satellites}\\
dw0238+3805 & 39.6712 & 38.085 & -13.4$^{**}$ & -13.6$^{**}$ & 0.57$^{**}$ & 656.2$^{**}$  \\
dw0239+3910 & 39.842 & 39.1729 & -7.74$\pm$0.24 & -8.01$\pm$0.24 & 0.66$\pm$0.14 & 227.6$\pm$22.7  \\
dw0241+3852 & 40.3364 & 38.8667 & -8.73$\pm$0.42 & -8.99$\pm$0.42 & 0.64$\pm$0.13 & 310.2$\pm$42.0  \\
dw0241+3829 & 40.4759 & 38.4982 & -10.6$\pm$0.13 & -10.85$\pm$0.13 & 0.64$\pm$0.02 & 300.1$\pm$30.1  \\
dw0242+3757 & 40.5923 & 37.9567 & -7.88$\pm$0.28 & -8.24$\pm$0.29 & 0.9$\pm$0.12 & 124.6$\pm$19.7  \\
dw0243+3915 & 40.9792 & 39.2558 & -11.13$\pm$0.13 & -11.43$\pm$0.13 & 0.74$\pm$0.04 & 313.7$\pm$47.2  \\
\enddata
\tablecomments{Confirmed and possible satellites in the NGC 1023 system. The $V$ band photometry is converted from our photometry, as described in \citet{LV_cat}. The asterisks mark systems which were not well fit by a S\'{e}rsic profile and the photometry might be biased. The photometry for dw0240+3903 comes from \citet{trentham2009}.}

\end{deluxetable*}

\begin{deluxetable*}{ccccccc}
\tablecaption{NGC 1156 Satellites\label{tab:ngc1156_sats}}

\tablewidth{\textwidth}

\tablehead{ 
\colhead{Name} & \colhead{RA} & \colhead{Dec}  & \colhead{$M_g$} & \colhead{$M_V$} & \colhead{$g-r$} & \colhead{$r_e$} \\
\colhead{} & \colhead{deg} &\colhead{deg} & \colhead{} & \colhead{} & \colhead{} & \colhead{pc} } 

\startdata
NGC 1156 & 44.9263 & 25.2378    & -- & -18.3 & -- & -- \\
\hline
\multicolumn{7}{c}{Confirmed Satellites}\\
\hline
\multicolumn{7}{c}{Possible Satellites}\\
dw0300+2514 & 45.0739 & 25.2485 & -10.39$\pm$0.5 & -10.66$\pm$0.5 & 0.46$\pm$0.11 & 213.7$\pm$31.0  \\
dw0301+2446 & 45.3848 & 24.7827 & -10.36$\pm$0.35 & -10.76$\pm$0.37 & 0.69$\pm$0.21 & 387.7$\pm$49.8  \\
\enddata
\tablecomments{Confirmed and possible satellites in the NGC 1156 system.}

\end{deluxetable*}

\begin{deluxetable*}{ccccccc}
\tablecaption{NGC 2903 Satellites\label{tab:ngc2903_sats}}

\tablewidth{\textwidth}

\tablehead{ 
\colhead{Name} & \colhead{RA} & \colhead{Dec}  & \colhead{$M_g$} & \colhead{$M_V$} & \colhead{$g-r$} & \colhead{$r_e$} \\
\colhead{} & \colhead{deg} &\colhead{deg} & \colhead{} & \colhead{} & \colhead{} & \colhead{pc} } 

\startdata
NGC 2903 & 143.0421 & 21.5008 & -- & -20.47 & -- & -- \\
\hline
\multicolumn{7}{c}{Confirmed Satellites}\\
dw0930+2143 & 142.6665 & 21.7244 & -10.86$\pm$0.08 & -11.01$\pm$0.09 & 0.26$\pm$0.03 & 297.6$\pm$21.5  \\
UGC 5086 & 143.2036 & 21.4654 & -13.79$\pm$0.08 & -14.13$\pm$0.08 & 0.58$\pm$0.03 & 653.6$\pm$30.3  \\
\hline
\multicolumn{7}{c}{Possible Satellites}\\
dw0933+2114 & 143.3685 & 21.2334 & -8.1$\pm$0.49 & -8.43$\pm$0.49 & 0.56$\pm$0.1 & 186.7$\pm$24.1  \\
dw0934+2204 & 143.592 & 22.0815 & -10.1$\pm$0.1 & -10.28$\pm$0.12 & 0.31$\pm$0.09 & 143.9$\pm$10.8  \\
\enddata
\tablecomments{Confirmed and possible satellites in the NGC 2903 system.}

\end{deluxetable*}

\begin{deluxetable*}{ccccccc}
\tablecaption{NGC 4258 Satellites\label{tab:ngc4258_sats}}

\tablewidth{\textwidth}

\tablehead{ 
\colhead{Name} & \colhead{RA} & \colhead{Dec}  & \colhead{$M_g$} & \colhead{$M_V$} & \colhead{$g-r$} & \colhead{$r_e$} \\
\colhead{} & \colhead{deg} &\colhead{deg} & \colhead{} & \colhead{} & \colhead{} & \colhead{pc} } 

\startdata
NGC 4258 & 184.7396 & 47.3040 & -- & -20.94 & -- & -- \\
\hline
\multicolumn{7}{c}{Confirmed Satellites}\\
NGC 4248 & 184.46 & 47.409 & -16.57$\pm$0.02 & -16.86$\pm$0.02 & 0.51$\pm$0.01 & 1824.1$\pm$17.1  \\
LVJ1218+4655 & 184.5462 & 46.9169 & -12.8$\pm$0.02 & -12.93$\pm$0.02 & 0.22$\pm$0.01 & 564.8$\pm$20.4  \\
dw1219+4743 & 184.7771 & 47.7308 & -10.76$\pm$0.16 & -11.0$\pm$0.16 & 0.41$\pm$0.04 & 361.4$\pm$43.8  \\
UGC 7356 & 184.7879 & 47.0897 & -14.0$\pm$0.09 & -14.32$\pm$0.1 & 0.53$\pm$0.05 & 896.4$\pm$60.2  \\
dw1220+4729$^\dagger$ & 185.1279 & 47.4909 & -9.16$\pm$0.35 & -9.33$\pm$0.35 & 0.28$\pm$0.12 & 479.8$\pm$68.2  \\
dw1220+4649 & 185.2287 & 46.8304 & -10.47$\pm$0.13 & -10.76$\pm$0.14 & 0.5$\pm$0.05 & 402.9$\pm$23.6  \\
dw1223+4739 & 185.9428 & 47.6589 & -11.27$\pm$0.09 & -11.54$\pm$0.09 & 0.45$\pm$0.04 & 601.6$\pm$80.9  \\
\hline
\multicolumn{7}{c}{Possible Satellites}\\
dw1218+4623 & 184.5111 & 46.3846 & -7.44$\pm$0.52 & -7.73$\pm$0.53 & 0.49$\pm$0.13 & 259.7$\pm$60.5  \\
dw1220+4922 & 185.0597 & 49.3809 & -9.34$\pm$0.09 & -9.59$\pm$0.09 & 0.44$\pm$0.05 & 201.3$\pm$9.7  \\
dw1220+4748 & 185.2326 & 47.8164 & -7.31$\pm$0.38 & -7.52$\pm$0.39 & 0.36$\pm$0.13 & 159.2$\pm$32.2  \\
dw1223+4848 & 185.8031 & 48.8156 & -8.39$\pm$0.22 & -8.71$\pm$0.22 & 0.55$\pm$0.07 & 164.1$\pm$21.7  \\
\enddata
\tablecomments{Confirmed and possible satellites in the NGC 4258 system. Satellites marked with $\dagger$ are below the $\mu_{0,V}=26.5$ mag arcsec$^{-2}$ surface brightness limit we assume in throughout this work.}

\end{deluxetable*}

\begin{deluxetable*}{ccccccc}
\tablecaption{NGC 4565 Satellites\label{tab:ngc4565_sats}}

\tablewidth{\textwidth}

\tablehead{ 
\colhead{Name} & \colhead{RA} & \colhead{Dec}  & \colhead{$M_g$} & \colhead{$M_V$} & \colhead{$g-r$} & \colhead{$r_e$} \\
\colhead{} & \colhead{deg} &\colhead{deg} & \colhead{} & \colhead{} & \colhead{} & \colhead{pc} } 

\startdata
NGC 4565 &  189.0866 & 25.9877 & -- & -21.8 & -- & -- \\
\hline
\multicolumn{7}{c}{Confirmed Satellites}\\
dw1234+2531 & 188.5971 & 25.5193 & -13.73$\pm$0.03 & -14.03$\pm$0.03 & 0.51$\pm$0.01 & 1148.5$\pm$28.2  \\
NGC 4562 & 188.8977 & 25.852 & -16.88$\pm$0.01 & -17.15$\pm$0.01 & 0.46$\pm$0.01 & 2043.7$\pm$8.0  \\
IC 3571 & 189.0836 & 26.084 & -13.8$^{**}$ & -13.9$^{**}$ & 0.16$^{**}$ & 491.1$^{**}$  \\
dw1237+2602 & 189.2551 & 26.0357 & -12.41$\pm$0.06 & -12.64$\pm$0.06 & 0.39$\pm$0.01 & 484.5$\pm$33.6  \\
\hline
\multicolumn{7}{c}{Possible Satellites}\\
dw1233+2535 & 188.2961 & 25.5987 & -11.73$\pm$0.07 & -11.97$\pm$0.07 & 0.4$\pm$0.01 & 269.3$\pm$11.5  \\
dw1233+2543 & 188.3267 & 25.7263 & -9.81$\pm$0.1 & -10.01$\pm$0.1 & 0.35$\pm$0.03 & 238.9$\pm$11.6  \\
dw1234+2627 & 188.6042 & 26.4542 & -8.53$\pm$0.26 & -8.8$\pm$0.26 & 0.47$\pm$0.09 & 212.4$\pm$33.1  \\
dw1234+2618 & 188.7399 & 26.314 & -10.13$\pm$0.06 & -10.32$\pm$0.06 & 0.33$\pm$0.03 & 267.8$\pm$10.0  \\
dw1235+2616 & 188.8438 & 26.2717 & -9.84$\pm$0.13 & -10.15$\pm$0.13 & 0.52$\pm$0.03 & 259.1$\pm$51.5  \\
dw1235+2534 & 188.9066 & 25.5702 & -8.45$\pm$0.21 & -8.66$\pm$0.22 & 0.36$\pm$0.09 & 246.2$\pm$45.2  \\
dw1235+2637 & 188.9252 & 26.6208 & -8.7$\pm$0.31 & -8.74$\pm$0.34 & 0.07$\pm$0.25 & 387.9$\pm$97.8  \\
dw1235+2609 & 188.9799 & 26.1654 & -7.64$\pm$0.21 & -7.87$\pm$0.22 & 0.4$\pm$0.08 & 172.8$\pm$30.7  \\
dw1236+2616 & 189.0247 & 26.2735 & -7.5$\pm$0.15 & -7.78$\pm$0.16 & 0.48$\pm$0.06 & 172.4$\pm$10.8  \\
dw1236+2603 & 189.1049 & 26.0552 & -8.93$\pm$0.19 & -9.09$\pm$0.2 & 0.28$\pm$0.07 & 258.2$\pm$41.7  \\
dw1236+2634 & 189.2448 & 26.5782 & -9.2$\pm$0.18 & -9.5$\pm$0.18 & 0.51$\pm$0.04 & 272.2$\pm$29.8  \\
dw1237+2605 & 189.3614 & 26.0855 & -10.6$\pm$0.32 & -10.85$\pm$0.32 & 0.42$\pm$0.07 & 761.0$\pm$194.6  \\
dw1237+2637 & 189.4278 & 26.6253 & -10.15$\pm$0.08 & -10.46$\pm$0.08 & 0.53$\pm$0.06 & 281.9$\pm$32.5  \\
dw1237+2631 & 189.4777 & 26.5188 & -7.86$\pm$0.36 & -8.12$\pm$0.37 & 0.44$\pm$0.09 & 155.6$\pm$18.6  \\
dw1238+2610 & 189.6651 & 26.1669 & -8.39$\pm$0.29 & -8.65$\pm$0.29 & 0.44$\pm$0.09 & 270.7$\pm$29.1  \\
\enddata
\tablecomments{Confirmed and possible satellites in the NGC 4565 system.}

\end{deluxetable*}

\begin{deluxetable*}{ccccccc}
\tablecaption{NGC 4631 Satellites\label{tab:ngc4631_sats}}

\tablewidth{\textwidth}

\tablehead{ 
\colhead{Name} & \colhead{RA} & \colhead{Dec}  & \colhead{$M_g$} & \colhead{$M_V$} & \colhead{$g-r$} & \colhead{$r_e$} \\
\colhead{} & \colhead{deg} &\colhead{deg} & \colhead{} & \colhead{} & \colhead{} & \colhead{pc} } 

\startdata
\multicolumn{7}{c}{Confirmed Satellites}\\
NGC 4631 & 190.5334 &  32.5415 & -- & -20.24 & -- & -- \\
\hline
NGC 4656 & 190.4985 & 32.5739 & -- & -18.9 & -- & --  \\
dw1239+3230 & 189.7705 & 32.5043 & -10.26$\pm$0.09 & -10.43$\pm$0.09 & 0.29$\pm$0.03 & 286.8$\pm$18.0  \\
dw1239+3251 & 189.8318 & 32.8612 & -9.31$\pm$0.31 & -9.65$\pm$0.31 & 0.58$\pm$0.11 & 490.4$\pm$98.9  \\
dw1240+3216 & 190.2209 & 32.282 & -10.35$\pm$0.1 & -10.64$\pm$0.1 & 0.5$\pm$0.05 & 311.5$\pm$20.0  \\
dw1240+3247 & 190.2451 & 32.7897 & -13.28$\pm$0.64 & -13.61$\pm$0.67 & 0.57$\pm$0.33 & 2549.6$\pm$684.5  \\
dw1241+3251 & 190.4463 & 32.8573 & -13.65$\pm$0.05 & -13.74$\pm$0.05 & 0.14$\pm$0.02 & 644.4$\pm$33.3  \\
NGC 4627 & 190.4985 & 32.5739 & -16.5$^{**}$ & -16.7$^{**}$ & 0.37$^{**}$ & 973.9$^{**}$  \\
dw1242+3237$^\dagger$ & 190.5256 & 32.6203 & -10.47$\pm$0.43 & -10.71$\pm$0.44 & 0.4$\pm$0.17 & 660.6$\pm$99.1  \\
dw1242+3158 & 190.6309 & 31.9693 & -10.22$\pm$0.1 & -10.51$\pm$0.1 & 0.5$\pm$0.05 & 295.2$\pm$22.0  \\
dw1243+3228 & 190.8537 & 32.4819 & -12.62$\pm$0.03 & -12.88$\pm$0.03 & 0.45$\pm$0.01 & 593.9$\pm$10.7  \\
\hline
\multicolumn{7}{c}{Possible Satellites}\\
\enddata
\tablecomments{Confirmed and possible satellites in the NGC 4631 system. Photometry for NGC 4565 comes from \citet{galex2007}. Satellites marked with $\dagger$ are below the $\mu_{0,V}=26.5$ mag arcsec$^{-2}$ surface brightness limit we assume in throughout this work.}

\end{deluxetable*}

\begin{deluxetable*}{ccccccc}
\tablecaption{NGC 5023 Satellites\label{tab:ngc5023_sats}}

\tablewidth{\textwidth}

\tablehead{ 
\colhead{Name} & \colhead{RA} & \colhead{Dec}  & \colhead{$M_g$} & \colhead{$M_V$} & \colhead{$g-i$} & \colhead{$r_e$} \\
\colhead{} & \colhead{deg} &\colhead{deg} & \colhead{} & \colhead{} & \colhead{} & \colhead{pc} } 

\startdata
NGC 5023 & 198.0525 & 44.0412 & -- & -14.9 & -- & -- \\
\hline
\multicolumn{7}{c}{Confirmed Satellites}\\
\hline
\multicolumn{7}{c}{Possible Satellites}\\
dw1314+4420 & 198.6437 & 44.3341 & -6.67$\pm$0.16 & -6.91$\pm$0.18 & 0.58$\pm$0.18 & 96.4$\pm$8.2  \\
\enddata
\tablecomments{Confirmed and possible satellites in the NGC 5023 system.}

\end{deluxetable*}

\begin{deluxetable*}{ccccccc}
\tablecaption{M51 Satellites\label{tab:m51_sats}}

\tablewidth{\textwidth}

\tablehead{ 
\colhead{Name} & \colhead{RA} & \colhead{Dec}  & \colhead{$M_g$} & \colhead{$M_V$} & \colhead{$g-r$} & \colhead{$r_e$} \\
\colhead{} & \colhead{deg} &\colhead{deg} & \colhead{} & \colhead{} & \colhead{} & \colhead{pc} } 

\startdata
M51 & 202.4696 & 47.1952 & -- & -21.38 & -- & -- \\
\hline
\multicolumn{7}{c}{Confirmed Satellites}\\
NGC 5195 & 202.4696 & 47.1952 & -- & -20.2$^{**}$ & -- & --  \\
NGC 5229 & 203.5127 & 47.9124 & -15.95$\pm$0.01 & -16.18$\pm$0.01 & 0.4$\pm$0.01 & 1543.2$\pm$3.9  \\
\hline
\multicolumn{7}{c}{Possible Satellites}\\
dw1327+4637 & 201.7945 & 46.6323 & -8.3$\pm$0.24 & -8.64$\pm$0.24 & 0.57$\pm$0.11 & 195.6$\pm$42.7  \\
dw1327+4626 & 201.9725 & 46.4406 & -8.92$\pm$0.2 & -9.14$\pm$0.2 & 0.38$\pm$0.03 & 163.6$\pm$31.7  \\
dw1328+4703 & 202.1027 & 47.0649 & -9.27$\pm$0.11 & -9.62$\pm$0.11 & 0.59$\pm$0.04 & 279.9$\pm$17.5  \\
dw1330+4731$^\dagger$ & 202.6405 & 47.5264 & -9.67$\pm$0.15 & -9.89$\pm$0.16 & 0.37$\pm$0.11 & 519.0$\pm$50.8  \\
dw1331+4654 & 202.7839 & 46.9076 & -7.45$\pm$0.08 & -7.77$\pm$0.1 & 0.55$\pm$0.08 & 129.4$\pm$4.1  \\
dw1331+4648 & 202.7983 & 46.8158 & -9.02$\pm$0.17 & -9.28$\pm$0.17 & 0.44$\pm$0.03 & 286.4$\pm$44.6  \\
\enddata
\tablecomments{Confirmed and possible satellites in the M51 system. Photometry for NGC 5195 comes from \citet{galex2007}. Satellites marked with $\dagger$ are below the $\mu_{0,V}=26.5$ mag arcsec$^{-2}$ surface brightness limit we assume in throughout this work.}

\end{deluxetable*}

\begin{deluxetable*}{ccccccc}
\tablecaption{M104 Satellites\label{tab:m104_sats}}

\tablewidth{\textwidth}

\tablehead{ 
\colhead{Name} & \colhead{RA} & \colhead{Dec}  & \colhead{$M_g$} & \colhead{$M_V$} & \colhead{$g-i$} & \colhead{$r_e$} \\
\colhead{} & \colhead{deg} &\colhead{deg} & \colhead{} & \colhead{} & \colhead{} & \colhead{pc} } 

\startdata
M104 & 189.9976 & -11.6231 & -- & -22.02 & -- & -- \\
\hline
\multicolumn{7}{c}{Confirmed Satellites}\\
dw1237-1125 & 189.2986 & -11.4331 & -11.62$\pm$0.06 & -12.02$\pm$0.1 & 0.99$\pm$0.23 & 463.5$\pm$17.9  \\
dw1239-1159 & 189.7866 & -11.9876 & -11.0$\pm$0.22 & -11.21$\pm$0.22 & 0.52$\pm$0.08 & 653.8$\pm$117.3  \\
dw1239-1152 & 189.7881 & -11.8763 & -8.09$\pm$0.22 & -8.29$\pm$0.23 & 0.49$\pm$0.1 & 229.0$\pm$33.4  \\
dw1239-1143 & 189.8136 & -11.7189 & -13.38$\pm$0.03 & -13.7$\pm$0.03 & 0.8$\pm$0.01 & 577.7$\pm$11.9  \\
dw1239-1113 & 189.8851 & -11.2253 & -11.9$\pm$0.27 & -12.23$\pm$0.27 & 0.84$\pm$0.13 & 799.9$\pm$119.3  \\
dw1239-1120 & 189.9628 & -11.342 & -10.49$\pm$0.1 & -10.73$\pm$0.1 & 0.59$\pm$0.03 & 323.2$\pm$29.2  \\
dw1239-1144 & 189.9839 & -11.7479 & -12.57$\pm$0.29 & -12.85$\pm$0.3 & 0.7$\pm$0.14 & 1039.4$\pm$208.2  \\
dw1240-1118 & 190.0351 & -11.309 & -14.0$\pm$0.03 & -14.32$\pm$0.03 & 0.81$\pm$0.01 & 697.1$\pm$16.9  \\
dw1240-1140 & 190.0737 & -11.679 & -10.64$\pm$0.46 & -11.01$\pm$0.46 & 0.94$\pm$0.06 & 606.6$\pm$130.9  \\
dw1241-1131 & 190.2617 & -11.5289 & -10.12$\pm$0.17 & -10.44$\pm$0.18 & 0.8$\pm$0.11 & 423.3$\pm$40.8  \\
dw1241-1153 & 190.3006 & -11.8915 & -11.57$\pm$0.22 & -11.86$\pm$0.22 & 0.72$\pm$0.06 & 706.4$\pm$104.5  \\
dw1241-1155 & 190.3298 & -11.9318 & -12.42$\pm$0.11 & -12.72$\pm$0.11 & 0.75$\pm$0.06 & 786.1$\pm$54.6  \\
\hline
\multicolumn{7}{c}{Possible Satellites}\\
dw1238-1208 & 189.5927 & -12.1357 & -7.27$\pm$0.15 & -7.52$\pm$0.21 & 0.61$\pm$0.37 & 155.8$\pm$10.0  \\
dw1238-1116 & 189.6297 & -11.2735 & -8.83$\pm$0.34 & -9.0$\pm$0.35 & 0.4$\pm$0.2 & 270.6$\pm$51.3  \\
dw1238-1122 & 189.64 & -11.368 & -12.3$^{**}$ & -12.6$^{**}$ & 0.84$^{**}$ & 648.9$^{**}$  \\
dw1238-1102 & 189.7429 & -11.0361 & -9.17$\pm$0.25 & -9.38$\pm$0.26 & 0.5$\pm$0.16 & 241.1$\pm$26.1  \\
dw1239-1154 & 189.843 & -11.907 & -8.47$\pm$0.45 & -8.76$\pm$0.46 & 0.73$\pm$0.18 & 386.1$\pm$60.9  \\
dw1239-1118 & 189.9059 & -11.309 & -8.31$\pm$0.13 & -8.54$\pm$0.14 & 0.58$\pm$0.1 & 184.3$\pm$18.5  \\
dw1239-1106 & 189.9242 & -11.1006 & -9.02$\pm$0.18 & -9.27$\pm$0.19 & 0.61$\pm$0.09 & 272.7$\pm$39.9  \\
dw1241-1123 & 190.2895 & -11.3989 & -8.82$\pm$0.38 & -9.14$\pm$0.38 & 0.8$\pm$0.09 & 565.5$\pm$116.6  \\
dw1241-1105 & 190.2927 & -11.0973 & -8.01$\pm$0.11 & -8.4$\pm$0.11 & 0.99$\pm$0.06 & 107.0$\pm$10.7  \\
dw1242-1116 & 190.6826 & -11.2745 & -11.81$\pm$0.22 & -12.05$\pm$0.22 & 0.58$\pm$0.07 & 1127.6$\pm$213.2  \\
dw1242-1129 & 190.7067 & -11.4894 & -8.82$\pm$0.23 & -9.11$\pm$0.24 & 0.73$\pm$0.13 & 150.5$\pm$22.3  \\
dw1243-1137 & 190.8249 & -11.6257 & -8.34$\pm$0.19 & -8.75$\pm$0.2 & 1.03$\pm$0.07 & 203.7$\pm$36.1  \\
\enddata
\tablecomments{Confirmed and possible satellites in the M104 system.}

\end{deluxetable*}

\subsection{Surveyed Systems from the Literature}
Table \ref{tab:mw_sats} - \ref{tab:m94_sats} lists the members of the previously surveyed systems. Positions, distances, and luminosities are given for all satellites, along with references.

\begin{deluxetable*}{cccccc}
\tablecaption{MW Satellites\label{tab:mw_sats}}

\tablewidth{\textwidth}

\tablehead{ \multicolumn{6}{c}{MW, $M_*=5\times10^{10}$\msun}\\ \hline
\colhead{Name} & \colhead{RA} & \colhead{Dec}  & \colhead{$m-M$} & \colhead{$M_V$} & \colhead{Source} \\
\colhead{} & \colhead{deg} &\colhead{deg} & \colhead{mag} & \colhead{} & \colhead{}} 

\startdata
MW	&	--	&	--	&	--	&	-21.4	&	--	\\
LMC	&	05:23:34&	-69:45:22&	$18.477\pm0.026$	&	-18.1	&	1,12,1\\
SMC	&	00:52:44&	-72:49:43&	$18.91\pm0.1$	&	-16.7	&	1,13,1\\
Sgr	&	18:55:19&	-30:32:43&	$17.13\pm0.11$	&	-13.5	&	1,2,14\\
Fornax	&	02:39:59&	-34:26:57&	$20.72\pm0.04$	&	-13.5	&	1,3,1\\
Leo 1	&	10:08:28&	+12:18:23&	$22.15\pm0.1$	&	-11.8	&	1,4,14\\
Sculptor&	01:00:09&	-33:42:33&	$19.64\pm0.13$	&	-10.8	&	1,5,14\\
Leo 2	&	11:13:28&	+22:09:06&	$21.76\pm0.13$	&	-9.7	&	1,6,14\\
Sextans*	&	10:13:03&	-01:36:53&	$19.67\pm0.15$	&	-8.7	&	1,7,14\\
Ursa Minor&	15:09:08&	+67:13:21&	$19.40\pm0.11$	&	-9.0	&	1,8,1\\
Carina	&	06:41:36&	-50:57:58&	$20.08\pm0.08$	&	-9.4	&	1,9,14\\
Draco	&	17:20:12&	+57:54:55&	$19.49\pm0.17$	&	-8.7	&	1,10,14\\
CVn 1*	&	13:28:03&	+33:33:21 &	$21.62\pm0.05$	&	-8.8	&	1,11,14\\
\enddata
\tablecomments{Known satellites of the MW. Satellites marked with * are below the $\mu_{0,V}=26.5$ mag arcsec$^{-2}$ surface brightness limit we assume in throughout this work. Distance modulus is given relative to the sun (not the galactic center). Sources for position, distance, and luminosity (in order): sources 1-\citet{mcconnachie2012}, 2-\citet{hama2016}, 3-\citet{rizzi2007}, 4-\citet{stetson2014}, 5-\citet{mv2016,piet2008}, 6-\citet{bella2005,gull2008}, 7-\citet{mateo95}, 8-\citet{carrera2002,bella2002}, 9-\citet{coppola2015, vivas2013}, 10-\citet{bonanos2004, kin2008}, 11-\citet{kuehn2008}, 12-\citet{piet2019}, 13-\citet{hilditch2005}, 14-\citet{munoz2018}}

\end{deluxetable*}
\begin{deluxetable*}{cccccc}
\tablecaption{M31 Satellites\label{tab:m31_sats}}

\tablewidth{\textwidth}

\tablehead{ \multicolumn{6}{c}{M31, $M_*=10.3\times10^{10}$\msun}\\ \hline
\colhead{Name} & \colhead{RA} & \colhead{Dec}  & \colhead{$D_\odot$} & \colhead{$M_V$} & \colhead{Source} \\
\colhead{} & \colhead{deg} &\colhead{deg} & \colhead{Mpc} & \colhead{} & \colhead{}} 

\startdata
M31	&	00:42:44&	+41:16:09&	0.780		&		-22	&	1,2,3	\\
M33	&	01:33:50&	+30:39:37&	0.821		&		-18.8	&	1,2,1	\\
NGC 205	&	00:40:22&	+41:41:07&	0.824$\pm$0.027	&		-16.5	&	1,4,1	\\	
M32	&	00:42:41&	+40:51:55&	0.781$\pm$0.02	&		-16.3	&	1,5,1	\\	
NGC 147	&	00:33:12&	+48:30:32&	0.713		&		-15.8	&	1,2,1	\\
IC 10	&	00:20:17&	+59:18:14&	0.798$\pm$0.029	&		-15.0	&	1,6,1	\\
NGC 185	&	00:38:58&	+48:20:15&	0.619		&		-15.5	&	1,2,1	\\
AndVII	&	23:26:31&	+50:40:33&	0.763$\pm$0.035	&		-13.2	&	1,4,1	\\
AndXXXII&	00:35:59.4& 	+51:33:35&	$0.871^{0.018}_{0.016}$	&	-12.5	&	1,8,1	\\	
AndII	&	01:16:29.8&	+33:25:09&	0.679$\pm$0.040		&	-12.7	&	1,9,1	\\	
AndI	&	00:45:39.8&	+38:02:28&	0.791$\pm$0.050		&	-12.0	&	1,9,1	\\	
AndXXXI	&	22:58:16.3&	+41:17:28&	$0.794^{0.018}_{0.013}$	&	-11.8	&	1,8,1	\\	
AndIII	&	00:35:33.8&	+36:29:52&	0.745$\pm$0.039		&	-10.2	&	1,9,1	\\
AndXXIII*&	01:29:21.8 &	+38:43:8&	$0.809^{0.022}_{0.01}$	&	-9.9	&	1,8,1	\\	
AndVI	&	23:51:46.3&	+24:34:57&	0.783$\pm$0.025		&	-11.5	&	1,4,1	\\
AndXXI*	&	23:54:47.7&	+42:28:15&	$0.851^{0.019}_{0.011}$	&	-9.2	&	1,8,1	\\	
AndXXV*	&	00:30:8.9&	+46:51:7&	$0.832^{0.021}_{0.015}$	&	-9.2	&	1,8,1	\\
LGS3	&	01:3:55.0&	+21:53:6&	0.769$\pm$0.023		&	-10.1	&	1,4,1	\\
AndXV	&	01:14:18.7&	+38:7:3	&	0.766$\pm$0.042		&	-8.4	&	1,9,1	\\	
AndV	&	01:10:17.1 &	+47:37:41&	0.774$\pm$0.028		&	-9.5	&	1,4,1	\\
AndXIX*	&	00:19:32.1&	+35:2:37&	0.805			&	-10.1	&	1,2,1	\\
AndXIV*	&	00:51:35.0&	+29:41:49&	$0.847^{0.021}_{0.015}$	&	-8.8	&	1,8,1	\\	
AndXVII	&	00:37:7.0&	+44:19:20&	$0.866^{0.025}_{0.013}$	&	-8.1	&	1,8,1	\\
AndXXIX	&	23:58:55.6&	+30:45:20&	$0.820^{0.017}_{0.015}$	&	-8.5	&	1,8,1	\\	
AndIX*	&	00:52:53.0&	+43:11:45&	$0.769^{0.021}_{0.012}$	&	-8.8	&	1,8,1	\\
AndXXX	&	00:36:34.9&	+49:38:48&	$0.628^{0.016}_{0.015}$	&	-8.	&	1,8,1	\\	
AndXXIV	&	01:18:30.0&	+46:21:58&	$0.724^{0.099}_{0.081}$	&	-8.0	&	1,8,1	\\	
AndXXXIII&	03:1:23.6&	+40:59:18&	$0.755^{0.018}_{0.009}$	&	-10.2	&	1,8,1	\\	
AndXXVIII&	22:32:41.2&	+31:12:58&	0.769$\pm$0.038		&	-8.8	&	1,9,1	\\	
AndXVIII&	00:2:14.5&	+45:5:20&	$1.219^{0.029}_{0.013}$	&	-9.2	&	1,8,1	\\
\enddata
\tablecomments{Known satellites of M31 with $M_V<-8$. Satellites marked with * are below the $\mu_{0,V}=26.5$ mag arcsec$^{-2}$ surface brightness limit we assume throughout this work. For the distances without errorbars, \citet{conn2012} provides the entire distance posterior which is what is used. For these cases, the median distance is reported. Sources for position, distance, and luminosity (in order): 1-\citet{mcconnachie2018}, 2-\citet{conn2012}, 3-\citet{sick2015}, 4-\citet{mcconnachie2005}, 5-\citet{watkins2013, tonry2001, jensen2003, monachesi2011, sarajedini2012, fiorentino2012}, 6-\citet{sanna2008}, 7-\citet{richardson2011}, 8-\citet{weisz2019}, 9-\citet{mv2017}, }

\end{deluxetable*}
\begin{deluxetable}{cccccc}
\tablecaption{M81 Satellites\label{tab:m81_sats}}

\tablewidth{\textwidth}

\tablehead{ \multicolumn{6}{c}{M81, $M_*=5\times10^{10}$\msun}\\ \hline
\colhead{Name} & \colhead{RA} & \colhead{Dec}  & \colhead{$D_\odot$} & \colhead{$M_V$} & \colhead{Source} \\
\colhead{} & \colhead{deg} &\colhead{deg} & \colhead{Mpc} & \colhead{} & \colhead{}} 

\startdata
M81	&	09:55:33.2&	+69:03:55&	3.69	&	-21.1	&	1,3,2	\\
M82	&	09:55:52.4&	+69:40:47&	3.61	&	-19.75	&	1,3,2\\
NGC 3077&	10:03:19.1&	+68:44:02&	3.82	&	-17.93	&	1,3,2\\
NGC 2976&	09:47:15.5&	+67:54:59&	3.66	&	-17.83	&	1,3,2\\
IC 2574	&	10:28:23.6&	+68:24:43&	3.93	&	-17.19	&	1,3,2\\
DDO 82	&	10:30:36.58&	+70:37:06&	3.93	&	-15.06	&	1,3,2\\
KDG 61	&	9:57:02.7&	+68:35:30&	3.66	&	-13.4	&	1,3,4\\
BK5N	&	10:04:40.3&	+68:15:20&	3.70	&	-11.23	&	1,3,4\\
IKN	&	10:08:05.9&	+68:23:57&	3.75	&	-14.3	&	1,3,4\\
FM1	&	9:45:10.0&	+68:45:54&	3.78	&	-11.3	&	1,3,1\\
KDG 64	&	10:07:01.9&	+67:49:39&	3.75	&	-13.3	&	1,3,4\\
F8D1	&	09:44:47.1&	+67:26:19&	3.75	&	-12.8	&	1,3,1\\
d0944p69&	09:44:22.5&	+69:12:40&	3.84	&	-6.4	&	1,3,1\\
d1014p68*&	10:14:55.8&	+68:45:27&	3.84	&	-9.0	&	1,3,1\\
KK77	&	9:50:10.0&	+67:30:24&	3.80	&	-12.6	&	1,3,1\\
d1006p67&	10:06:46.2&	+67:12:04&	3.61	&	-9.4	&	1,3,1\\
d0939p71&	09:39:15.9&	+71:18:42&	3.65	&	-9.0	&	1,3,1\\
KDG 63	&	10:05:07.3&	+66:33:18&	3.65	&	-12.6	&	1,3,1\\
d0958p66&	09:58:48.5&	+66:50:59&	3.82	&	-12.8	&	1,3,1\\
ddo78	&	10:26:27.9&	+67:39:24&	3.48	&	-12.4	&	1,3,1\\
d1028p70&	10:28:39.7&	+70:14:01&	3.84	&	-12.0	&	1,3,1\\
d1015p69&	10:15:06.9&	+69:02:15&	4.07	&	-8.4	&	1,3,1\\
d0955p70*&	09:55:13.6&	+70:24:29&	3.45	&	-9.4	&	1,3,1\\
d1041p70&	10:41:16.8&	+70:09:03&	3.70	&	-8.9	&	1,3,1\\
HS117	&	10:21:25.2&	+71:06:58&	3.96	&	-11.7	&	1,3,1\\
d0944p71&	09:44:34.4&	+71:28:57&	3.47	&	-12	&	1,3,1\\
d1012p64&	10:12:48.4&	+64:06:27&	3.7	&	-12.9	&	1,3,1\\
d0926p70&	09:26:27.9&	+70:30:24&	3.4	&	-9.4	&	1,3,1\\
Ho1	&	09:40:32.3&	+71:11:11&	4.02	&	-14.2	&	1,3,1\\
BK6N	&	10:34:31.9&	+66:00:42&	3.31	&	-11.3	&	1,3,1\\
d0934p70&	09:34:03.7&	+70:12:57&	3.02	&	-9.0	&	1,3,1\\
\enddata
\tablecomments{Known satellites of M81. Satellites marked with * are below the $\mu_{0,V}=26.5$ mag arcsec$^{-2}$ surface brightness limit we assume in throughout this work. Sources for position, distance, and luminosity (in order): 1-\citet{chiboucas2013}, 2-\citet{galex2007}, 3-\citet{karachentsev}, 4-\citet{okamoto2019}}

\end{deluxetable}
\begin{deluxetable*}{cccccc}
\tablecaption{CenA Satellites\label{tab:cena_sats}}

\tablewidth{\textwidth}

\tablehead{ \multicolumn{6}{c}{CenA, $M_*=8\times10^{10}$\msun}\\ \hline
\colhead{Name} & \colhead{RA} & \colhead{Dec}  & \colhead{$D_\odot$} & \colhead{$M_V$} & \colhead{Source} \\
\colhead{} & \colhead{deg} &\colhead{deg} & \colhead{Mpc} & \colhead{} & \colhead{}} 

\startdata
CenA		&201.3650 & 	-43.0191&	3.77		&	-21.04  &	1,1,1 \\
KK189		&198.1875&	-41.8319&	4.23		&	-11.2&		1,1,1 \\
ESO269-066	&198.2875 &	-44.8900&	3.75		&	-14.1&		1,1,1 \\
NGC 5011C	&198.2958 &	-43.2656&	3.73		&	-13.9&		1,1,1 \\
CenA-Dw11	&199.4550 &	-42.9269&	$3.52\pm0.35$	&	-9.4	&	2,2,2 \\
CenA-Dw5	&199.9667 &	-41.9936&	$3.61\pm0.33$	&	-8.2	&	2,2,2 \\
KK196		&200.4458 &	-45.0633&	3.96		&	-12.5&		1,1,1 \\
KK197		&200.5042&	-42.5356&	3.84		&	-12.6&		1,1,1 \\
KKs55		&200.5500 &	-42.7308&	3.85		&	-12.4&		1,1,1 \\
CenA-Dw10*	&200.6214 &	-39.8839&	$3.27\pm0.44$	&	-7.8	&	2,2,2 \\
dw1322-39	&200.6558 &	-39.9084&	$2.95\pm0.05$	&	-10.0&		1,1,1 \\
CenA-Dw4	&200.7583 &	-41.7861&	$4.09\pm0.26$	&	-9.9	&	2,2,2 \\
dw1323-40b	&201.0000 &	-40.8367&	$3.91\pm0.6$	 &	-9.9	&	1,1,1 \\
dw1323-40	&201.2421 &	-40.7622&	$3.73\pm0.15$	&	-10.4&		1,1,1 \\
CenA-Dw6	&201.4875 &	-41.0942&	$4.04\pm0.20 $	&	-9.1	&	2,2,2 \\
CenA-Dw7	&201.6167 &	-43.5567&	$4.11\pm0.27 $	&	-9.9	&	2,2,2 \\
ESO324-024	&201.9042 &	-41.4806&	3.78		&	-15.5&		1,1,1 \\
KK203		&201.8667 &	-45.3525&	3.78		&	-10.5&		1,1,1 \\
dw1329-45	&202.3121 &	-45.1767&	$2.90\pm0.12$	&	-8.4	&	1,1,1 \\
CenA-Dw2	&202.4875 &	-41.8731&	$4.14\pm0.23$	&	-9.7	&	2,2,2 \\
CenA-Dw1	&202.5583  &	-41.8933&	$3.91\pm0.12$	&	-13.8&		2,2,2 \\
CenA-Dw3	&202.5875 &	-42.19255&	$3.88\pm0.16$	&	-13.1&		2,2,2 \\
CenA-Dw9*	&203.2542 &	-42.5300&	$3.81\pm0.36$	&	-9.1	&	2,2,2 \\
CenA-Dw8*	&203.3917 &	-41.6078&	$3.47\pm0.33$	&	-9.7	&	2,2,2 \\
dw1336-44	&204.2033 &	-43.8578&	$3.50\pm0.28$	&	-8.6	&	1,1,1 \\
NGC5237		&204.4083 &	-42.8475&	3.33		&	-15.3&		1,1,1 \\
KKs57		&205.4083 &	-42.5819&	3.83 		&	-10.6&		1,1,1 \\
dw1341-43	&205.4221 &	-44.4485&	$3.53\pm0.02$	&	-10.1&		1,1,1 \\
dw1342-43	&205.7029 &	-43.8561&	$2.90\pm0.14$	&	-9.8	&	1,1,1 \\
KK213		&205.8958 &	-43.7692&	3.77		&	-10.0&		1,1,1 \\
\enddata
\tablecomments{Known satellites of CenA. Satellites marked with * are below the $\mu_{0,V}=26.5$ mag arcsec$^{-2}$ surface brightness limit we assume in throughout this work. Sources for position, distance, and luminosity (in order): 1-\citet{muller2019}, 2-\citet{crnojevic2019} }

\end{deluxetable*}
\begin{deluxetable}{cccccc}
\tablecaption{M101 Satellites\label{tab:m101_sats}}

\tablewidth{\textwidth}

\tablehead{ \multicolumn{6}{c}{M101, $M_*=4\times10^{10}$\msun}\\ \hline
\colhead{Name} & \colhead{RA} & \colhead{Dec}  & \colhead{$D_\odot$} & \colhead{$M_V$} & \colhead{Source} \\
\colhead{} & \colhead{deg} &\colhead{deg} & \colhead{Mpc} & \colhead{} & \colhead{}} 

\startdata
M101		&14:03:12.5	&+54:20:56	&$6.52\pm0.19$		&-21.1		&1,3,1\\
NGC 5474	&14:05:01.6	&+53:39:44 	&$6.82\pm0.41$		&-18.24		&1,1,1\\
NGC 5477	&14:05:33.3	&+54:27:40	&$6.77\pm0.40$		&-15.37		&1,1,1\\
HolmIV		&13:54:45.7	&+53:54:03	&$6.93\pm0.48$		&-15.98		&1,1,1\\
DF1		&14:03:45.0	&+53:56:40	&$6.37\pm0.35$		&-11.5		&5,5,5\\
DF2		&14:08:37.5	&+54:19:31	&$6.87\pm0.26$		&-10.4		&5,5,5\\
DF3		&14:03:05.7	&+53:36:56	&$6.52\pm0.26$		&-10.1		&5,5,5\\
dwa		&14:06:49.9	&+53:44:30	&$6.83\pm0.27$		&-9.5		&2,4,4\\
dw9		&13:55:44.8	&+55:08:46	&$7.34\pm0.38$		&-8.2		&2,4,4\\
UGC8882		& 13:57:14.7 &	+54:06:03	&$7.0\pm0.5$		&-14.59		&6,6,6\\
\enddata
\tablecomments{Known satellites of M101. Sources for position, distance, and luminosity (in order): 1-\citet{t15}, 2-\citet{bennet2017}, 3-\citet{beaton2019}, 4-\citet{bennet2019}, 5-\citet{danieli101}, 6-Current work }

\end{deluxetable}
\begin{deluxetable}{cccccc}
\tablecaption{M94 Satellites\label{tab:m94_sats}}

\tablewidth{\textwidth}

\tablehead{ \multicolumn{6}{c}{M94, $M_*=3\times10^{10}$\msun}\\ \hline
\colhead{Name} & \colhead{RA} & \colhead{Dec}  & \colhead{$D_\odot$} & \colhead{$M_V$} & \colhead{Source} \\
\colhead{} & \colhead{deg} &\colhead{deg} & \colhead{Mpc} & \colhead{} & \colhead{}} 

\startdata
M94	&	12:50:53.1	&+41:07:13&	$4.2$		&	-19.95 	& 1,1,2 \\
dw1	&	12:55:02.5	&+40:35:22&	$4.1\pm0.2$	&	-10.1	& 3,3,3 \\
dw2	&	12:51:04.4	&+41:38:10&	$4.7\pm0.3$	&	-9.7	& 3,3,3 \\
\enddata
\tablecomments{Known satellites of M94. Sources for position, distance, and luminosity (in order): 1-\citet{karachentsev}, 2-\citet{galex2007}, 3-\citet{smercina2018} }

\end{deluxetable}

\section{Luminosity Function Checks with the ELVIS Simulation Suite}
\label{app:elvis}

\begin{figure*}
\includegraphics[width=\textwidth]{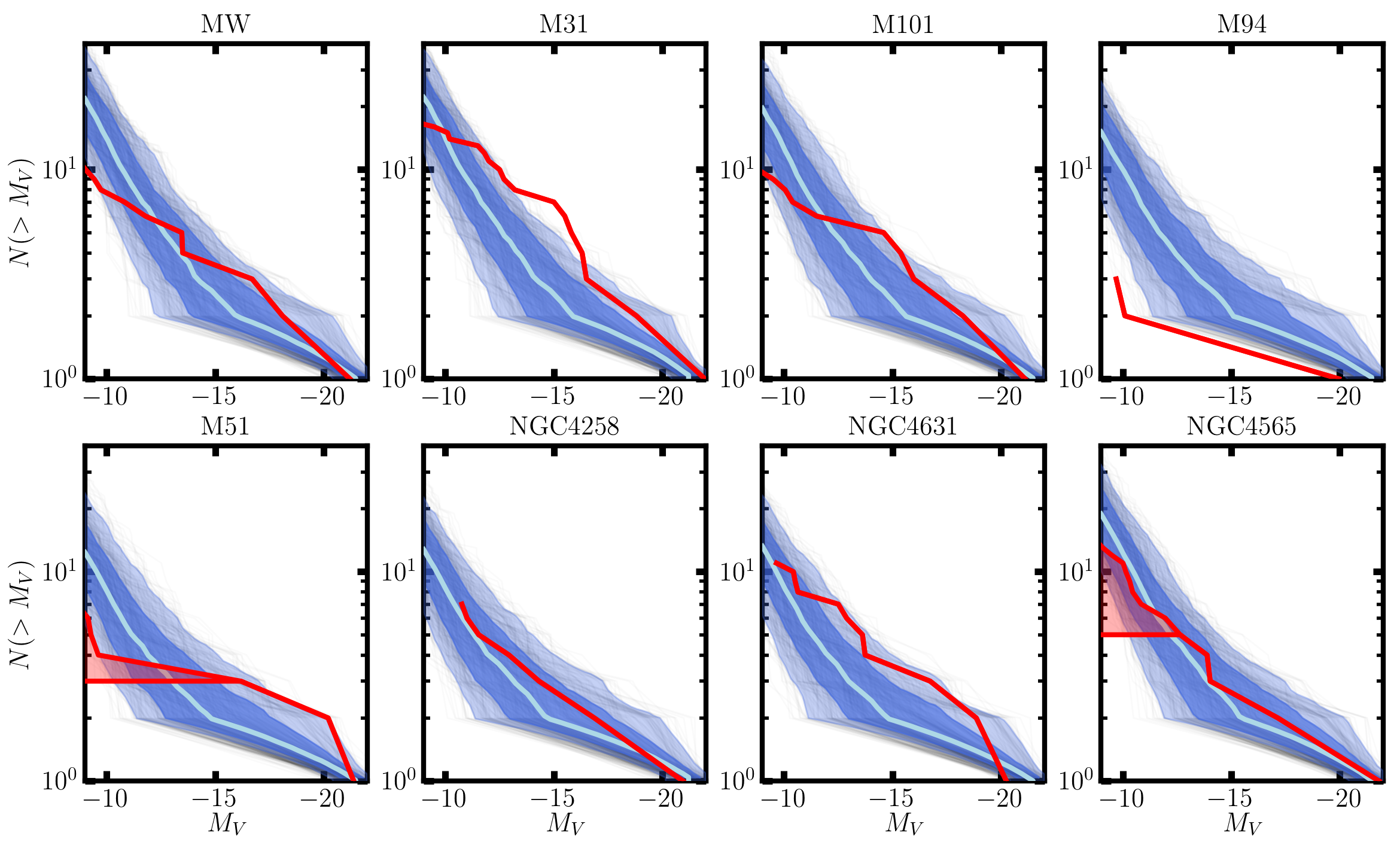}
\caption{The cumulative luminosity functions for the 8 'MW-sized' hosts that have been well surveyed for satellites (red). The thin black lines show the predicted LFs from the abundance matching model described in the text combined with the ELVIS zoom DMO simulation. The ELVIS hosts have halo mass in the range $1-3\times 10^{12}$ \msun. The blue regions show the $\pm 1,2 \sigma$ spread in the models. The luminosity completeness is different for each host but is  $M_V\lesssim-9$ in all cases. For each host, the model satellite systems have been forward modeled considering the survey area selection function for that specific host. }
\label{fig:mw_lf_comp_elvis}
\end{figure*}

Figure \ref{fig:mw_lf_comp_elvis} shows the analogous plot to Figure \ref{fig:mw_lf_comp} for the ELVIS \citep{elvis} zoom DMO simulation. The observed `MW-sized' host satellite systems are compared with those predicted by the ELVIS DMO simulations combined with our fiducial SHMR. We consider all ELVIS hosts here which are evenly distributed in halo mass between $1-3\times 10^{12}$ \msun. Note that this is quite different from the way we select TNG halos in the main text to have roughly the same stellar mass as the observed LV hosts. The thin black lines show the predicted LFs for the ELVIS hosts mock-observed from a random direction and forward modelled through the survey area selection function for a specific host, as described in \S\ref{sec:results}. For each of the 48 ELVIS hosts, 10 random directions are used. 

The ELVIS LFs are noticeably richer in satellites than the LFs predicted by TNG using the same SHMR. As explained in the main text this is because the ELVIS hosts are, on average, more massive than the TNG `MW-like' hosts. A secondary, smaller effect is that the ELVIS subhalos do not experience any disruption by a central disk while the TNG subhalos do.

\end{document}